\documentclass[aps,prd,showpacs,notitlepage,nofootinbib,preprintnumbers,amsmath,amssymb]{revtex4-1}

\usepackage{graphics,graphicx}
\usepackage{dcolumn}
\usepackage{bm}
\usepackage{epsfig}
\usepackage[usenames]{color}
\usepackage{hyperref} 
\usepackage{mathbbol}
\usepackage{epstopdf}
\usepackage{simplewick}
\usepackage[utf8]{inputenc} 
\usepackage{slashed}
\usepackage{stackrel}
\usepackage{mathrsfs}
\usepackage{xcolor}
\usepackage{cancel}
\usepackage{mathtools}
\usepackage{amsmath,amssymb,amsfonts,mathrsfs,bbold}
\usepackage[most]{tcolorbox}

\usepackage{physics}
\usepackage{subcaption}

\usepackage{ragged2e}
\DeclareCaptionJustification{justified}{\justifying}

\usepackage[normalem]{ulem}

\def\eq#1{{Eq.~(\ref{#1})}}
\def\fig#1{{Fig.~\ref{#1}}}
\newcommand{\ben}{\begin{eqnarray*}}
\newcommand{\een}{\end{eqnarray*}}
\newcommand{\un}[1]{\underline{#1}}

\newcommand{\pd}{\partial}

\newcommand{\thalf}{\tfrac{1}{2}}
\newcommand{\llangle}{\Big\langle \!\! \Big\langle}
\newcommand{\rrangle}{\Big\rangle \!\! \Big\rangle}

\newcommand{\as}{\alpha_s}

\newcommand{\dhd}{{\textstyle d}
\lower.03ex\hbox{\kern-0.38em$^{\scriptstyle-}$}\kern-0.05em{}}
\newcommand{\dbar}{{\textstyle \delta}
\lower.03ex\hbox{\kern-0.38em$^{\scriptstyle-}$}\kern-0.05em{}}

\newcommand{\ul}[1]{\underline{#1}}

\newcommand{\tord}{\textrm{T} \:}
\newcommand{\atord}{\bar{\textrm{T}} \:}

\makeatletter
\DeclareRobustCommand{\cev}[1]{%
  {\mathpalette\do@cev{#1}}%
}
\newcommand{\do@cev}[2]{%
  \vbox{\offinterlineskip
    \sbox\z@{$\m@th#1 x$}%
    \ialign{##\cr
      \hidewidth\reflectbox{$\m@th#1\vec{}\mkern4mu$}\hidewidth\cr
      \noalign{\kern-\ht\z@}
      $\m@th#1#2$\cr
    }%
  }%
}
\makeatother

\begin{document}

\title{T-odd Leading-Twist Quark TMDs at Small $x$}

\author{Yuri V. Kovchegov} 
         \email[Email: ]{kovchegov.1@osu.edu}
         \affiliation{Department of Physics, The Ohio State
           University, Columbus, OH 43210, USA}

\author{M. Gabriel Santiago}
  \email[Email: ]{gsantiago@sura.org}
	\affiliation{Center for Nuclear Femtography, SURA,
           1201 New York Ave. NW, Washington, DC 20005 USA}

\begin{abstract}
We study the small-$x$ asymptotics of the flavor non-singlet T-odd leading-twist quark transverse momentum dependent parton distributions (TMDs), the Sivers and Boer-Mulders functions. While the leading eikonal small-$x$ asymptotics of the quark Sivers function is given by the spin-dependent odderon \cite{Boer:2015pni,Dong:2018wsp}, we are interested in revisiting the sub-eikonal correction considered by us earlier in \cite{Kovchegov:2021iyc}. We first simplify the expressions for both TMDs at small Bjorken $x$ and then construct small-$x$ evolution equations for the resulting operators in the large-$N_c$ limit, with $N_c$ the number of quark colors. For both TMDs, the evolution equations resum all powers of the double-logarithmic parameter $\as \, \ln^2 (1/x)$, where $\as$ is the strong coupling constant, which is assumed to be small. Solving these evolution equations numerically (for the Sivers function) and analytically (for the Boer-Mulders function) we arrive at the following leading small-$x$ asymptotics of these TMDs at large $N_c$:
\begin{align}
f_{1 \: T}^{\perp \: NS} (x \ll 1 ,k_T^2) & =  C_O (x, k_T^2) \, \frac{1}{x} + C_1 (x, k_T^2) \, \left( \frac{1}{x} \right)^{3.4 \, \sqrt{\frac{\as \, N_c}{4 \pi}}} ,  \notag \\
 h_1^{\perp \, \textrm{NS}} (x \ll 1, k_T^2) & = C (x, k_T^2) \left( \frac{1}{x} \right)^{-1}. \notag
\end{align}
The functions $C_O (x, k_T^2)$, $C_1 (x, k_T^2)$, and $C (x, k_T^2)$ can be readily obtained in our formalism: they are mildly $x$-dependent and do not strongly affect the 
power-of-$x$ asymptotics shown above. The function $C_O$, along with the $1/x$ factor, arises from the odderon exchange. For the sub-eikonal contribution to the quark Sivers function (the term with $C_1$), our result shown above supersedes the one obtained in \cite{Kovchegov:2021iyc} due to the new contributions identified recently in \cite{Cougoulic:2022gbk}.
\end{abstract}

\pacs{12.38.-t, 12.38.Bx, 12.38.Cy}

\maketitle

\tableofcontents


\section{Introduction}

The problem of understanding the small Bjorken $x$ behavior of the transverse momentum dependent parton distribution functions (TMD PDFs or TMDs for short) received considerable attention in recent years \cite{Dominguez:2011wm, Dominguez:2011br, Kovchegov:2012ga,Kovchegov:2013cva,Zhou:2013gsa, Altinoluk:2014oxa, Boer:2015pni,Kovchegov:2015zha,Kovchegov:2015pbl, Dumitru:2015gaa, Szymanowski:2016mbq, Hatta:2016aoc, Hatta2016a,Hatta:2016khv,Boer:2016bfj,Balitsky:2016dgz, Kovchegov:2016zex, Kovchegov:2016weo, Kovchegov:2017jxc, Kovchegov:2017lsr, Dong:2018wsp, Kovchegov:2018zeq, Kovchegov:2018znm, Chirilli:2018kkw, Altinoluk:2019wyu,Kovchegov:2019rrz, Boussarie:2019icw, Cougoulic:2019aja, Kovchegov:2020hgb, Cougoulic:2020tbc, Altinoluk:2020oyd, Kovchegov:2020kxg, Chirilli:2021lif, Altinoluk:2021lvu, Adamiak:2021ppq, Kovchegov:2021lvz,Bondarenko:2021rbp,Abir:2021kma, Cougoulic:2022gbk}. The theoretical appeal of this problem is due to the fact that the small-$x$ asymptotics of TMDs cannot be obtained from the Collins-Soper-Sterman (CSS) evolution equations \cite{Collins:1981uw,Collins:1981uk,Collins:1981va,Collins:1984kg,Collins:1989gx}, which re-sum logarithms of the momentum scale $Q^2$ rather than logarithms of $x$. The formalism frequently employed to obtain the small-$x$ asymptotics of TMDs and related observables is the saturation/color glass condensate (CGC) framework \cite{Gribov:1984tu,Iancu:2003xm,Weigert:2005us,JalilianMarian:2005jf,Gelis:2010nm,Albacete:2014fwa,Kovchegov:2012mbw,Morreale:2021pnn}, in which the unpolarized TMDs, also known as unintegrated quark and gluon distributions, emerge quite naturally \cite{Jalilian-Marian:1997xn,Kovchegov:1998bi,Mueller:1999wm,Braun:2000bh,Kovchegov:2001sc,Kharzeev:2003wz,Hautmann:2007cx}. The saturation approach had to be extended to include sub-eikonal and sub-sub-eikonal operators in order to study spin-dependent TMDs at small $x$ \cite{Altinoluk:2014oxa,Kovchegov:2015pbl,Balitsky:2016dgz, Hatta:2016aoc, Kovchegov:2016zex, Kovchegov:2016weo, Kovchegov:2017jxc, Kovchegov:2017lsr, Kovchegov:2018znm,Chirilli:2018kkw, Altinoluk:2019wyu, Kovchegov:2019rrz, Cougoulic:2019aja, Kovchegov:2020hgb, Cougoulic:2020tbc, Altinoluk:2020oyd, Chirilli:2021lif, Adamiak:2021ppq, Altinoluk:2021lvu, Kovchegov:2021lvz, Kovchegov:2021iyc, Cougoulic:2022gbk}, opening up an exciting brand-new research direction of light-cone operator treatment (LCOT) for small-$x$ physics beyond the eikonal approximation. On the phenomenological side, the possibility of measuring various leading-twist TMDs at small $x$ at the future Electron-Ion Collider (EIC) \cite{Accardi:2012qut,Boer:2011fh,Proceedings:2020eah,AbdulKhalek:2021gbh} would allow us to test the predictions of the above-mentioned theoretical developments in the future by comparing the beyond-eikonal small-$x$ formalism with the data.     

The goal of this paper is to study the small-$x$ asymptotics of the T-odd leading-twist quark TMDs, the Sivers \cite{Sivers:1989cc,Sivers:1990fh} and Boer-Mulders \cite{Boer:1997nt} functions. The Sivers function at small $x$ has been extensively studied before \cite{Boer:2015pni,Szymanowski:2016mbq,Dong:2018wsp}, with the leading small-$x$ asymptotics shown to arise from the so-called spin-dependent odderon exchange \cite{Boer:2015pni,Szymanowski:2016mbq}. Our goal here is to revisit the sub-eikonal correction to the eikonal odderon contribution. (In our notation, the sub-eikonal terms are suppressed by one power of $x$ compared to the eikonal ones, the sub-sub-eikonal terms are suppressed by two powers of $x$, etc.) The sub-eikonal correction for the Sivers function has been recently calculated in \cite{Kovchegov:2021iyc}, employing the formalism developed in \cite{Kovchegov:2015pbl,Kovchegov:2017lsr, Kovchegov:2018znm, Kovchegov:2018zeq} for small-$x$ helicity and transversity evolution. However, the calculation in \cite{Kovchegov:2021iyc} did not include mixing of different types of sub-eikonal gluon operators and was, therefore, incomplete: here we will complete that calculation.

Since the calculation of  \cite{Kovchegov:2021iyc}, a new sub-eikonal operator was found to contribute to helicity evolution in \cite{Cougoulic:2022gbk}: this operator mixes with the evolution for other helicity operators, modifying the evolution equations and the resulting small-$x$ helicity asymptotics. More specifically, in the sub-eikonal gluon sector, the evolution for helicity TMDs and PDFs had been described by the $F^{12}$ component of the gluon field strength tensor \cite{Kovchegov:2015pbl,Kovchegov:2017lsr, Kovchegov:2018znm, Chirilli:2021lif}, which entered the calculation with the helicity-dependent prefactor $\sigma \, \delta_{\sigma, \sigma'}$ in the Brodsky-Lepage (BL) spinor basis \cite{Lepage:1980fj} for the projectile quark (and with a similar expression for the projectile gluon). However, mixing with another sub-eikonal gluon operator, $\cev{D}^i  \vec{D}^i$, made out of the right- and left-acting covariant derivatives $\vec{D}^i = \vec{\pd}^i - i g A^i$, $\cev{D}^i = \cev{\pd}^i + i g A^i$ with the transverse gluon field $A^i$, $i=1,2$, was neglected in \cite{Kovchegov:2015pbl,Kovchegov:2017lsr, Kovchegov:2018znm}, since this latter operator came in with a prefactor $\delta_{\sigma, \sigma'}$, which was (perhaps naively) interpreted as independent of helicity. As it turned out \cite{Cougoulic:2022gbk}, the small-$x$ evolution of the two operators ($F^{12}$ and $\cev{D}^i  \vec{D}^i$) mixes, such that the contribution of the $\cev{D}^i  \vec{D}^i$ operator is important for helicity evolution. In our recent calculation of the sub-eikonal correction to the quark Sivers function \cite{Kovchegov:2021iyc}, an opposite approximation was made: the evolution of the $\cev{D}^i  \vec{D}^i$ operator, that coupled to the Sivers function at the sub-eikonal order, was calculated without taking its mixing with the $F^{12}$ operator into account, resulting in incomplete evolution equations and asymptotics. In addition, the calculations of \cite{Kovchegov:2015pbl,Kovchegov:2017lsr, Kovchegov:2018znm, Kovchegov:2021iyc} were missing the sub-eikonal contribution to the propagator $\contraction{}{a^+ \:}{}{a^+} a^+ \:a^+$ of the ``quantum" gluon field $a^\mu$ in the shock wave, which is important for the evolution of the $\cev{D}^i  \vec{D}^i$ operator. Below we will reinstate the missing contributions in the sub-eikonal small-$x$ evolution of the flavor non-singlet Sivers function, obtaining novel double-logarithmic evolution equations which re-sum powers of $\as \, \ln^2 (1/x)$ with $\as$ the strong coupling constant. (See \cite{Kirschner:1983di,Kirschner:1994rq,Kirschner:1994vc,Bartels:1995iu,Bartels:1996wc,Griffiths:1999dj,Itakura:2003jp,Bartels:2003dj} for resummations of this parameter for other parton distributions and related  observables.) The equations are derived in the limit of the large number of colors $N_c$. By solving these equations numerically, we find the resulting sub-eikonal small-$x$ asymptotics of the flavor non-singlet Sivers function to be
\begin{align}\label{Sivers_asympt}
    f_{1 \: T}^{\perp \: NS} (x \ll 1 ,k_T^2) \bigg|_{\textrm{sub-eikonal}} \sim  \left( \frac{1}{x} \right)^{3.4 \, \sqrt{\frac{\as \, N_c}{4 \pi}}} .
\end{align}
The power of $1/x$ in \eq{Sivers_asympt} is quite large numerically (even for moderately low $\as \in [0.2 , 0.3]$) and is close to one, which is the exact power of $1/x$ in the eikonal odderon exchange, $\sim 1/x$  \cite{Bartels:1999yt,Kovchegov:2003dm,Kovchegov:2012rz,Caron-Huot:2013fea,Brower:2008cy,Avsar:2009hc,Brower:2014wha} (possibly to all orders in the coupling). This means that the $x$-dependence of the sub-eikonal correction \eqref{Sivers_asympt} is very close to that of the leading eikonal spin-dependent odderon $\sim 1/x$ contribution to the Sivers function \cite{Boer:2015pni,Szymanowski:2016mbq,Dong:2018wsp}: this, potentially, makes an experimental detection of the spin-dependent odderon rather difficult. 

We also apply our technique to determining the small-$x$ behavior of the flavor non-singlet Boer-Mulders function, which has not been studied yet in the literature to the best of our knowledge. It turns out that  the Boer-Mulders function is sub-sub-eikonal at small $x$, suppressed by two powers of $x$ compared to the eikonal TMDs and PDFs, akin to the transversity distribution \cite{Kirschner:1996jj,Kovchegov:2018zeq}. Similar to the case of Sivers function, we construct small-$x$ evolution equations in the double-logarithmic approximation (DLA) re-summing powers of $\as \, \ln^2 (1/x)$. Somewhat surprisingly, the solution of these equations yields no correction to the sub-sub-eikonal power of $x$, such that we obtain 
\begin{align}\label{BM_asympt}
    h_1^{\perp \, \textrm{NS}} (x \ll 1, k_T^2) \sim \left( \frac{1}{x} \right)^{-1 + {\cal O} (\as)}
\end{align}
for the flavor non-singlet Boer-Mulders function, no order-$\sqrt{\as}$ correction to the power.

The paper is structured as follows. We analyse the sub-eikonal contribution to the quark Sivers function in Sec.~\ref{sec:Sivers_all}, concentrating on the flavor non-singlet channel. Following the approach developed in \cite{Kovchegov:2015pbl,Kovchegov:2017lsr, Kovchegov:2018znm, Kovchegov:2021iyc, Cougoulic:2022gbk}, we first simplify the Sivers function at small $x$ and then construct the DLA evolution equations for the relevant operators (in the large-$N_c$ limit). Numerical solution of these equations yields the asymptotics in \eq{Sivers_asympt} above. The Boer-Mulders function is studied in Sec.~\ref{sec:evolution_BM} along the similar lines. An analytic solution of the obtained DLA evolution equations  results in the asymptotics \eqref{BM_asympt}. We conclude in Sec.~\ref{sec:conc} by summarizing our results and speculating that the leading small-$x$ asymptotics of T-odd TMDs may receive no corrections to the integer powers of $x$ to all orders in the strong coupling $\as$.  


\section{The Sivers Function}
\label{sec:Sivers_all}

\subsection{The Eikonal Sivers Function}
\label{sec:Sivers_eik}

We begin our discussion of the Sivers function by reviewing the eikonal contribution due to the spin-dependent odderon exchange constructed in \cite{Kovchegov:2021iyc} (see also \cite{Dong:2018wsp}). The eikonal contribution is given by the diagrams B and C from Fig.~5 of \cite{Kovchegov:2021iyc} (or by the same diagrams B and C in the classification of \cite{Kovchegov:2018znm}). The resulting standard linear combination of the unpolarized quark TMD $f_1^q$ and the quark Sivers function $f_{1 \: T}^{\perp \: q}$ is \cite{Kovchegov:2021iyc}
\begin{align}\label{ff1}
& \left[ f_1^q (x,k_T^2) -  \frac{\un{k} \times \underline{S}_P}{M_P} f_{1 \: T}^{\perp \: q} (x,k_T^2) \right]_\textrm{eikonal} = \frac{4 p_1^+}{(2 \pi)^3} \int d^2 {\zeta_{\perp}}  d^2 {w_{\perp}} \frac{d^2 {k_{1 \perp}} d {k_1^-}}{(2\pi)^3} e^{i (\underline{k}_1 + \underline{k}) \cdot (\un{w} - \un{\zeta})} \theta (k_1^-) \\
& \times \left\{ \frac{\un{k} \cdot \un{k}_1}{(x p_1^+ k_1^- + \underline{k}_1^2 ) (x p_1^+ k_1^- + \underline{k}^2)} \,  \Big{\langle} \tord \tr \left[ V_{\underline{\zeta}} \, V_{{\un w}}^\dagger \right] + \atord \tr \left[ V_{\underline{\zeta}} \, V_{{\un w}}^\dagger \right] \Big{\rangle} + \frac{\un{k}_1^2}{(x p_1^+ k_1^- + \underline{k}_1^2 )^2}  \, \Big{\langle} \tord \tr \left[ V_{\underline{\zeta}} \, V_{{\un w}}^\dagger \right]  \Big{\rangle}  \right\}. \notag
\end{align}
We are working in the frame where the proton is moving in the light-cone plus direction with the large $p_1^+$ momentum component. Our 4-vectors are $x^\mu = (x^+, x^-, {\un x})$ where $x^\pm = t \pm z$ and ${\un x} = (x^1, x^2) = (x, y)$. The magnitude of the transverse momentum vector is  labeled $k_T = |{\un k}|$. The cross-product is defined by ${\un u} \times {\un v} \equiv u^1 \, v^2 - u^2 \, v^1$.

The fundamental light-cone Wilson lines are denoted as 
\begin{align}\label{Vline}
V_{\un{x}} [x^-_f,x^-_i] = \mathcal{P} \exp \left[ \frac{ig}{2} \int\limits_{x^-_i}^{x^-_f} \dd{x}^- A^+ (0^+, x^-, \un{x}) \right]
\end{align}
with the infinite Wilson lines $V_{\un{x}} \equiv V_{\un{x}} [\infty,-\infty]$. Here $\mathcal{P}$ is the path-ordering operator, $A^{\mu} = \sum_a A^{a \, \mu} \, t^a$ is the background gluon field with the fundamental SU($N_c$) generators $t^a$, and $g$ is the strong coupling constant. The trace operators ``tr" in \eq{ff1} apply to the fundamental indices. Equation \eqref{ff1} also contains the time-ordering ($\tord$) and anti-time ordering ($\atord$) operators. The angle brackets $\langle \ldots \rangle$ denote the standard small-$x$ averaging, applied here in the transversely-polarized proton state \cite{McLerran:1993ni,McLerran:1993ka,McLerran:1994vd,Kovchegov:1996ty,Balitsky:1997mk,Balitsky:1998ya,Kovchegov:2019rrz}. The proton has mass $M_P$ and spin ${\un S}_P$. Throughout the paper, we will work in the $A^- =0$ light-cone gauge of the projectile. 

Below we will also employ the adjoint light-cone Wilson lines
\begin{align}\label{Uline}
U_{\un{x}} [x^-_f,x^-_i] = \mathcal{P} \exp \left[ \frac{ig}{2} \int\limits_{x^-_i}^{x^-_f} \dd{x}^- {\cal A}^+ (0^+, x^-, \un{x}) \right]
\end{align}
with $U_{\un{x}} \equiv U_{\un{x}} [\infty,-\infty]$. The gluon field ${\cal A}^{\mu} = \sum_a {A}^{a \, \mu} \, T^a$ is now in the adjoint representation with the adjoint SU($N_c$) generators $T^a$ defined by $(T^a)_{bc} = - i f^{abc}$. 

The Sivers function is extracted from the right-hand side of \eq{ff1} by identifying the terms which change sign under the ${\un k} \to - {\un k}$ replacement. While the calculation in \cite{Kovchegov:2021iyc} employed the quark Sivers function, here we would like to distinguish the flavor-singlet and non-singlet Sivers functions. To this end, we define these functions by
\begin{subequations}\label{Sivers_S_NS}
\begin{align}
&  f_{1 \: T}^{\perp \: S} (x,k_T^2) =  \sum_f \left[ f_{1 \: T}^{\perp \: q_f} (x,k_T^2) +  f_{1 \: T}^{\perp \: {\bar q}_f} (x,k_T^2) \right], \label{Sivers_S} \\
&  f_{1 \: T}^{\perp \: NS} (x,k_T^2) =  f_{1 \: T}^{\perp \: q} (x,k_T^2) -  f_{1 \: T}^{\perp \: {\bar q}} (x,k_T^2), \label{Sivers_NS}
\end{align}
\end{subequations}
with the same definition for the flavor-singlet and non-singlet unpolarized TMDs (and for the Boer-Mulders function below). Here $f_{1 \: T}^{\perp \: {\bar q}}$ is the anti-quark Sivers function. At the eikonal level, the anti-quark distribution can be obtained by interchanging $V \leftrightarrow V^\dagger$ in \eq{ff1}, 
\begin{align}\label{ff2}
& \left[ f_1^{\bar q} (x,k_T^2) -  \frac{\un{k} \times \underline{S}_P}{M_P} f_{1 \: T}^{\perp \: {\bar q}} (x,k_T^2) \right]_\textrm{eikonal} = \frac{4 p_1^+}{(2 \pi)^3} \int d^2 {\zeta_{\perp}}  d^2 {w_{\perp}} \frac{d^2 {k_{1 \perp}} d {k_1^-}}{(2\pi)^3} e^{i (\underline{k}_1 + \underline{k}) \cdot (\un{w} - \un{\zeta})} \theta (k_1^-) \\
& \times \left\{ \frac{\un{k} \cdot \un{k}_1}{(x p_1^+ k_1^- + \underline{k}_1^2 ) (x p_1^+ k_1^- + \underline{k}^2)} \,  \Big{\langle} \tord \tr \left[ V_{\underline{\zeta}}^\dagger \, V_{{\un w}} \right] + \atord \tr \left[ V_{\underline{\zeta}}^\dagger \, V_{{\un w}} \right] \Big{\rangle} + \frac{\un{k}_1^2}{(x p_1^+ k_1^- + \underline{k}_1^2 )^2}  \, \Big{\langle} \tord \tr \left[ V_{\underline{\zeta}}^\dagger \, V_{{\un w}} \right]  \Big{\rangle}  \right\}. \notag
\end{align}

Employing Eqs.~\eqref{ff1} and \eqref{ff2} in \eq{Sivers_S} we see that the flavor-singlet contribution is
\begin{align}\label{ff_S}
& \left[ f_1^S (x,k_T^2) -  \frac{\un{k} \times \underline{S}_P}{M_P} f_{1 \: T}^{\perp \: S} (x,k_T^2) \right]_\textrm{eikonal} = \frac{4 p_1^+ \, N_f}{(2 \pi)^3} \int d^2 {\zeta_{\perp}}  d^2 {w_{\perp}} \frac{d^2 {k_{1 \perp}} d {k_1^-}}{(2\pi)^3} e^{i (\underline{k}_1 + \underline{k}) \cdot (\un{w} - \un{\zeta})} \theta (k_1^-) \\
& \times \left\{ \frac{\un{k} \cdot \un{k}_1}{(x p_1^+ k_1^- + \underline{k}_1^2 ) (x p_1^+ k_1^- + \underline{k}^2)} \,  \Big{\langle} \tord \tr \left[ V_{\underline{\zeta}} \, V_{{\un w}}^\dagger \right] + \tord \tr \left[ V_{\underline{\zeta}}^\dagger \, V_{{\un w}} \right] + \atord \tr \left[ V_{\underline{\zeta}} \, V_{{\un w}}^\dagger \right] + \atord \tr \left[ V_{\underline{\zeta}}^\dagger \, V_{{\un w}} \right]  \Big{\rangle} \right. \notag \\ 
& \left. + \frac{\un{k}_1^2}{(x p_1^+ k_1^- + \underline{k}_1^2 )^2}  \, \Big{\langle} \tord \tr \left[ V_{\underline{\zeta}} \, V_{{\un w}}^\dagger \right] +  \tord \tr \left[ V_{\underline{\zeta}}^\dagger \, V_{{\un w}} \right]  \Big{\rangle}  \right\}, \notag
\end{align}
where $N_f$ is the number of quark flavors.
Since the right-hand side of \eq{ff_S} does not change the sign under ${\un k} \to - {\un k}$, it does not contribute to the Sivers function. Thus, the flavor-singlet Sivers function in this eikonal approximation is zero, 
\begin{align}\label{S0}
    f_{1 \: T}^{\perp \: S} (x,k_T^2) =0.
\end{align}

The flavor non-singlet part of Eqs.~\eqref{ff1} and \eqref{ff2} is obtained by using \eq{Sivers_NS}, which yields
\begin{align}\label{ff_NS}
& \left[ f_1^{NS} (x,k_T^2) -  \frac{\un{k} \times \underline{S}_P}{M_P} f_{1 \: T}^{\perp \: NS} (x,k_T^2) \right]_\textrm{eikonal} = \frac{4 p_1^+}{(2 \pi)^3} \int d^2 {\zeta_{\perp}}  d^2 {w_{\perp}} \frac{d^2 {k_{1 \perp}} d {k_1^-}}{(2\pi)^3} e^{i (\underline{k}_1 + \underline{k}) \cdot (\un{w} - \un{\zeta})} \theta (k_1^-) \\
& \times \left\{ \frac{\un{k} \cdot \un{k}_1}{(x p_1^+ k_1^- + \underline{k}_1^2 ) (x p_1^+ k_1^- + \underline{k}^2)} \,  \Big{\langle} \tord \tr \left[ V_{\underline{\zeta}} \, V_{{\un w}}^\dagger \right] - \tord \tr \left[ V_{\underline{\zeta}}^\dagger \, V_{{\un w}} \right] + \atord \tr \left[ V_{\underline{\zeta}} \, V_{{\un w}}^\dagger \right] - \atord \tr \left[ V_{\underline{\zeta}}^\dagger \, V_{{\un w}} \right]  \Big{\rangle} \right. \notag \\ 
& \left. + \frac{\un{k}_1^2}{(x p_1^+ k_1^- + \underline{k}_1^2 )^2}  \, \Big{\langle} \tord \tr \left[ V_{\underline{\zeta}} \, V_{{\un w}}^\dagger \right] -  \tord \tr \left[ V_{\underline{\zeta}}^\dagger \, V_{{\un w}} \right]  \Big{\rangle}  \right\}. \notag
\end{align}
The right-hand side of \eq{ff_NS} is anti-symmetric under ${\un k} \to - {\un k}$ and, therefore, only contributes to the Sivers function, such that  
\begin{align}\label{siv}
 - \frac{\un{k} \times \underline{S}_P}{M_P} f_{1 \: T}^{\perp \: NS} (x,k_T^2) \Big|_\textrm{eikonal} & = \frac{8 i \, N_c \, p_1^+}{(2 \pi)^3} \int d^2 {\zeta_{\perp}}  d^2 {w_{\perp}} \frac{d^2 {k_{1 \perp}} d{k_1^-}}{(2\pi)^3} e^{i (\underline{k}_1 + \underline{k}) \cdot (\un{w} - \un{\zeta})} \theta (k_1^-) \\
& \times \left[ \frac{2 \, \un{k} \cdot \un{k}_1}{(x p_1^+ k_1^- + \underline{k}_1^2 ) (x p_1^+ k_1^- + \underline{k}^2)} + \frac{\un{k}_1^2}{(x p_1^+ k_1^- + \underline{k}_1^2 )^2}  \right] \, \mathcal{O}_{\un{\zeta} \un{w}}, \notag
\end{align} 
where the odderon exchange dipole amplitude is defined by  \cite{Hatta:2005as,Kovchegov:2003dm}
\begin{align}\label{Odderon}
    \mathcal{O}_{\un{\zeta} \un{w}} = \frac{1}{2 i N_c} \Big{\langle}  \tord \tr [ V_{\underline{\zeta}}  V_{\underline{w}}^{\dagger }] - \tord \tr [ V_{\underline{w}}  V_{\underline{\zeta}}^{\dagger }] \Big{\rangle} .
\end{align}

Equation~\eqref{siv} gives us the contribution of the spin-dependent odderon \cite{Boer:2015pni,Szymanowski:2016mbq,Dong:2018wsp} to the flavor non-singlet Sivers function. The expression \eqref{siv} was shown to be non-zero in \cite{Kovchegov:2021iyc}. Comparing \eq{S0} to \eq{siv} we conclude that the spin-dependent odderon only contributes to the flavor non-singlet Sivers function, with the flavor-singlet Sivers function being zero at this eikonal order. The situation is completely opposite to the unpolarized or helicity distributions at small $x$, where the flavor-singlet distributions \cite{Jalilian-Marian:1997xn,Kovchegov:1998bi,Mueller:1999wm,Braun:2000bh,Kovchegov:2001sc,Kharzeev:2003wz,Bartels:1996wc,Cougoulic:2022gbk} typically dominate (in magnitude) over the flavor non-singlet ones \cite{Bartels:1995iu,Itakura:2003jp,Kovchegov:2016zex}: the small-$x$ flavor non-singlet Sivers function dominates over the flavor-singlet one.\footnote{The authors would like to thank Markus Diehl for pointing this feature out to them.} 

Below we will study the sub-eikonal correction to the flavor non-singlet Sivers function.


\subsection{The Sub-Eikonal Sivers Function}
\label{sec:Sivers}

\begin{figure}[h]
\centering
\includegraphics[width=0.5\linewidth]{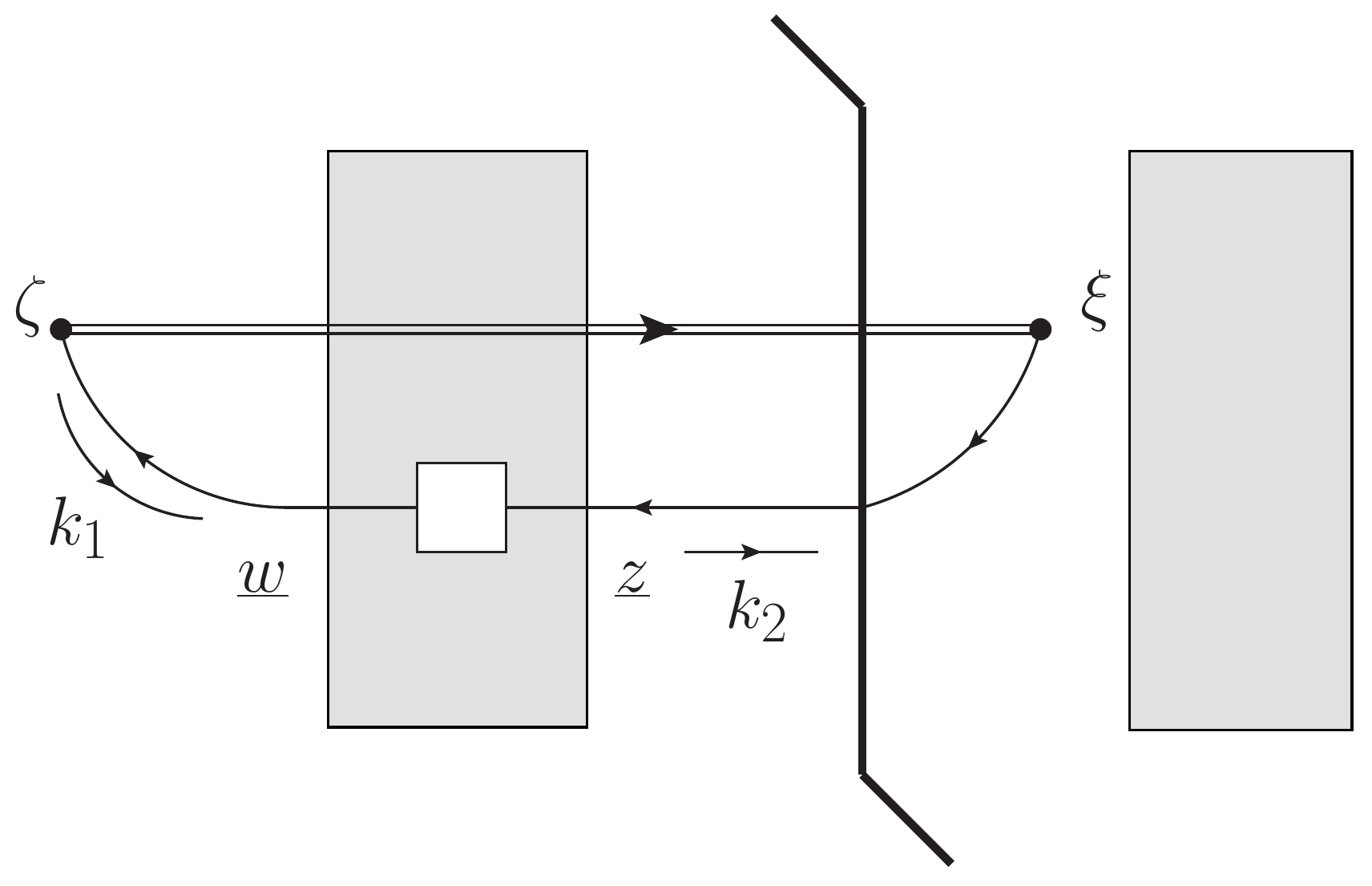}  
\caption{Diagram B with momentum and position variables shown in detail. The anti-quark with polarization $\chi_1$ propagates from the position $\zeta$ to $\un{w}$ with momentum $k_1$, undergoes a sub-eikonal interaction with the proton which changes its transverse position from $\un{w}$ on the left of the shock wave (the left shaded rectangle) to $\un{z}$ on the right of the shock wave and its polarization from $\chi_1$ to $\chi_2$. The anti-quark then propagates from $\un{z}$ to the position $\xi$ with momentum $k_2$. The shock wave is denoted by the shaded rectangle, while the sub-eikonal interaction with the shock wave is denoted by the white box.}
\label{FIG:diagbdet}
\end{figure}

To calculate the sub-eikonal correction to the quark Sivers function we employ the analysis carried out earlier in \cite{Kovchegov:2018znm,Kovchegov:2021iyc}. The sub-eikonal contribution comes only from the diagram B (and its complex conjugate) in the classification of  \cite{Kovchegov:2018znm,Kovchegov:2021iyc}, which is shown in detail in \fig{FIG:diagbdet}.

Following \cite{Kovchegov:2021iyc}, we start with Eq.~(46) in that reference, which gives the contribution of the diagram B and its complex conjugate in the notation introduced in \cite{Kovchegov:2018znm}, 
\begin{align}
\label{sivwsum}
B + \mbox{c.c.} & \, = \frac{2 p_1^+}{2 (2 \pi)^3} \sum_{\bar{q}} \int\limits_{-\infty}^0 d{\zeta^-} \int\limits_0^{\infty} d{\xi^-} \int d^2 {\zeta_{\perp}}  d^2 {\xi_{\perp}} e^{i k \cdot (\zeta - \xi)} \Big[\frac{\gamma^+}{2} \Big]_{\alpha \beta} \Big{\langle} \bar{\psi}_{\alpha} (\xi) V_{\underline{\xi}} [\xi^-,\infty] | \bar{q} \rangle \langle \bar{q} | V_{\underline{\zeta}} [\infty, \zeta^-] \psi_{\beta} (\zeta) \Big{\rangle} + c.c. \notag \\
=& -\frac{2 p_1^+}{2 (2 \pi)^3} \int d^2 {\zeta_{\perp}}  d^2 {w_{\perp}} \frac{d^2 {k_{1 \perp}} d{k_1^-}}{(2\pi)^3} e^{i (\underline{k}_1 + \underline{k}) \cdot (\un{w} - \un{\zeta})} \theta (k_1^-) \, \frac{1}{(x p_1^+ k_1^- + \underline{k}_1^2 ) (x p_1^+ k_1^- + \underline{k}^2)}  \notag \\
\times & \sum_{\chi_1 , \chi_2} \bar{v}_{\chi_2} (k_2) \frac{\gamma^+}{2} v_{\chi_1}(k_1)  \, \Big{\langle} \tord V_{\underline{\zeta}}^{ij} \, \bar{v}_{\chi_1} (k_1) \left( \hat{V}_{{\un w}}^\dagger \right)^{ji}  v_{\chi_2} (k_2) \Big{\rangle} \Bigg|_{k_2^- = k_1^-, k_1^2 =0, k_2^2 =0, \un{k}_2 = - \un{k}} + \mbox{c.c.}.
\end{align}
The momentum and position variables employed in \eq{sivwsum} are detailed in \fig{FIG:diagbdet}.

We are working with the spinors in the transverse spin basis \cite{Kovchegov:2012ga}
\begin{align}\label{chi_def}
u_\chi \equiv \frac{1}{\sqrt{2}} \, \left[ u_+ + \chi \, u_- \right], \ \ \ v_\chi \equiv \frac{1}{\sqrt{2}} \, \left[ v_+ + \chi \, v_- \right],
\end{align}
where $\chi = \pm 1$ and  the helicity-basis $\pm$-reversed Brodsky-Lepage (BL) spinors \cite{Lepage:1980fj} are \cite{Kovchegov:2018znm,Kovchegov:2018zeq}
\begin{align}\label{anti-BLspinors}
u_\sigma (p) = \frac{1}{\sqrt{p^-}} \, [p^- + m \, \gamma^0 +  \gamma^0 \, {\un \gamma} \cdot {\un p} ] \,  \rho (\sigma), \ \ \ v_\sigma (p) = \frac{1}{\sqrt{p^-}} \, [p^- - m \, \gamma^0 +  \gamma^0 \, {\un \gamma} \cdot {\un p} ] \,  \rho (-\sigma),
\end{align}
with $p^\mu = \left( \frac{{\un p}^2+ m^2}{p^-}, p^-, {\un p} \right)$ and
\begin{align}
  \rho (+1) \, = \, \frac{1}{\sqrt{2}} \, \left(
  \begin{array}{c}
      1 \\ 0 \\ -1 \\ 0
  \end{array}
\right), \ \ \ \rho (-1) \, = \, \frac{1}{\sqrt{2}} \, \left(
  \begin{array}{c}
        0 \\ 1 \\ 0 \\ 1
  \end{array}
\right) .
\end{align}
In what follows below, except for Appendix~\ref{sec:app_mass}, we assume the quarks to be massless, $m=0$. 

Equation \eqref{sivwsum} was written in \cite{Kovchegov:2018znm} for the forward scattering on the target shock wave which does not change the transverse plane position of the anti-quark. To account for recoil associated with the change of this position we replace the interaction with the shock wave by \cite{Kovchegov:2021iyc,Cougoulic:2022gbk} (for massless quarks)
\begin{align}\label{V_repl}
\bar{v}_{\chi_1} (k_1) \left( \hat{V}_{{\un w}}^\dagger \right)^{ji}  v_{\chi_2} (k_2)  \to 2 \sqrt{k_1^- \, k_2^-}  \,  \int d^2 z_\perp \, \left( V^{\textrm{pol} \, \dagger}_{{\un z}, {\un w}; \chi_2 , \chi_1} \right)^{ji} 
\end{align}
along with $e^{i \un{k} \cdot (\un{w} - \un{\zeta})} \to e^{i \un{k} \cdot (\un{z} - \un{\zeta})}$ in \eq{sivwsum}. Employing 
\begin{align}\label{vG+v+}
\bar{v}_{\chi_2} (k_2) \frac{\gamma^+}{2} v_{\chi_1}(k_1) 
= \frac{1}{ \sqrt{k_1^- k_2^-}} \big[ \delta_{\chi_1,\chi_2}  \, \underline{k}_1 \cdot \underline{k}_2 - i  \, \delta_{\chi_1, -\chi_2} \, \underline{k}_2 \times \underline{k}_1 \big] 
\end{align}
we get the following sub-eikonal  contribution to the quark Sivers function
\begin{align}
\label{sivwsum2}
 - \frac{\un{k} \times \underline{S}_P}{M_P}   f_{1 \: T}^{\perp \: q} (x,k_T^2) \Big|_\textrm{sub-eikonal} & 
\subset   \frac{2 p_1^+}{(2 \pi)^3} \int d^2 {\zeta_{\perp}}  d^2 {w_{\perp}} d^2 z_\perp \frac{d^2 {k_{1 \perp}} d{k_1^-}}{(2\pi)^3}  \theta (k_1^-) \, \frac{e^{i \underline{k}_1 \cdot (\un{w} - \un{\zeta}) + i \underline{k} \cdot (\un{z} - \un{\zeta})}}{(x p_1^+ k_1^- + \underline{k}_1^2 ) (x p_1^+ k_1^- + \underline{k}^2)}  \notag \\
\times & \sum_{\chi_1 , \chi_2} \big[ \delta_{\chi_1,\chi_2}  \, \underline{k}_1 \cdot \underline{k} {\color{red}{-}} i  \, \delta_{\chi_1, -\chi_2} \, \underline{k} \times \underline{k}_1 \big]   \, \Big{\langle} \tord  \tr \left[ V_{\underline{\zeta}} \, V^{\textrm{pol} \, \dagger}_{{\un z}, {\un w}; \chi_2 , \chi_1} \right] \Big{\rangle} + \mbox{c.c.} .
\end{align}

Here the object $V^{\textrm{pol}}_{{\un z}, {\un w}; \chi_2 , \chi_1}$ is the $S$-matrix of a massless quark scattering on background gluon and quark fields at the sub-eikonal order, which was calculated in \cite{Balitsky:2015qba,Altinoluk:2020oyd,Kovchegov:2021iyc,Kovchegov:2018znm,Chirilli:2018kkw} to be
\begin{align}\label{Vphase&mag}
V_{\un{x}, \underline{y}; \chi', \chi}^{\textrm{pol}} = \delta_{\chi, \chi'} \, V_{\un{x}, \un{y}}^{\textrm{phase}} + \delta_{\chi, - \chi'} \, V_{\un{x}, \un{y}}^{\textrm{mag}}
\end{align}
where 
\begin{subequations}\label{V_subeik_1}
\begin{align}
& V_{\un{x}, \un{y}}^{\textrm{phase}} = - \frac{i \, p_1^+}{2 \, s}  \int\limits_{-\infty}^{\infty} d{z}^- d^2 z \ V_{\un{x}} [ \infty, z^-] \, \delta^2 (\un{x} - \un{z}) \, \cev{D}_z^i  \vec{D}^i_z \, V_{\un{y}} [ z^-, -\infty] \, \delta^2 (\un{y} - \un{z}) \\ 
& - \frac{g^2 \, p_1^+}{4 \, s} \, \delta^2 (\un{x} - \un{y})  \, \int\limits_{-\infty}^{\infty} d{z}_1^- \int\limits_{z_1^-}^\infty d z_2^-  \ V_{\un{x}} [ \infty, z_2^-] \,  t^b \, \psi_{\beta} (z_2^-,\un{x}) \, U_{\un{x}}^{ba} [z_2^-,z_1^-] \, \left[ \frac{\gamma^+}{2}  \right]_{\alpha \beta} \bar{\psi}_\alpha (z_1^-,\un{x}) \, t^a \, V_{\un{x}} [ z_1^-, -\infty] , \notag \\
& V_{\un{x}, \underline{y}}^{\textrm{mag}} = \frac{i \, g \, p_1^+}{2 \, s} \delta^2 (\un{x} - \un{y}) \,  \int\limits_{-\infty}^{\infty} d{z}^- \ V_{\un{x}} [ \infty, z^-] \, F^{12} \, V_{\un{x}} [ z^-, -\infty] \label{V_subeik_1b} \\ 
& - \frac{g^2 \, p_1^+}{4 \, s} \, \delta^2 (\un{x} - \un{y})  \, \int\limits_{-\infty}^{\infty} d{z}_1^- \int\limits_{z_1^-}^\infty d z_2^-  \ V_{\un{x}} [ \infty, z_2^-] \,  t^b \, \psi_{\beta} (z_2^-,\un{x}) \, U_{\un{x}}^{ba} [z_2^-,z_1^-] \, \left[ \frac{\gamma^+ \gamma^5}{2} \right]_{\alpha \beta} \bar{\psi}_\alpha (z_1^-,\un{x}) \, t^a \, V_{\un{x}} [ z_1^-, -\infty] . \notag 
\end{align}
\end{subequations}
Here $\psi, {\bar \psi}$ are the background quark and anti-quark fields. Note that the sign of the last term in \eq{V_subeik_1b} is different from that in \cite{Kovchegov:2018znm,Kovchegov:2021iyc}, but is in agreement with \cite{Kovchegov:2018zeq,Cougoulic:2022gbk}, due to the sign error in the two former references. Once more, our right- and left-acting fundamental covariant derivatives are defined by
$\vec{D}^i = \vec{\pd}_i - i g A^i, \cev{D}^i = \cev{\pd}_i + i g A^i$. The center of mass energy squared for the quark--target scattering is denoted by $s$. 

Using Eqs.~\eqref{Vphase&mag} and \eqref{V_subeik_1} in \eq{sivwsum2} and integrating over $\un z$ yields
\begin{align}
\label{sivwsum3}
& - \frac{\un{k} \times \underline{S}_P}{M_P}   f_{1 \: T}^{\perp \: q} (x,k_T^2) \Big|_\textrm{sub-eikonal} 
\subset   \frac{4 p_1^+}{(2 \pi)^3} \int d^2 {\zeta_{\perp}}  d^2 {w_{\perp}} \frac{d^2 {k_{1 \perp}} d{k_1^-}}{(2\pi)^3}  \theta (k_1^-) \, \frac{e^{i (\un{k} + \underline{k}_1) \cdot (\un{w} - \un{\zeta}) }}{(x p_1^+ k_1^- + \underline{k}_1^2 ) (x p_1^+ k_1^- + \underline{k}^2)}  \notag \\
\times & \left[ \underline{k}_1 \cdot \underline{k} \, \Big{\langle} \tord  \tr \left[ V_{\underline{\zeta}} \, V^{\textrm{phase} \, \dagger}_{{\un w}; \un{k}, \un{k}_1} \right] + \atord \tr \left[ V_{\underline{w}}^\dagger \, V^{\textrm{phase}}_{{\un \zeta}; \un{k}, \un{k}_1} \right] \Big{\rangle} -  i  \, \underline{k} \times \underline{k}_1 \, \Big{\langle} \tord  \tr \left[ V_{\underline{\zeta}} \, V^{\textrm{mag} \, \dagger}_{{\un w}} \right] - \atord \tr  \left[ V_{\underline{w}}^\dagger \, V^{\textrm{mag}}_{{\un \zeta}} \right] \Big{\rangle}  \right] ,
\end{align}
where now
\begin{subequations}\label{V_subeik_2}
\begin{align}
& V_{\un{x}; \un{k}, \un{k}_1}^{\textrm{phase}} = \frac{i \, p_1^+}{8 \, s}  \int\limits_{-\infty}^{\infty} d{z}^- \ V_{\un{x}} [ \infty, z^-] \, \left[ \vec{D}^i_x -  \cev{D}_x^i + i k_1^i - i k^i \right]^2  \, V_{\un{x}} [ z^-, -\infty]  \label{V_subeik_2a} \\ 
& - \frac{g^2 \, p_1^+}{4 \, s} \, \int\limits_{-\infty}^{\infty} d{z}_1^- \int\limits_{z_1^-}^\infty d z_2^-  \ V_{\un{x}} [ \infty, z_2^-] \,  t^b \, \psi_{\beta} (z_2^-,\un{x}) \, U_{\un{x}}^{ba} [z_2^-,z_1^-] \, \left[ \frac{\gamma^+}{2}  \right]_{\alpha \beta} \bar{\psi}_\alpha (z_1^-,\un{x}) \, t^a \, V_{\un{x}} [ z_1^-, -\infty] , \notag \\
& V_{\un{x}}^{\textrm{mag}} = \frac{i \, g \, p_1^+}{2 \, s}  \int\limits_{-\infty}^{\infty} d{z}^- \ V_{\un{x}} [ \infty, z^-] \, F^{12} \, V_{\un{x}} [ z^-, -\infty] \\ 
& - \frac{g^2 \, p_1^+}{4 \, s} \, \int\limits_{-\infty}^{\infty} d{z}_1^- \int\limits_{z_1^-}^\infty d z_2^-  \ V_{\un{x}} [ \infty, z_2^-] \,  t^b \, \psi_{\beta} (z_2^-,\un{x}) \, U_{\un{x}}^{ba} [z_2^-,z_1^-] \, \left[ \frac{\gamma^+ \gamma^5}{2} \right]_{\alpha \beta} \bar{\psi}_\alpha (z_1^-,\un{x}) \, t^a \, V_{\un{x}} [ z_1^-, -\infty] . \notag 
\end{align}
\end{subequations}

Further, rewriting 
\begin{align}\label{Vkk1}
V_{\un{x}; \un{k}, \un{k}_1}^{\textrm{phase}} = (k -k_1)^i \, V_{\un{x}}^i + V_{\un{x}; \un{k}, \un{k}_1}^{[2]}
\end{align}
with
\begin{subequations}\label{ViV2}
\begin{align}
& V_{\un{x}}^i = \frac{p_1^+}{4 \, s}  \int\limits_{-\infty}^{\infty} d{z}^- \ V_{\un{x}} [ \infty, z^-] \, \left[ \vec{D}^i_x -  \cev{D}_x^i \right]  \, V_{\un{x}} [ z^-, -\infty], \label{ViV2a} \\
& V_{\un{x}; \un{k}, \un{k}_1}^{[2]} = \frac{i \, p_1^+}{8 \, s}  \int\limits_{-\infty}^{\infty} d{z}^- \ V_{\un{x}} [ \infty, z^-] \, \left[ (\vec{D}^i_x -  \cev{D}_x^i)^2 - (k_1^i -  k^i)^2 \right]  \, V_{\un{x}} [ z^-, -\infty]  \label{ViV2b} \\ 
& - \frac{g^2 \, p_1^+}{4 \, s} \, \int\limits_{-\infty}^{\infty} d{z}_1^- \int\limits_{z_1^-}^\infty d z_2^-  \ V_{\un{x}} [ \infty, z_2^-] \,  t^b \, \psi_{\beta} (z_2^-,\un{x}) \, U_{\un{x}}^{ba} [z_2^-,z_1^-] \, \left[ \frac{\gamma^+}{2}  \right]_{\alpha \beta} \bar{\psi}_\alpha (z_1^-,\un{x}) \, t^a \, V_{\un{x}} [ z_1^-, -\infty] , \notag
\end{align}
\end{subequations}
we recast \eq{sivwsum3} as
\begin{align}
\label{sivwsum4}
& - \frac{\un{k} \times \underline{S}_P}{M_P}   f_{1 \: T}^{\perp \: q} (x,k_T^2) \Big|_\textrm{sub-eikonal} 
\subset   \frac{4 p_1^+}{(2 \pi)^3} \int d^2 {\zeta_{\perp}}  d^2 {w_{\perp}} \frac{d^2 {k_{1 \perp}} d{k_1^-}}{(2\pi)^3}  \theta (k_1^-) \, \frac{e^{i (\un{k} + \underline{k}_1) \cdot (\un{w} - \un{\zeta}) }}{(x p_1^+ k_1^- + \underline{k}_1^2 ) (x p_1^+ k_1^- + \underline{k}^2)}  \notag \\
\times & \left[ \underline{k}_1 \cdot \underline{k} \, (k -k_1)^i \,  \Big{\langle} \tord  \tr \left[ V_{\underline{\zeta}} \, V^{i \, \dagger}_{{\un w}} \right] + \atord \tr \left[ V_{\underline{w}}^\dagger \, V^{i}_{{\un \zeta}} \right] \Big{\rangle} + \underline{k}_1 \cdot \underline{k} \, \Big{\langle} \tord  \tr \left[ V_{\underline{\zeta}} \, V^{[2] \, \dagger}_{{\un w}; \un{k}, \un{k}_1} \right] + \atord \tr \left[ V_{\underline{w}}^\dagger \, V^{[2]}_{{\un \zeta}; \un{k}, \un{k}_1} \right] \Big{\rangle} \right. \notag \\ 
& \left. -  i  \, \underline{k} \times \underline{k}_1 \, \Big{\langle} \tord  \tr \left[ V_{\underline{\zeta}} \, V^{\textrm{mag} \, \dagger}_{{\un w}} \right] - \atord \tr  \left[ V_{\underline{w}}^\dagger \, V^{\textrm{mag}}_{{\un \zeta}} \right] \Big{\rangle}  \right] .
\end{align}
Equation \eqref{sivwsum4} should be compared to Eqs.~(76) and (79) in \cite{Kovchegov:2021iyc}: the latter are missing the contribution of $V^\textrm{mag}$. As mentioned before, this contribution was neglected in \cite{Kovchegov:2021iyc} due to it being incorrectly understood in that reference (and, earlier, in \cite{Kovchegov:2015pbl,Kovchegov:2017lsr, Kovchegov:2018znm}) as a helicity-only dependent operator, which should not couple to the unpolarized quark entering the Sivers function. We see from \eq{sivwsum4} that this operator indeed does enter the expression for the quark Sivers function. Additionally, Eqs.~\eqref{ViV2} should be compared to Eqs.~(77) in  \cite{Kovchegov:2021iyc}: in the latter only terms linear in $\un A$ were retained, and all other terms were neglected. As we will see below (cf. \cite{Cougoulic:2022gbk}), the $\un A$-independent parts of the operators in Eqs.~\eqref{ViV2} also contribute to the sub-eikonal evolution we will obtain. (Note that the $\un A$-dependent part of $V_{\un{x}}^i$ in \eq{ViV2a} is different by a minus sign from that in Eq.~(77a) of  \cite{Kovchegov:2021iyc}: this is just a difference in conventions.)

Let us repeat the calculation for the anti-quarks. For the (forward version of the) diagram B from \fig{FIG:diagbdet} (plus its complex conjugate) with the anti-quark line replaced by a quark line we write
\begin{align}
\label{sivwsum_qbar}
B + \mbox{c.c.} & \, = -\frac{2 p_1^+}{2 (2 \pi)^3} \int d^2 {\zeta_{\perp}}  d^2 {w_{\perp}} \frac{d^2 {k_{1 \perp}} d{k_1^-}}{(2\pi)^3} e^{i (\underline{k}_1 + \underline{k}) \cdot (\un{w} - \un{\zeta})} \theta (k_1^-) \, \frac{1}{(x p_1^+ k_1^- + \underline{k}_1^2 ) (x p_1^+ k_1^- + \underline{k}^2)}  \notag \\
\times & \sum_{\chi_1 , \chi_2} \bar{u}_{\chi_1} (k_1) \frac{\gamma^+}{2} u_{\chi_2}(k_2)  \, \Big{\langle} \tord V_{\underline{\zeta}}^{\dagger \, ij} \, \bar{u}_{\chi_2} (k_2) \left( \hat{V}_{{\un w}} \right)^{ji}  u_{\chi_1} (k_1) \Big{\rangle} \Bigg|_{k_2^- = k_1^-, k_1^2 =0, k_2^2 =0, \un{k}_2 = - \un{k}} + \mbox{c.c.}
\end{align}
and replace 
\begin{align}
\bar{u}_{\chi_2} (k_2) \left( \hat{V}_{{\un w}} \right)^{ji}  u_{\chi_1} (k_1)  \to 2 \sqrt{k_1^- \, k_2^-}  \, \int d^2 z_\perp \, \left( V^{\textrm{pol}}_{{\un z}, {\un w}; \chi_2 , \chi_1} \right)^{ji} 
\end{align}
along with $e^{i \un{k} \cdot (\un{w} - \un{\zeta})} \to e^{i \un{k} \cdot (\un{z} - \un{\zeta})}$. Similar to the above, employing (again, for massless quarks)
\begin{align}\label{vG+v}
\bar{u}_{\chi_1} (k_1) \frac{\gamma^+}{2} u_{\chi_2}(k_2) 
=  \frac{1}{ \sqrt{k_1^- k_2^-}} \big[ \delta_{\chi_1,\chi_2}  \, \underline{k}_1 \cdot \underline{k}_2 + i  \, \delta_{\chi_1, -\chi_2} \, \underline{k}_1 \times \underline{k}_2 \big] 
\end{align}
we arrive at the sub-eikonal contribution to the anti-quark Sivers function 
\begin{align}
\label{sivwsum2_qbar}
 - \frac{\un{k} \times \underline{S}_P}{M_P}   f_{1 \: T}^{\perp \: {\bar q}} (x,k_T^2) \Big|_\textrm{sub-eikonal} & 
\subset   \frac{2 p_1^+}{(2 \pi)^3} \int d^2 {\zeta_{\perp}}  d^2 {w_{\perp}} d^2 z_\perp \frac{d^2 {k_{1 \perp}} d{k_1^-}}{(2\pi)^3}  \theta (k_1^-) \, \frac{e^{i \underline{k}_1 \cdot (\un{w} - \un{\zeta}) + i \underline{k} \cdot (\un{z} - \un{\zeta})}}{(x p_1^+ k_1^- + \underline{k}_1^2 ) (x p_1^+ k_1^- + \underline{k}^2)}  \notag \\
\times & \sum_{\chi_1 , \chi_2} \big[ \delta_{\chi_1,\chi_2}  \, \underline{k}_1 \cdot \underline{k} - i  \, \delta_{\chi_1, -\chi_2} \, \underline{k} \times \underline{k}_1 \big]   \, \Big{\langle} \tord  \tr \left[ V^{\textrm{pol}}_{{\un z}, {\un w}; \chi_2 , \chi_1} \, V_{\underline{\zeta}}^\dagger \right] \Big{\rangle} + \mbox{c.c.} .
\end{align}
Employing Eqs.~\eqref{Vphase&mag}, \eqref{V_subeik_1}, \eqref{V_subeik_2}, and \eqref{Vkk1} we integrate over $\un z$ obtaining
\begin{align}
\label{sivwsum4_qbar}
& - \frac{\un{k} \times \underline{S}_P}{M_P}   f_{1 \: T}^{\perp \: {\bar q}} (x,k_T^2) \Big|_\textrm{sub-eikonal} 
\subset   \frac{4 p_1^+}{(2 \pi)^3} \int d^2 {\zeta_{\perp}}  d^2 {w_{\perp}} \frac{d^2 {k_{1 \perp}} d{k_1^-}}{(2\pi)^3}  \theta (k_1^-) \, \frac{e^{i (\un{k} + \underline{k}_1) \cdot (\un{w} - \un{\zeta}) }}{(x p_1^+ k_1^- + \underline{k}_1^2 ) (x p_1^+ k_1^- + \underline{k}^2)}  \notag \\
\times & \left[ \underline{k}_1 \cdot \underline{k} \, (k -k_1)^i \,  \Big{\langle} \tord  \tr \left[ V^{i}_{{\un w}} \, V_{\underline{\zeta}}^\dagger \right] + \atord \tr \left[ V^{i  \, \dagger}_{{\un \zeta}} \, V_{\underline{w}} \right] \Big{\rangle} + \underline{k}_1 \cdot \underline{k} \, \Big{\langle} \tord  \tr \left[ V^{[2]}_{{\un w}; \un{k}, \un{k}_1} \, V_{\underline{\zeta}}^\dagger \right] + \atord \tr \left[ V^{[2] \, \dagger}_{{\un \zeta}; \un{k}, \un{k}_1} \, V_{\underline{w}}\right] \Big{\rangle} \right. \notag \\ 
& \left. - i  \, \underline{k} \times \underline{k}_1 \, \Big{\langle} \tord  \tr \left[V^{\textrm{mag} }_{{\un w}} \, V_{\underline{\zeta}}^\dagger \right] - \atord \tr  \left[ V^{\textrm{mag} \, \dagger}_{{\un \zeta}} \, V_{\underline{w}} \right] \Big{\rangle}  \right] .
\end{align}

Using Eqs.~\eqref{sivwsum4} and \eqref{sivwsum4_qbar} in Eqs.~\eqref{Sivers_S_NS} and taking the small-$x$ limit we obtain the expressions for the flavor-singlet and non-singlet sub-eikonal quark Sivers functions, 
\begin{subequations}\label{F_S&NS1}
\begin{align}
& - \frac{\un{k} \times \underline{S}_P}{M_P}   f_{1 \: T}^{\perp \: S} (x,k_T^2) \Big|_\textrm{sub-eikonal} 
\subset   \frac{4 p_1^+}{(2 \pi)^3} \sum_f \int d^2 {\zeta_{\perp}}  d^2 {w_{\perp}} \frac{d^2 {k_{1 \perp}} d{k_1^-}}{(2\pi)^3}  \theta (k_1^-) \, \frac{e^{i (\un{k} + \underline{k}_1) \cdot (\un{w} - \un{\zeta}) }}{\underline{k}_1^2 \,  \underline{k}^2}  \notag \\
& \times \left[ \underline{k}_1 \cdot \underline{k} \, (k -k_1)^i \,  \Big{\langle} \tord  \tr \left[ V_{\underline{\zeta}} \, V^{i \, \dagger}_{{\un w}} \right] + \atord \tr \left[ V_{\underline{w}}^\dagger \, V^{i}_{{\un \zeta}} \right]  + \tord  \tr \left[ V^{i}_{{\un w}} \, V_{\underline{\zeta}}^\dagger \right] + \atord \tr \left[ V^{i  \, \dagger}_{{\un \zeta}} \, V_{\underline{w}} \right] \Big{\rangle} \right. \notag \\ 
& \left. + \underline{k}_1 \cdot \underline{k} \, \Big{\langle} \tord  \tr \left[ V_{\underline{\zeta}} \, V^{[2] \, \dagger}_{{\un w}; \un{k}, \un{k}_1} \right] + \atord \tr \left[ V_{\underline{w}}^\dagger \, V^{[2]}_{{\un \zeta}; \un{k}, \un{k}_1} \right]  + \tord  \tr \left[ V^{[2]}_{{\un w}; \un{k}, \un{k}_1} \, V_{\underline{\zeta}}^\dagger \right] + \atord \tr \left[ V^{[2] \, \dagger}_{{\un \zeta}; \un{k}, \un{k}_1} \, V_{\underline{w}}\right] \Big{\rangle} \right. \notag \\ 
& \left. -  i  \, \underline{k} \times \underline{k}_1 \, \Big{\langle} \tord  \tr \left[ V_{\underline{\zeta}} \, V^{\textrm{mag} \, \dagger}_{{\un w}} \right] - \atord \tr  \left[ V_{\underline{w}}^\dagger \, V^{\textrm{mag}}_{{\un \zeta}} \right] +  \tord  \tr \left[V^{\textrm{mag} }_{{\un w}} \, V_{\underline{\zeta}}^\dagger \right] -  \atord \tr  \left[ V^{\textrm{mag} \, \dagger}_{{\un \zeta}} \, V_{\underline{w}} \right] \Big{\rangle}  \right] , \\
& - \frac{\un{k} \times \underline{S}_P}{M_P}   f_{1 \: T}^{\perp \: NS} (x,k_T^2) \Big|_\textrm{sub-eikonal} 
\subset   \frac{4 p_1^+}{(2 \pi)^3} \int d^2 {\zeta_{\perp}}  d^2 {w_{\perp}} \frac{d^2 {k_{1 \perp}} d{k_1^-}}{(2\pi)^3}  \theta (k_1^-) \, \frac{e^{i (\un{k} + \underline{k}_1) \cdot (\un{w} - \un{\zeta}) }}{\underline{k}_1^2 \, \underline{k}^2}  \notag \\
& \times  \left[ \underline{k}_1 \cdot \underline{k} \, (k -k_1)^i \,  \Big{\langle} \tord  \tr \left[ V_{\underline{\zeta}} \, V^{i \, \dagger}_{{\un w}} \right] + \atord \tr \left[ V_{\underline{w}}^\dagger \, V^{i}_{{\un \zeta}} \right]  - \tord  \tr \left[ V^{i}_{{\un w}} \, V_{\underline{\zeta}}^\dagger \right] - \atord \tr \left[ V^{i  \, \dagger}_{{\un \zeta}} \, V_{\underline{w}} \right] \Big{\rangle} \right. \notag \\ 
& \left. + \underline{k}_1 \cdot \underline{k} \, \Big{\langle} \tord  \tr \left[ V_{\underline{\zeta}} \, V^{[2] \, \dagger}_{{\un w}; \un{k}, \un{k}_1} \right] + \atord \tr \left[ V_{\underline{w}}^\dagger \, V^{[2]}_{{\un \zeta}; \un{k}, \un{k}_1} \right]  - \tord  \tr \left[ V^{[2]}_{{\un w}; \un{k}, \un{k}_1} \, V_{\underline{\zeta}}^\dagger \right] - \atord \tr \left[ V^{[2] \, \dagger}_{{\un \zeta}; \un{k}, \un{k}_1} \, V_{\underline{w}}\right] \Big{\rangle} \right. \notag \\ 
& \left. -  i  \, \underline{k} \times \underline{k}_1 \, \Big{\langle} \tord  \tr \left[ V_{\underline{\zeta}} \, V^{\textrm{mag} \, \dagger}_{{\un w}} \right] - \atord \tr  \left[ V_{\underline{w}}^\dagger \, V^{\textrm{mag}}_{{\un \zeta}} \right] -  \tord  \tr \left[V^{\textrm{mag} }_{{\un w}} \, V_{\underline{\zeta}}^\dagger \right] +  \atord \tr  \left[ V^{\textrm{mag} \, \dagger}_{{\un \zeta}} \, V_{\underline{w}} \right] \Big{\rangle}  \right] .
\end{align}
\end{subequations}
The expressions \eqref{F_S&NS1} are inclusive, in the sense that they are not equalities: just as in the eikonal case, only the ${\un k} \to - {\un k}$ anti-symmetric parts of their right-hand sides contribute to the Sivers functions-containing expression on the left. Unlike the eikonal case considered above, for which all flavors' contributions were identical such that we replaced $\sum_f \to N_f$ in arriving at \eq{ff_S}, at the sub-eikonal level different flavors may have different initial conditions, making the corresponding correlators different (see \cite{Adamiak:2021ppq}). Therefore, we explicitly keep the sum over flavors in Eqs.~\eqref{F_S&NS1}.

Define the new scattering amplitudes
\begin{subequations}\label{FS_defs}
\begin{align}
& F^{S \, i}_{ {\un w}, {\un \zeta}} (z) = \frac{1}{2 N_c} \, \sum_f \, \mbox{Re} \, \llangle \tord  \tr \left[ V_{\underline{\zeta}} \, V^{i \, \dagger}_{{\un w}} \right] + \tord  \tr \left[ V^{i}_{{\un w}} \, V_{\underline{\zeta}}^\dagger \right]  \rrangle, \\
& F^{S \, [2]}_{ {\un w}, {\un \zeta}} (z) = \frac{1}{2 N_c} \, \sum_f \, \mbox{Im} \, \llangle \tord  \tr \left[ V_{\underline{\zeta}} \, V^{[2] \, \dagger}_{{\un w}; \un{k}, \un{k}_1} \right] + \tord  \tr \left[ V^{[2]}_{{\un w}; \un{k}, \un{k}_1} \, V_{\underline{\zeta}}^\dagger \right]  \rrangle , \\ 
& F^{S \, \textrm{mag}}_{{\un w}, {\un \zeta}} (z) = \frac{1}{2 N_c} \, \sum_f \, \mbox{Re} \, \llangle \tord  \tr \left[ V_{\underline{\zeta}} \, V^{\textrm{mag} \, \dagger}_{{\un w}} \right] +  \tord  \tr \left[V^{\textrm{mag} }_{{\un w}} \, V_{\underline{\zeta}}^\dagger \right]  \rrangle ,
\end{align}
\end{subequations}
and
\begin{subequations}\label{FNS_defs}
\begin{align}
& F^{NS \, i}_{ {\un w}, {\un \zeta}} (z) = \frac{1}{2 N_c} \, \mbox{Re} \, \llangle \tord  \tr \left[ V_{\underline{\zeta}} \, V^{i \, \dagger}_{{\un w}} \right] - \tord  \tr \left[ V^{i}_{{\un w}} \, V_{\underline{\zeta}}^\dagger \right]  \rrangle, \\
& F^{NS \, [2]}_{ {\un w}, {\un \zeta}} (z) = \frac{1}{2 N_c} \, \mbox{Im} \, \llangle \tord  \tr \left[ V_{\underline{\zeta}} \, V^{[2] \, \dagger}_{{\un w}; \un{k}, \un{k}_1} \right] - \tord  \tr \left[ V^{[2]}_{{\un w}; \un{k}, \un{k}_1} \, V_{\underline{\zeta}}^\dagger \right]  \rrangle , \label{F2NS_def} \\ 
& F^{NS \, \textrm{mag}}_{{\un w}, {\un \zeta}} (z) = \frac{1}{2 N_c} \, \mbox{Re} \, \llangle \tord  \tr \left[ V_{\underline{\zeta}} \, V^{\textrm{mag} \, \dagger}_{{\un w}} \right] -  \tord  \tr \left[V^{\textrm{mag} }_{{\un w}} \, V_{\underline{\zeta}}^\dagger \right]  \rrangle ,
\end{align}
\end{subequations}
with the energy-rescaled averaging in the proton state defined by \cite{Kovchegov:2015pbl}
\begin{align}\label{sub_eik_braket}
\llangle \ldots \rrangle \equiv z s \, \Big\langle \ldots \Big\rangle = p_1^+ k_1^-  \, \Big\langle \ldots \Big\rangle
\end{align}
and with $z = k_1^- /p_2^-$ the fraction of some projectile's momentum $p_2^-$ carried by the polarized (sub-eikonal) line in the dipole and $s = p_1^+ \, p_2^-$. 

To study the properties of the newly-defined amplitudes, let us remember that we are looking for their contributions to the Sivers functions which describe a transversely polarized target proton. This means that the amplitudes should depend on ${\un S}_P$, the transverse spin of the proton. In addition, the structure of Eqs.~\eqref{F_S&NS1} dictates that $F^i$ and $F^{[2]}$ should contain the two-dimensional Levi-Civita symbol $\epsilon^{ij}$ ($\epsilon^{12} = - \epsilon^{21} = 1$, $\epsilon^{11} = \epsilon^{22} =0$), while $F^\textrm{mag}$ should not, since $\epsilon^{ij}$ is present on the left-hand side of Eqs.~\eqref{F_S&NS1} in $\un{k} \times \underline{S}_P = \epsilon^{ij} k^i S_P^j$. We thus conclude, both for the flavor singlet and non-singlet amplitudes, that the impact-parameter integrated amplitudes contributing to the Sivers function(s) should be
\begin{subequations}\label{decomp}
\begin{align}
& \int d^2 b_\perp F^i_{10} = \epsilon^{ij} \, S_P^j \, x_{10}^2 \, F_A (x_{10}^2, z) + x_{10}^i \, {\un x}_{10} \times {\un S}_P \, F_B (x_{10}^2, z) +  \epsilon^{ij} \, x_{10}^j \, {\un x}_{10} \cdot {\un S}_P \, F_C (x_{10}^2, z), \label{decompA} \\
& \int d^2 b_\perp F^{[2]}_{10} = {\un x}_{10} \times {\un S}_P \, F^{[2]} (x_{10}^2, z), \label{decompF2} \\
& \int d^2 b_\perp F^\textrm{mag}_{10} = {\un x}_{10} \cdot {\un S}_P \, F_\textrm{mag} (x_{10}^2, z).
\end{align}
\end{subequations}
Here ${\un b} = ({\un x}_1 + {\un x}_0)/2$ and the integrals over $\un b$ are performed while keeping the dipole separation ${\un x}_{10} \equiv {\un x}_1 - {\un x}_0$ fixed. (Here and below $x_{10} = |{\un x}_{10}|$.)  We also use an abbreviated notation where $F_{10} = F_{{\un x}_1, {\un x}_0}$ for all dipole amplitudes. Equations \eqref{decomp} define the new amplitudes $F_A, F_B, F_C, F^{[2]}$ and $F_\textrm{mag}$. An important consequence of Eqs.~\eqref{decomp} is that the impact parameter integral of $F^i_{10}$ is an even function of ${\un x}_{10}$, while the impact parameter integrals of $F^{[2]}_{10}$ and $F^\textrm{mag}_{10}$ are odd functions of ${\un x}_{10}$.

Since the relations \eqref{decomp} are valid before we apply the Re and Im operations in the definitions \eqref{FS_defs} and \eqref{FNS_defs}, we can rewrite Eqs.~\eqref{F_S&NS1} as
\begin{subequations}\label{F_S&NS2}
\begin{align}
& - \frac{\un{k} \times \underline{S}_P}{M_P}   f_{1 \: T}^{\perp \: S} (x,k_T^2) \Big|_\textrm{sub-eikonal} 
=   \frac{16 N_c}{(2 \pi)^3} \int d^2 x_{10}   \, \frac{d^2 {k_{1 \perp}} }{(2\pi)^3}  \, \frac{e^{i (\un{k} + \underline{k}_1) \cdot \un{x}_{10} }}{\underline{k}_1^2 \,  \underline{k}^2}  \int\limits_\frac{\Lambda^2}{s}^1 \frac{dz}{z} \notag \\
& \times  \left\{ \underline{k}_1 \cdot \underline{k} \, (k -k_1)^i \, \left[ \epsilon^{ij} \, S_P^j \, x_{10}^2 \, F_A^S (x_{10}^2, z) + x_{10}^i \, {\un x}_{10} \times {\un S}_P \, F_B^S (x_{10}^2, z) +  \epsilon^{ij} \, x_{10}^j \, {\un x}_{10} \cdot {\un S}_P \, F_C^S (x_{10}^2, z) \right] \right. \notag \\ 
& \left. + i \, \underline{k}_1 \cdot \underline{k} \ {\un x}_{10} \times {\un S}_P \, F^{S \, [2]} (x_{10}^2, z) -  i \, \underline{k} \times \underline{k}_1 \, {\un x}_{10} \cdot {\un S}_P \, F^{S}_\textrm{mag} (x_{10}^2, z) \right\} , \\
& - \frac{\un{k} \times \underline{S}_P}{M_P}   f_{1 \: T}^{\perp \: NS} (x,k_T^2) \Big|_\textrm{sub-eikonal} 
=   \frac{16 N_c}{(2 \pi)^3} \int d^2 x_{10}  \,  \frac{d^2 {k_{1 \perp}} }{(2\pi)^3}  \, \frac{e^{i (\un{k} + \underline{k}_1) \cdot {\un x}_{10}}}{\underline{k}_1^2 \, \underline{k}^2}  \int\limits_\frac{\Lambda^2}{s}^1 \frac{dz}{z} \notag \\
& \times  \left\{ \underline{k}_1 \cdot \underline{k} \, (k -k_1)^i \, \left[ \epsilon^{ij} \, S_P^j \, x_{10}^2 \, F_A^{NS} (x_{10}^2, z) + x_{10}^i \, {\un x}_{10} \times {\un S}_P \, F_B^{NS} (x_{10}^2, z) +  \epsilon^{ij} \, x_{10}^j \, {\un x}_{10} \cdot {\un S}_P \, F_C^{NS} (x_{10}^2, z) \right] \right. \notag \\ 
& \left. + i \, \underline{k}_1 \cdot \underline{k} \ {\un x}_{10} \times {\un S}_P \, F^{NS \, [2]} (x_{10}^2, z) -  i \, \underline{k} \times \underline{k}_1 \, {\un x}_{10} \cdot {\un S}_P \, F^{NS}_\textrm{mag} (x_{10}^2, z) \right\}. \label{F_S&NS2b}
\end{align}
\end{subequations}
Here $\Lambda$ is the scale characterizing the target (proton), and may be taken to be an infrared (IR) cutoff. The essential difference of Eqs.~\eqref{F_S&NS2}, as compared to Eqs.~(82) and (108) in \cite{Kovchegov:2021iyc}, is in the presence of the amplitudes $F_B$, $F_C$ and $F_\textrm{mag}$. Inspired by the lowest non-trivial order initial conditions for the amplitude $F^i$, the decomposition \eqref{decompA} was assumed in \cite{Kovchegov:2021iyc} to contain only $F_A$ in Eq.~(108) of \cite{Kovchegov:2021iyc}. These omissions are corrected in the Eqs.~\eqref{F_S&NS2} above. 

We see that both the flavor-singlet and non-singlet Sivers functions depend on five different amplitudes, $F_A, F_B, F_C$, $F^{[2]}$, and $F_\textrm{mag}$. To construct the small-$x$ asymptotics of the Sivers function, we need to derive evolution equations for all these five amplitudes, and solve them. We are going to do this next.


\subsection{Evolution of the Flavor Non-Singlet Sivers Function}
\label{sec:evolution}

In preparation for taking the large-$N_c$ limit,  let us consider only the gluon contribution to small-$x$ evolution of the flavor non-singlet amplitudes $F^i_{10} (z)$, $F^{\textrm{mag}}_{10} (z)$ and $F^{[2]}_{10} (z)$. In this Subsection we construct the large-$N_c$ DLA evolution equations for $F^i_{10} (z)$ and $F^{\textrm{mag}}_{10} (z)$, that is, for $F^{NS}_A, F^{NS}_B, F^{NS}_C$, and $F^{NS}_\textrm{mag}$. These equations do not mix with $F^{[2]}_{10} (z)$. We will then show that the evolution for $F^{NS \, [2]}_{10} (z)$ decouples and does not mix with the other amplitudes: combined with zero initial conditions, this will allow us to conclude that $F^{NS \, [2]}_{10} (z) =0$, in agreement with  \cite{Kovchegov:2021iyc}.

Keeping only the gluon parts of the operators, we write
\begin{subequations}\label{Vi+mag}
\begin{align}
& V_{\un{x}}^i = \frac{p_1^+}{4 \, s}  \int\limits_{-\infty}^{\infty} d{z}^- \ V_{\un{x}} [ \infty, z^-] \, \left[ \vec{D}^i_x -  \cev{D}_x^i \right]  \, V_{\un{x}} [ z^-, -\infty], \label{Vi+maga} \\
& V_{\un{x}}^{\textrm{mag}} = \frac{i \, g \, p_1^+}{2 \, s}  \,  \int\limits_{-\infty}^{\infty} d{z}^- \ V_{\un{x}} [ \infty, z^-] \, F^{12} \, V_{\un{x}} [ z^-, -\infty]  . \label{Vi+magb}
\end{align}
\end{subequations}
Incidentally, these are the same operators as for (the gluon sector of) helicity evolution \cite{Cougoulic:2022gbk}. Therefore, we can utilize the recently-constructed evolution equations from \cite{Cougoulic:2022gbk} to derive the evolution of the operators \eqref{Vi+mag}. Let us note once again that we will be working in $A^- =0$ light-cone gauge.

\subsubsection{Evolution for $V^i$}

We first rewrite the operator in \eq{Vi+maga} as 
\begin{align}\label{Vi2}
V_{\un{z}}^{i} & \equiv \frac{p_1^+}{4 s} \, \int\limits_{-\infty}^{\infty} d {z}^- \, V_{\un{z}} [ \infty, z^-] \, \left[ {D}^i (z^-, \un{z}) - \cev{D}^i (z^-, \un{z}) \right]  \, V_{\un{z}} [ z^-, -\infty]  \notag \\ & = - i g \, \frac{p_1^+}{2 s} \, \int\limits_{-\infty}^{\infty} d {z}^- \, V_{\un{z}} [ \infty, z^-] \, \left[ \frac{1}{2} \, z^- {\pd}^i A^+ (z^-, \un{z}) + A^i (z^-, \un{z}) \right] \, V_{\un{z}} [ z^-, -\infty].
\end{align}
As usual for the background field method \cite{Abbott:1980hw,Abbott:1981ke,Balitsky:1995ub}, to evolve an operator we rewrite the gluon field as a sum of a (classical) background field and a quantum field, $A^\mu = A^\mu_\textrm{background} + a^\mu$, and integrate out the quantum field $a^\mu$. To accomplish the latter we will need the sub-eikonal propagator in the shock wave \cite{Cougoulic:2022gbk}
\begin{align}\label{aaa}
& \frac{1}{4} \int\limits_{-\infty}^0 dx_{2'}^-  
\int\limits_0^\infty dx_2^- \left\{ \Big[ \frac{x_{2'}^-}{2} {\pd}^i a^{+ \, a} (x_{2'}^- , \ul{x}_1)
\contraction[2ex]
{}
{+}{a^{i \, a}
(x_{2'}^- , \ul{x}_1) \Big]}
{a}
+ a^{i \, a} (x_{2'}^- , \ul{x}_1) \Big]
a^{+ \, b} (x_2^- , \ul{x}_0) +
\Big[ \frac{x_2^-}{2} {\pd}^i a^{+ \, b} (x_2^- , \ul{x}_1)
\contraction[2ex]
{}
{+}{a^{i \, b}
(x_2^- , \ul{x}_1) \Big] }
{a}
+ a^{i \, b} (x_2^- , \ul{x}_1) \Big] 
a^{+ \, a} (x_{2'}^- , \ul{x}_0) \right\}
\notag \\ & = \frac{1}{8 \pi^3} \, \int\limits_0^{p_2^-} d k^- \int d^2 x_2 \, \Bigg\{\left[ \frac{\epsilon^{ij} x_{20}^j}{x_{20}^2} - 2 x_{21}^i \frac{{\un x}_{21} \times {\un x}_{20}}{x_{21}^2 \, x_{20}^2} \right] \, \left( U_{{\ul 2}}^{\textrm{mag}} \right)^{b a} \\ 
& +  \left[ \delta^{ij} \left( 2 \frac{{\un x}_{20} \cdot {\un x}_{21}}{x_{20}^2 \, x_{21}^2} + \frac{1}{x_{20}^2} \right) + 2 \frac{x_{21}^i \, x_{20}^j}{x_{21}^2 \, x_{20}^2} \left( 2 \frac{{\un x}_{20} \cdot {\un x}_{21}}{x_{20}^2} + 1 \right) - 2 \frac{x_{21}^i \, x_{21}^j}{x_{21}^2 \, x_{20}^2} \left( 2 \frac{{\un x}_{20} \cdot {\un x}_{21}}{x_{21}^2} + 1 \right) - 2 \frac{x_{20}^i \, x_{20}^j}{x_{20}^4}  \right] \, \left( U_{{\ul 2}}^{j} \right)^{b a} \Bigg\}, \notag 
\end{align}
where
\begin{subequations}\label{Ui+mag}
\begin{align}
& U_{\un{x}}^i = \frac{p_1^+}{4 \, s}  \int\limits_{-\infty}^{\infty} d{z}^- \ U_{\un{x}} [ \infty, z^-] \, \left[ \vec{\mathscr{D}}^i_x -  \cev{\mathscr{D}}_x^i \right]  \, U_{\un{x}} [ z^-, -\infty], \\
& U_{\un{x}}^{\textrm{mag}} = \frac{i \, g \, p_1^+}{s}  \,  \int\limits_{-\infty}^{\infty} d{z}^- \ U_{\un{x}} [ \infty, z^-] \, {\cal F}^{12} \, U_{\un{x}} [ z^-, -\infty]  
\end{align}
\end{subequations}
are the adjoint versions of Eqs.~\eqref{Vi+mag} with the adjoint right- and left-acting covariant derivatives $\vec{\mathscr{D}}^i = \vec{\pd}^i - i g {\cal A}^i$, $\cev{\mathscr{D}}^i = \cev{\pd}^i + i g {\cal A}^i$. The transverse separations are defined by ${\un x}_{ij} = {\un x}_i - {\un x}_j$ with the magnitude $x_{ij} = |{\un x}_{ij}|$.

Employing \eq{aaa} we obtain the evolution equation for the trace entering the amplitude $F^i_{10} (z)$, suppressing the time-ordering signs for brevity (cf. Eq.~(105) in \cite{Cougoulic:2022gbk}),
\begin{align}\label{Fi1}
& \frac{1}{2 N_c} \, \mbox{Re} \, \llangle \tr \left[ V_{\underline{0}} \, V^{i \, \dagger}_{{\un 1}} \right] \rrangle (z) = \frac{1}{2 N_c} \, \mbox{Re} \, \llangle \tr \left[ V_{\underline{0}} \, V^{i \, \dagger}_{{\un 1}} \right] \rrangle_0 (z) + \\
& + \frac{\as \, N_c}{4 \pi^2} \, \int\limits_{\frac{\Lambda^2}{s}}^z \frac{d z'}{z'} \, \int d^2 x_2 \Bigg\{ \left[ \frac{\epsilon^{ij} x_{21}^j}{x_{21}^2} - \frac{\epsilon^{ij} x_{20}^j}{x_{20}^2} + 2 x_{21}^i \frac{{\un x}_{21} \times {\un x}_{20}}{x_{21}^2 \, x_{20}^2} \right] \, \frac{1}{N_c^2} \, \mbox{Re} \, \llangle  \tr \left[ t^b V_{\ul 0} t^a V_{\un 1}^{\dagger} \right] \left(U_{{\ul 2}}^{\textrm{mag}} \right)^{b a} \rrangle (z') \notag \\ 
& + \left[ \delta^{ij} \left( \frac{3}{x_{21}^2} -  2 \, \frac{{\un x}_{20} \cdot {\un x}_{21}}{x_{20}^2 \, x_{21}^2} - \frac{1}{x_{20}^2} \right)  - 2 \frac{x_{21}^i \, x_{20}^j}{x_{21}^2 \, x_{20}^2} \left( 2 \frac{{\un x}_{20} \cdot {\un x}_{21}}{x_{20}^2} + 1 \right) + 2 \frac{x_{21}^i \, x_{21}^j}{x_{21}^2 \, x_{20}^2} \left( 2 \frac{{\un x}_{20} \cdot {\un x}_{21}}{x_{21}^2} + 1 \right) + 2 \frac{x_{20}^i \, x_{20}^j}{x_{20}^4} - 2 \frac{x_{21}^i \, x_{21}^j}{x_{21}^4}   \right] \notag \\
& \times \, \frac{1}{N_c^2} \, \mbox{Re} \, \llangle  \tr \left[ t^b V_{\ul 0} t^a V_{\un 1}^{\dagger} \right]  \left( U_{\un{2}}^{j} \right)^{b a}  \rrangle (z')  \Bigg\}  \notag \\
& + \frac{\as \, N_c}{2 \pi^2} \, \int\limits_{\frac{\Lambda^2}{s}}^z \frac{d z'}{z'} \, \int d^2 x_2 \, \frac{x_{10}^2}{x_{21}^2 \, x_{20}^2} \, \mbox{Re} \, \Bigg\{ \frac{1}{N_c^2} \, \llangle \tr \left[ t^b \, V_{\un 0} \, t^a \, V_{\un 1}^{i \, \dagger} \right] \, \left( U_{\un 2} \right)^{ba} \rrangle (z')  - \frac{C_F}{N_c^2} \, \llangle \tr \left[ V_{\un 0} \, V_{\un 1}^{i \, \dagger} \right] \rrangle (z' s) \Bigg\}  .  \notag
\end{align}
Here and below the sub(super)script $0$ denotes the inhomogeneous term in the evolution equations, also known as the initial condition. It can be constructed by performing the lowest-order perturbative calculation for the dipole amplitude involved.

To take the large-$N_c$ limit of \eq{Fi1} we employ the following relations \cite{Kovchegov:2018znm,Kovchegov:2021lvz,Cougoulic:2022gbk}, valid for the operators in Eqs.~\eqref{Vi+mag} and \eqref{Ui+mag}, 
\begin{subequations}
\begin{align}\label{Ui+mag_Fierz}
& \left( U_{\un{x}}^{\textrm{mag}} \right)^{b a} = 4 \, \tr \left[ t^b \, V_{\un{x}} \, t^a \, V_{\un{x}}^{\textrm{mag} \, \dagger} \right] + 4 \, \tr \left[ t^b \, V_{\un{x}}^{\textrm{mag}}  \, t^a \, V_{\un{x}}^\dagger \right], \\  
& \left( U_{\un{x}}^{i} \right)^{b a} = 2 \, \tr \left[ t^b \, V_{\un{x}} \, t^a \, V_{\un{x}}^{i \, \dagger} \right] + 2 \, \tr \left[ t^b \, V_{\un{x}}^{i }  \, t^a \, V_{\un{x}}^\dagger \right], \\
& \left( U_{\un{x}}\right)^{b a} = 2 \, \tr \left[ t^b \, V_{\un{x}} \, t^a \, V_{\un{x}}^{\dagger} \right] , \label{Ui+mag_Fierzc}
\end{align}
\end{subequations}
which yield, after applying the Fierz identity and neglecting the odderon contribution as a higher-order correction to the unpolarized $S$-matrix, 
\begin{subequations}\label{largeNc2}
\begin{align}
& \frac{1}{N_c^2} \llangle \textrm{T} \, \tr \left[ t^b V_{\ul 0} t^a V_{\un 1}^{\dagger} \right] \left(U_{{\ul 2}}^{\textrm{mag}} \right)^{b a}  \rrangle =  S_{20} (z) \, \frac{1}{N_c} \llangle \textrm{T} \, \tr \left[ V_{\un 1}^{\dagger} \, V_{\un 2}^{\textrm{mag}} \right] \rrangle +  S_{21} (z) \, \frac{1}{N_c} \llangle \textrm{T} \, \tr \left[ V_{\un 0} \, V_{\un 2}^{\textrm{mag} \, \dagger} \right] \rrangle , \\
& \frac{1}{N_c^2} \llangle  \textrm{T} \, \tr \left[ t^b V_{\ul 0} t^a V_{\un 1}^{\dagger} \right]  \left( U_{\un{2}}^{j} \right)^{b a}  \rrangle = S_{20} (z) \, \frac{1}{2 N_c} \llangle \textrm{T} \, \tr \left[ V_{\un 1}^{\dagger} \, V_{\un 2}^{j} \right] \rrangle +  S_{21} (z) \, \frac{1}{2 N_c} \llangle \textrm{T} \, \tr \left[ V_{\un 0} \, V_{\un 2}^{j \, \dagger} \right] \rrangle , \\
& \frac{1}{N_c^2} \, \llangle \tr \left[ t^b \, V_{\un 0} \, t^a \, V_{\un 1}^{i \, \dagger} \right] \, \left( U_{\un 2} \right)^{ba} \rrangle (z) = S_{20} (z) \, \frac{1}{2 N_c} \llangle \textrm{T} \, \tr \left[ V_{\un 1}^{i \, \dagger} \, V_{\un 2} \right] \rrangle (z) .
\end{align}
\end{subequations}

The ``standard" (eikonal) unpolarized dipole $S$-matrix is defined by \cite{Mueller:1994rr,Mueller:1994jq,Mueller:1995gb,Balitsky:1995ub,Balitsky:1998ya,Kovchegov:1999yj,Kovchegov:1999ua,Jalilian-Marian:1997dw,Jalilian-Marian:1997gr,Weigert:2000gi,Iancu:2001ad,Iancu:2000hn,Ferreiro:2001qy}
\begin{align}
S_{10} (zs) = \frac{1}{N_c} \, \left\langle \mbox{T} \, \tr \left[ V_{\un 1} \,  V_{\un 0}^{\dagger} \right] \right\rangle (zs).
\end{align}

Employing Eqs.~\eqref{largeNc2} in \eq{Fi1} we get
\begin{align}\label{Fi2}
& \frac{1}{2 N_c} \, \mbox{Re} \, \llangle \tr \left[ V_{\underline{0}} \, V^{i \, \dagger}_{{\un 1}} \right] \rrangle (z) = \frac{1}{2 N_c} \, \mbox{Re} \, \llangle \tr \left[ V_{\underline{0}} \, V^{i \, \dagger}_{{\un 1}} \right] \rrangle_0 (z) + \frac{\as \, N_c}{4 \pi^2} \, \int\limits_{\frac{\Lambda^2}{s}}^z \frac{d z'}{z'} \, \int d^2 x_2  \\
& \times \Bigg\{ \left[ \frac{\epsilon^{ij} x_{21}^j}{x_{21}^2} - \frac{\epsilon^{ij} x_{20}^j}{x_{20}^2} + 2 x_{21}^i \frac{{\un x}_{21} \times {\un x}_{20}}{x_{21}^2 \, x_{20}^2} \right] \mbox{Re} \left[ S_{20} (z') \frac{1}{N_c} \llangle \textrm{T} \tr \left[ V_{\un 1}^{\dagger} \, V_{\un 2}^{\textrm{mag}} \right] \rrangle (z') +  S_{21} (z') \frac{1}{N_c} \llangle \textrm{T} \tr \left[ V_{\un 0} \, V_{\un 2}^{\textrm{mag} \, \dagger} \right] \rrangle (z') \right] \notag \\ 
& + \left[ \delta^{ij} \left( \frac{3}{x_{21}^2} -  2 \, \frac{{\un x}_{20} \cdot {\un x}_{21}}{x_{20}^2 \, x_{21}^2} - \frac{1}{x_{20}^2} \right)  - 2 \frac{x_{21}^i \, x_{20}^j}{x_{21}^2 \, x_{20}^2} \left( 2 \frac{{\un x}_{20} \cdot {\un x}_{21}}{x_{20}^2} + 1 \right) + 2 \frac{x_{21}^i \, x_{21}^j}{x_{21}^2 \, x_{20}^2} \left( 2 \frac{{\un x}_{20} \cdot {\un x}_{21}}{x_{21}^2} + 1 \right) + 2 \frac{x_{20}^i \, x_{20}^j}{x_{20}^4} - 2 \frac{x_{21}^i \, x_{21}^j}{x_{21}^4}   \right] \notag \\
& \times \,  \mbox{Re} \, \left[ S_{20} (z') \, \frac{1}{2 N_c} \llangle \textrm{T} \, \tr \left[ V_{\un 1}^{\dagger} \, V_{\un 2}^{j} \right] \rrangle (z') +  S_{21} (z') \, \frac{1}{2 N_c} \llangle \textrm{T} \, \tr \left[ V_{\un 0} \, V_{\un 2}^{j \, \dagger} \right] \rrangle (z') \right]  \Bigg\}  \notag \\
& + \frac{\as \, N_c}{2 \pi^2} \, \int\limits_{\frac{\Lambda^2}{s}}^z \frac{d z'}{z'} \, \int d^2 x_2 \, \frac{x_{10}^2}{x_{21}^2 \, x_{20}^2} \, \mbox{Re} \, \Bigg\{ S_{20} (z') \, \frac{1}{2 N_c} \llangle \textrm{T} \, \tr \left[ V_{\un 1}^{i \, \dagger} \, V_{\un 2} \right] \rrangle (z')  - \frac{1}{2 N_c} \, \llangle \tr \left[ V_{\un 0} \, V_{\un 1}^{i \, \dagger} \right] \rrangle (z') \Bigg\}  .  \notag
\end{align}

We are interested in the behavior of the Sivers function outside of the saturation region. Therefore, for simplicity we linearize \eq{Fi2} by putting $S=1$. We obtain the following equation for the flavor non-singlet amplitude $F^{NS \, i}_{10}$ at large $N_c$:
\begin{align}\label{F_NS_1}
& F^{NS \, i}_{10} (z) = F^{NS \, (0) \, i}_{10} (z) + \frac{\as N_c}{4 \pi^2}  \int\limits_{\frac{\Lambda^2}{s}}^z \frac{d z'}{z'}  \int d^2 x_2 \, \Bigg\{ 2 \, \left[ \frac{\epsilon^{ij} x_{21}^j}{x_{21}^2} - \frac{\epsilon^{ij} x_{20}^j}{x_{20}^2} + 2 x_{21}^i \frac{{\un x}_{21} \times {\un x}_{20}}{x_{21}^2 \, x_{20}^2} \right]  \left[ - F^{NS \, \textrm{mag}}_{21} (z') +  F^{NS \, \textrm{mag}}_{20} (z') \right] \notag \\ 
& + \left[ \delta^{ij} \left( \frac{3}{x_{21}^2} -  2 \, \frac{{\un x}_{20} \cdot {\un x}_{21}}{x_{20}^2 \, x_{21}^2} - \frac{1}{x_{20}^2} \right)  - 2 \frac{x_{21}^i \, x_{20}^j}{x_{21}^2 \, x_{20}^2} \left( 2 \frac{{\un x}_{20} \cdot {\un x}_{21}}{x_{20}^2} + 1 \right) + 2 \frac{x_{21}^i \, x_{21}^j}{x_{21}^2 \, x_{20}^2} \left( 2 \frac{{\un x}_{20} \cdot {\un x}_{21}}{x_{21}^2} + 1 \right) + 2 \frac{x_{20}^i \, x_{20}^j}{x_{20}^4} - 2 \frac{x_{21}^i \, x_{21}^j}{x_{21}^4}   \right] \notag \\
& \times \,  \left[ - F^{NS \, j}_{21} (z') + F^{NS \, j}_{20} (z')  \right]  \Bigg\} + \frac{\as \, N_c}{2 \pi^2} \, \int\limits_{\frac{\Lambda^2}{s}}^z \frac{d z'}{z'} \, \int d^2 x_2 \, \frac{x_{10}^2}{x_{21}^2 \, x_{20}^2} \, \Bigg\{ F^{NS \, i}_{12} (z') - \Gamma^{NS \, i}_{10, 21} (z')  \Bigg\}  .  
\end{align}
Here $\Gamma^{NS \, i}_{10, 21} (z')$ is the ``neighbor" dipole amplitude \cite{Kovchegov:2015pbl,Kovchegov:2016zex,Kovchegov:2018znm} for $F^{NS \, i}_{10}$. Any of the amplitudes $F^i$ (and, hence, $F_A, F_B, F_C$) and $F^\textrm{mag}$ has a corresponding ``neighbor" amplitude $\Gamma^i$ ($\Gamma_A, \Gamma_B, \Gamma_C$) and $\Gamma^\textrm{mag}$. The neighbor dipole amplitude $\Gamma_{10,21} (z)$ is defined as the amplitude $F_{10} (z)$ with the $x^-$-lifetime cutoff on its evolution dependent on another dipole size, $x_{21}$. More specifically, the subsequent emissions in the dipole 10 described by the amplitude $\Gamma_{10,21} (z)$ are limited from above by the lifetime $z \, x_{21}^2$ of a neighbor dipole 21 (see \cite{Kovchegov:2016zex,Cougoulic:2019aja} for detailed discussions). Below we will also employ the impact-parameter integrated neighbor dipole amplitude
\begin{align}\label{neighbor_b}
    \Gamma  (x_{10}^2, x^2_{21}, z) = \int d^2 b_\perp \, \Gamma_{10,21} (z).
\end{align}

Integrating over impact parameters with the help of Eqs.~\eqref{decomp} we arrive at
\begin{align}\label{F_NS_2}
& \epsilon^{ij} \, S_P^j \, x_{10}^2 \, F_A^{NS} (x_{10}^2, z) + x_{10}^i \, {\un x}_{10} \times {\un S}_P \, F^{NS}_B (x_{10}^2, z) +  \epsilon^{ij} \, x_{10}^j \, {\un x}_{10} \cdot {\un S}_P \, F^{NS}_C (x_{10}^2, z) \notag \\ 
& = \epsilon^{ij} \, S_P^j \, x_{10}^2 \, F_A^{NS\, (0)} (x_{10}^2, z) + x_{10}^i \, {\un x}_{10} \times {\un S}_P \, F^{NS\, (0)}_B (x_{10}^2, z) +  \epsilon^{ij} \, x_{10}^j \, {\un x}_{10} \cdot {\un S}_P \, F^{NS\, (0)}_C (x_{10}^2, z)  \notag \\ 
& + \frac{\as N_c}{4 \pi^2}  \int\limits_{\frac{\Lambda^2}{s}}^z \frac{d z'}{z'} \int d^2 x_2 \, \Bigg\{ 2 \, \left[ \frac{\epsilon^{ij} x_{21}^j}{x_{21}^2} - \frac{\epsilon^{ij} x_{20}^j}{x_{20}^2} + 2 x_{21}^i \frac{{\un x}_{21} \times {\un x}_{20}}{x_{21}^2 \, x_{20}^2} \right] \notag \\ 
& \times \left[ - {\un x}_{21} \cdot {\un S}_P \, F^{NS \, \textrm{mag}} (x_{21}^2, z')  +  {\un x}_{20} \cdot {\un S}_P \, F^{NS \, \textrm{mag}} (x_{20}^2, z')   \right] \notag \\ 
& + \left[ \delta^{ij} \left( \frac{3}{x_{21}^2} -  2 \, \frac{{\un x}_{20} \cdot {\un x}_{21}}{x_{20}^2 \, x_{21}^2} - \frac{1}{x_{20}^2} \right)  - 2 \frac{x_{21}^i \, x_{20}^j}{x_{21}^2 \, x_{20}^2} \left( 2 \frac{{\un x}_{20} \cdot {\un x}_{21}}{x_{20}^2} + 1 \right) + 2 \frac{x_{21}^i \, x_{21}^j}{x_{21}^2 \, x_{20}^2} \left( 2 \frac{{\un x}_{20} \cdot {\un x}_{21}}{x_{21}^2} + 1 \right) + 2 \frac{x_{20}^i \, x_{20}^j}{x_{20}^4} - 2 \frac{x_{21}^i \, x_{21}^j}{x_{21}^4}   \right] \notag \\
& \times \,  \left[ - \epsilon^{jk} \, S_P^k \, x_{21}^2 \, F_A^{NS} (x_{21}^2, z') - x_{21}^j \, {\un x}_{21} \times {\un S}_P \, F^{NS}_B (x_{21}^2, z') -  \epsilon^{jk} \, x_{21}^k \, {\un x}_{21} \cdot {\un S}_P \, F^{NS}_C (x_{21}^2, z')  \right. \notag \\ 
& \left. + \epsilon^{jk} \, S_P^k \, x_{20}^2 \, F_A^{NS} (x_{20}^2, z') + x_{20}^j \, {\un x}_{20} \times {\un S}_P \, F^{NS}_B (x_{20}^2, z') +  \epsilon^{jk} \, x_{20}^k \, {\un x}_{20} \cdot {\un S}_P \, F^{NS}_C (x_{20}^2, z')  \right]  \Bigg\} \notag \\ 
& + \frac{\as \, N_c}{2 \pi^2} \, \int\limits_{\frac{\Lambda^2}{s}}^z \frac{d z'}{z'} \, \int d^2 x_2 \, \frac{x_{10}^2}{x_{21}^2 \, x_{20}^2} \, \Bigg\{ \epsilon^{ij} \, S_P^j \, x_{21}^2 \, F_A^{NS} (x_{21}^2, z') + x_{21}^i \, {\un x}_{21} \times {\un S}_P \, F^{NS}_B (x_{21}^2, z') +  \epsilon^{ij} \, x_{21}^j \, {\un x}_{21} \cdot {\un S}_P \, F^{NS}_C (x_{21}^2, z') \notag \\ 
&  - \epsilon^{ij} \, S_P^j \, x_{10}^2 \, \Gamma_A^{NS} (x_{10}^2, x_{21}^2, z') - x_{10}^i \, {\un x}_{10} \times {\un S}_P \, \Gamma^{NS}_B (x_{10}^2, x_{21}^2, z') -  \epsilon^{ij} \, x_{10}^j \, {\un x}_{10} \cdot {\un S}_P \, \Gamma^{NS}_C (x_{10}^2, x_{21}^2, z')  \Bigg\}  .  
\end{align}
Next we assume that all the amplitudes $F$ in \eq{F_NS_2} contain no integer powers of dipole sizes and that they contain only perturbatively small powers, $\sim \sqrt{\as}$ or $\sim \as$, of $x_{10}$ and $x_{21}$ and/or logarithms of these dipole sizes. We will extract the DLA version of \eq{F_NS_2} employing this approximation. The derivation of the double-logarithmic approximation of \eq{F_NS_2} is detailed in Appendix~\ref{sec:app_DLA}.
It leads to the following evolution equations:
\begin{subequations}\label{FABC}
\begin{align}
& F_A^{NS} (x_{10}^2, z) = F_A^{NS\, (0)} (x_{10}^2, z) + \frac{\as \, N_c}{4 \pi} \,   \int\limits_{\frac{\Lambda^2}{s}}^z \frac{d z'}{z'} \,  \int\limits_{\max \left[ x_{10}^2, \frac{1}{z' s} \right]}^{\frac{z}{z'} x_{10}^2 } \frac{d x^2_{21}}{x^2_{21}} \,  \left[ 6 \, F_A^{NS} (x^2_{21}, z') - F_B^{NS} (x^2_{21}, z')  + F_C^{NS} (x^2_{21}, z') \right], \\ 
& F^{NS}_B (x_{10}^2, z) = F^{NS\, (0)}_B (x_{10}^2, z) + \frac{\as \, N_c}{4 \pi} \, \int\limits_{\frac{\Lambda^2}{s}}^z \frac{d z'}{z'} \,  \int\limits_{\max \left[ x_{10}^2, \frac{1}{z' s} \right]}^{\frac{z}{z'} x_{10}^2} \frac{d x^2_{21}}{x^2_{21}} \, \left[ - 2 \, F_A^{NS} (x^2_{21}, z') + 5 \, F_B^{NS} (x^2_{21}, z') - F_C^{NS} (x^2_{21}, z') \right] ,  \\
& F^{NS}_C (x_{10}^2, z) = F^{NS\, (0)}_C (x_{10}^2, z) + \frac{\as \, N_c}{4 \pi} \,  \int\limits_{\frac{\Lambda^2}{s}}^z \frac{d z'}{z'} \,  \int\limits_{\max \left[ x_{10}^2, \frac{1}{z' s} \right] }^{\frac{z}{z'} x_{10}^2 } \frac{d x_{21}^2}{x_{21}^2} \, \left[ 2\, F^{NS \, \textrm{mag}} (x_{21}^2, z')  + 6 \, F_C^{NS} (x^2_{21}, z') \right]  .
\end{align}
\end{subequations}

\subsubsection{Evolution for $V^\textrm{mag}$}

We now move on to the evolution for $V^\textrm{mag}$. Employing the propagator \cite{Cougoulic:2022gbk}
\begin{align}\label{+perp_sub_eik_6}
& \frac{1}{4} \, \epsilon^{ki} \pd_1^k \left[ \int\limits_{-\infty}^0 dx_1^- \, 
\int\limits_0^\infty dx_0^- \, 
\contraction[2ex]
{}
{a}{_{\bot}^{i \, a}
(x_1^- , \ul{x}_1) \:}
{a}
\: 
a_{\bot}^{i \, a} (x_1^- , \ul{x}_1) \:
a^{+ \, b} (x_0^- , \ul{x}_0) +
\int\limits^{\infty}_0 dx_1^- \,  
\int\limits^0_{-\infty} dx_0^- \, 
\contraction[2ex]
{}
{a}{_{\bot}^{i \, b}
(x_1^- , \ul{x}_1) \:}
{a}
\: 
a_{\bot}^{i \, b} (x_1^- , \ul{x}_1) \:
a^{+ \, a} (x_0^- , \ul{x}_0) \right] = \\ 
& 
- \frac{1}{4 \pi^3} \, \int\limits_0^{p_2^-} d k^- \int d^2 x_2 \,
\Bigg\{ \frac{{\un x}_{21}}{x_{21}^2} \cdot \frac{{\un x}_{20}}{x_{20}^2} \, (U_{{\ul 2}}^{\textrm{mag}})^{b a} + \left[ \frac{\epsilon^{ij} \, (x_{20}^i + x_{21}^i)}{x_{20}^2 \, x_{21}^2}  + \frac{2 \, {\un x}_{21} \times {\un x}_{20}}{x_{20}^2 \, x_{21}^2} \left( \frac{x_{21}^j}{x_{21}^2} - \frac{x_{20}^j}{x_{20}^2}\right) \right] \, \left( U_{\un{2}}^{j} \right)^{b a}  \,   \Bigg\} \notag 
\end{align}
one can derive the following evolution equation (cf. Eq.~(94) in \cite{Cougoulic:2022gbk}, without the soft quark emission)
\begin{align}\label{Q_evol_main}
& \frac{1}{2 N_c} \, \llangle \tr \left[ V_{\ul 0} \, V_{\un 1}^{\textrm{mag} \, \dagger} \right]  \rrangle (zs) =   \frac{1}{2 N_c} \, \llangle \tr \left[ V_{\ul 0} \, V_{\un 1}^{\textrm{mag} \, \dagger} \right] \rrangle_0 (zs) \\
& + \frac{\as \, N_c}{2 \pi^2} \, \int\limits_{\frac{\Lambda^2}{s}}^z \frac{d z'}{z'} \, \int d^2 x_2 \Bigg\{ \left[ \frac{1}{x_{21}^2} -  \frac{{\un x}_{21}}{x_{21}^2} \cdot \frac{{\un x}_{20}}{x_{20}^2} \right] \, \frac{1}{N_c^2} \llangle  \tr \left[ t^b V_{\ul 0} t^a V_{\un 1}^{\dagger} \right] \left(U_{{\ul 2}}^{\textrm{mag}} \right)^{b a} \rrangle (z' s) \notag \\ 
& + \left[ 2 \frac{\epsilon^{ij} \, x_{21}^j}{x_{21}^4} - \frac{\epsilon^{ij} \, (x_{20}^j + x_{21}^j)}{x_{20}^2 \, x_{21}^2}  - \frac{2 \, {\un x}_{20} \times {\un x}_{21}}{x_{20}^2 \, x_{21}^2} \left( \frac{x_{21}^i}{x_{21}^2} - \frac{x_{20}^i}{x_{20}^2}\right) \right] \frac{1}{N_c^2} \llangle  \tr \left[ t^b V_{\ul 0} t^a V_{\un 1}^{\dagger} \right]  \left( U_{\un{2}}^{i} \right)^{b a}  \rrangle (z' s)  \Bigg\}  \notag \\
& + \frac{\as \, N_c}{2 \pi^2} \, \int\limits_{\frac{\Lambda^2}{s}}^z \frac{d z'}{z'} \, \int d^2 x_2 \, \frac{x_{10}^2}{x_{21}^2 \, x_{20}^2} \,  \Bigg\{ \frac{1}{N_c^2} \, \llangle \tr \left[ t^b \, V_{\un 0} \, t^a \, V_{\un 1}^{\textrm{mag} \, \dagger} \right] \, U_{\un 2}^{ba} \rrangle (z' s)  - \frac{C_F}{N_c^2} \, \llangle \tr \left[ V_{\un 0} \, V_{\un 1}^{\textrm{mag} \, \dagger} \right] \rrangle (z' s)  \Bigg\}  .  \notag
\end{align}

Employing Eqs.~\eqref{largeNc2} we write
\begin{align}\label{Q_evol_main_2}
& \frac{1}{2 N_c} \, \llangle \tr \left[ V_{\ul 0} \, V_{\un 1}^{\textrm{mag} \, \dagger} \right]  \rrangle (z) =   \frac{1}{2 N_c} \, \llangle \tr \left[ V_{\ul 0} \, V_{\un 1}^{\textrm{mag} \, \dagger} \right] \rrangle_0 (z) + \frac{\as \, N_c}{2 \pi^2} \, \int\limits_{\frac{\Lambda^2}{s}}^z \frac{d z'}{z'} \, \int d^2 x_2 \\
& \times \Bigg\{ \left[ \frac{1}{x_{21}^2} -  \frac{{\un x}_{21}}{x_{21}^2} \cdot \frac{{\un x}_{20}}{x_{20}^2} \right] \, \left[ S_{20} (z') \, \frac{1}{N_c} \llangle \textrm{T} \, \tr \left[ V_{\un 1}^{\dagger} \, V_{\un 2}^{\textrm{mag}} \right] \rrangle (z') +  S_{21} (z') \, \frac{1}{N_c} \llangle \textrm{T} \, \tr \left[ V_{\un 0} \, V_{\un 2}^{\textrm{mag} \, \dagger} \right] \rrangle (z') \right] \notag \\ 
& + \left[ 2 \frac{\epsilon^{ij} \, x_{21}^j}{x_{21}^4} - \frac{\epsilon^{ij} \, (x_{20}^j + x_{21}^j)}{x_{20}^2 \, x_{21}^2}  - \frac{2 \, {\un x}_{20} \times {\un x}_{21}}{x_{20}^2 \, x_{21}^2} \left( \frac{x_{21}^i}{x_{21}^2} - \frac{x_{20}^i}{x_{20}^2}\right) \right] \left[ S_{20} (z') \, \frac{1}{2 N_c} \llangle \textrm{T} \, \tr \left[ V_{\un 1}^{\dagger} \, V_{\un 2}^{i} \right] \rrangle (z') \right. \notag \\ 
& \left. +  S_{21} (z') \, \frac{1}{2 N_c} \llangle \textrm{T} \, \tr \left[ V_{\un 0} \, V_{\un 2}^{i \, \dagger} \right] \rrangle (z') \right] \Bigg\}  \notag \\
& + \frac{\as \, N_c}{2 \pi^2} \, \int\limits_{\frac{\Lambda^2}{s}}^z \frac{d z'}{z'} \, \int d^2 x_2 \, \frac{x_{10}^2}{x_{21}^2 \, x_{20}^2} \,  \Bigg\{ S_{20} (z') \, \frac{1}{2 N_c} \llangle \textrm{T} \, \tr \left[ V_{\un 1}^{\textrm{mag} \, \dagger} \, V_{\un 2} \right] \rrangle (z')   - \frac{C_F}{N_c^2} \, \llangle \tr \left[ V_{\un 0} \, V_{\un 1}^{\textrm{mag} \, \dagger} \right] \rrangle (z' s)  \Bigg\}  .  \notag
\end{align}

Putting $S=1$ to linearize \eq{Q_evol_main_2} and subtracting the ``complex conjugate" version of this equation while taking the Re part we obtain the equation for the flavor non-singlet amplitude at large $N_c$,
\begin{align}\label{F_mag_eq}
& F^{NS \, \textrm{mag}}_{10} (z) =  F^{NS \, \textrm{mag} \, (0) }_{10} (z)  \\
& + \frac{\as \, N_c}{2 \pi^2} \, \int\limits_{\frac{\Lambda^2}{s}}^z \frac{d z'}{z'} \, \int d^2 x_2 \Bigg\{ 2 \, \left[ \frac{1}{x_{21}^2} -  \frac{{\un x}_{21}}{x_{21}^2} \cdot \frac{{\un x}_{20}}{x_{20}^2} \right] \, \left[ - F^{NS \, \textrm{mag}}_{21} (z') +  F^{NS \, \textrm{mag}}_{20} (z') \right] \notag \\ 
& + \left[ 2 \frac{\epsilon^{ij} \, x_{21}^j}{x_{21}^4} - \frac{\epsilon^{ij} \, (x_{20}^j + x_{21}^j)}{x_{20}^2 \, x_{21}^2}  - \frac{2 \, {\un x}_{20} \times {\un x}_{21}}{x_{20}^2 \, x_{21}^2} \left( \frac{x_{21}^i}{x_{21}^2} - \frac{x_{20}^i}{x_{20}^2}\right) \right] \left[ - F^{NS \, i}_{21} (z') +  F^{NS \, i}_{20} (z')   \right] \Bigg\}  \notag \\
& + \frac{\as \, N_c}{2 \pi^2} \, \int\limits_{\frac{\Lambda^2}{s}}^z \frac{d z'}{z'} \, \int d^2 x_2 \, \frac{x_{10}^2}{x_{21}^2 \, x_{20}^2} \,  \Bigg\{ F^{NS \, \textrm{mag}}_{12} (z')   - \Gamma^{NS \, \textrm{mag}}_{10,21} (z')  \Bigg\}  .  \notag
\end{align}

Integrating \eq{F_mag_eq} over impact parameters we obtain 
\begin{align}\label{F_mag_eq_2}
& {\un x}_{10} \cdot {\un S}_P \, F^{NS \, \textrm{mag}} (x_{10}^2, z) =  {\un x}_{10} \cdot {\un S}_P \,  F^{NS \, \textrm{mag} \, (0) } (x_{10}^2, z)   \\
& + \frac{\as \, N_c}{2 \pi^2} \, \int\limits_{\frac{\Lambda^2}{s}}^z \frac{d z'}{z'} \, \int d^2 x_2 \Bigg\{ 2 \, \left[ \frac{1}{x_{21}^2} -  \frac{{\un x}_{21}}{x_{21}^2} \cdot \frac{{\un x}_{20}}{x_{20}^2} \right] \, \left[ - {\un x}_{21} \cdot {\un S}_P \, F^{NS \, \textrm{mag}} (x_{21}^2, z') +  {\un x}_{20} \cdot {\un S}_P \, F^{NS \, \textrm{mag}} (x_{20}^2, z') \right] \notag \\ 
& + \left[ 2 \frac{\epsilon^{ij} \, x_{21}^j}{x_{21}^4} - \frac{\epsilon^{ij} \, (x_{20}^j + x_{21}^j)}{x_{20}^2 \, x_{21}^2}  - \frac{2 \, {\un x}_{20} \times {\un x}_{21}}{x_{20}^2 \, x_{21}^2} \left( \frac{x_{21}^i}{x_{21}^2} - \frac{x_{20}^i}{x_{20}^2}\right) \right] \notag \\
& \times \,  \left[ - \epsilon^{ik} \, S_P^k \, x_{21}^2 \, F_A^{NS} (x_{21}^2, z') - x_{21}^i \, {\un x}_{21} \times {\un S}_P \, F^{NS}_B (x_{21}^2, z') -  \epsilon^{ik} \, x_{21}^k \, {\un x}_{21} \cdot {\un S}_P \, F^{NS}_C (x_{21}^2, z')  \right. \notag \\ 
& \left. + \epsilon^{ik} \, S_P^k \, x_{20}^2 \, F_A^{NS} (x_{20}^2, z') + x_{20}^i \, {\un x}_{20} \times {\un S}_P \, F^{NS}_B (x_{20}^2, z') +  \epsilon^{ik} \, x_{20}^k \, {\un x}_{20} \cdot {\un S}_P \, F^{NS}_C (x_{20}^2, z')  \right]  \Bigg\} \notag \\ 
& + \frac{\as \, N_c}{2 \pi^2} \, \int\limits_{\frac{\Lambda^2}{s}}^z \frac{d z'}{z'} \, \int d^2 x_2 \, \frac{x_{10}^2}{x_{21}^2 \, x_{20}^2} \,  \Bigg\{ - {\un x}_{21} \cdot {\un S}_P \, F^{NS \, \textrm{mag}} (x_{21}^2, z')    - {\un x}_{10} \cdot {\un S}_P \, \Gamma^{NS \, \textrm{mag}} (x_{10}^2, x_{21}^2,  z')  \Bigg\}  .  \notag
\end{align}
The DLA version of this equation is obtained in Appendix~\ref{sec:app_DLA}, again assuming that all dipole amplitudes involved contain no order-one powers of the dipole sizes they depend on. The DLA and large-$N_c$ evolution equation reads
\begin{align}\label{F_mag_eq_4}
& F^{NS \, \textrm{mag}} (x_{10}^2, z) =  F^{NS \, \textrm{mag} \, (0) } (x_{10}^2, z)   \\
& + \frac{\as \, N_c}{2 \pi} \!\!\!\!\! \int\limits_{\max \left\{ \frac{\Lambda^2}{s}, \frac{1}{s x_{10}^2} \right\}}^z \!\!\!\!\! \frac{d z'}{z'} \, \int\limits^{x_{10}^2}_\frac{1}{z' s} \frac{d x^2_{21}}{x^2_{21}} \, \left[ \Gamma^{NS \, \textrm{mag}} (x_{10}^2, x^2_{21}, z') + 2 \, \Gamma^{NS}_A (x_{10}^2, x^2_{21}, z') - \Gamma^{NS}_B (x_{10}^2, x^2_{21}, z') + 3 \, \Gamma^{NS}_C (x_{10}^2, x^2_{21}, z') \right] . \notag 
\end{align}


\subsubsection{Evolution Equations}

Let us summarize the DLA large-$N_c$ evolution equations we obtained. We have Eqs.~\eqref{FABC} and \eqref{F_mag_eq_4}, which we list here together,
\begin{subequations}\label{FFFF}
\begin{tcolorbox}[ams align]
& F_A^{NS} (x_{10}^2, z) = F_A^{NS\, (0)} (x_{10}^2, z) + \frac{\as \, N_c}{4 \pi} \int\limits_{\frac{\Lambda^2}{s}}^z \frac{d z'}{z'} \!\!\!\!\!\!   \int\limits_{\max \left[ x_{10}^2, \frac{1}{z' s} \right]}^{\frac{z}{z'} x_{10}^2} \frac{d x^2_{21}}{x^2_{21}} \,  \left[ 6 \, F_A^{NS} (x^2_{21}, z') - F_B^{NS} (x^2_{21}, z')  + F_C^{NS} (x^2_{21}, z') \right] , \\ 
& F^{NS}_B (x_{10}^2, z) = F^{NS\, (0)}_B (x_{10}^2, z) + \frac{\as N_c}{4 \pi} \int\limits_{\frac{\Lambda^2}{s}}^z \frac{d z'}{z'} \!\!\!\!\!\!\!  \int\limits_{\max \left[ x_{10}^2, \frac{1}{z' s} \right]}^{\frac{z}{z'} x_{10}^2} \!\!\!\!\!   \frac{d x^2_{21}}{x^2_{21}} \left[ - 2 \, F_A^{NS} (x^2_{21}, z') + 5 \, F_B^{NS} (x^2_{21}, z') - F_C^{NS} (x^2_{21}, z') \right],  \\
& F^{NS}_C (x_{10}^2, z) = F^{NS\, (0)}_C (x_{10}^2, z) + \frac{\as N_c}{4 \pi}   \int\limits_{\frac{\Lambda^2}{s}}^z \frac{d z'}{z'} \!\!\!   \int\limits_{\max \left[ x_{10}^2, \frac{1}{z' s} \right] }^{\frac{z}{z'} x_{10}^2} \frac{d x_{21}^2}{x_{21}^2} \left[ 2\, F^{NS \, \textrm{mag}} (x_{21}^2, z')  + 6 \, F_C^{NS} (x^2_{21}, z') \right]  , \\ 
& F^{NS \, \textrm{mag}} (x_{10}^2, z) =  F^{NS \, \textrm{mag} \, (0) } (x_{10}^2, z)   + \frac{\as N_c}{2 \pi}   \int\limits_{\max \left\{ \frac{\Lambda^2}{s}, \frac{1}{s x_{10}^2} \right\}}^z  \frac{d z'}{z'} \ \int\limits^{x_{10}^2}_\frac{1}{z' s} \frac{d x^2_{21}}{x^2_{21}} \\
& \hspace*{3cm} \times \left[ \Gamma^{NS \, \textrm{mag}} (x_{10}^2, x^2_{21}, z') + 2 \, \Gamma^{NS}_A (x_{10}^2, x^2_{21}, z') - \Gamma^{NS}_B (x_{10}^2, x^2_{21}, z') + 3 \, \Gamma^{NS}_C (x_{10}^2, x^2_{21}, z') \right] . \notag 
\end{tcolorbox}
\end{subequations}

It is not difficult to construct ``mirror" evolution equations for the ``neighbor" amplitudes \cite{Kovchegov:2015pbl,Kovchegov:2016zex,Kovchegov:2018znm,Kovchegov:2021lvz,Kovchegov:2021iyc}, 
\begin{subequations}\label{Gamma4}
\begin{tcolorbox}[ams align]
& \Gamma_A^{NS} (x_{10}^2, x_{21}^2, z') = F_A^{NS\, (0)} (x_{10}^2, z') + \frac{\as \, N_c}{4 \pi} \,   \int\limits_{\frac{\Lambda^2}{s}}^{z' \frac{x_{21}^2}{x_{10}^2}} \frac{d z''}{z''} \,  \int\limits_{\max \left[ x_{10}^2, \frac{1}{z'' s} \right]}^{\frac{z'}{z''} x_{21}^2} \frac{d x^2_{32}}{x^2_{32}} \\ & \hspace*{8cm} \times \,  \left[ 6 \, F_A^{NS} (x^2_{32}, z'') - F_B^{NS} (x^2_{32}, z'')  + F_C^{NS} (x^2_{32}, z'') \right], \notag  \\ 
& \Gamma_B^{NS} (x_{10}^2, x_{21}^2, z') = F^{NS\, (0)}_B (x_{10}^2, z') + \frac{\as \, N_c}{4 \pi} \,  \int\limits_{\frac{\Lambda^2}{s}}^{z' \frac{x_{21}^2}{x_{10}^2}} \frac{d z''}{z''} \,  \int\limits_{\max \left[ x_{10}^2, \frac{1}{z'' s} \right]}^{\frac{z'}{z''} x_{21}^2} \frac{d x^2_{32}}{x^2_{32}} \\ & \hspace*{8cm} \times \, \left[ - 2 \, F_A^{NS} (x^2_{23}, z'') + 5 \, F_B^{NS} (x^2_{32}, z'') - F_C^{NS} (x^2_{32}, z'') \right] , \notag  \\
&  \Gamma_C^{NS} (x_{10}^2, x_{21}^2, z') = F^{NS\, (0)}_C (x_{10}^2, z') + \frac{\as \, N_c}{4 \pi} \,  \int\limits_{\frac{\Lambda^2}{s}}^{z' \frac{x_{21}^2}{x_{10}^2}} \frac{d z''}{z''} \,  \int\limits_{\max \left[ x_{10}^2, \frac{1}{z'' s} \right]}^{\frac{z'}{z''} x_{21}^2} \frac{d x^2_{32}}{x^2_{32}} \\ 
& \hspace*{8cm} \times \, \left[ 2\, F^{NS \, \textrm{mag}} (x_{32}^2, z'')  + 6 \, F_C^{NS} (x^2_{32}, z'') \right], \notag \\ 
& \Gamma^{NS \, \textrm{mag}} (x_{10}^2, x_{21}^2 , z') =  F^{NS \, \textrm{mag} \, (0) } (x_{10}^2, z') + \frac{\as \, N_c}{2 \pi} \, \int\limits_{\max \left\{ \frac{\Lambda^2}{s}, \frac{1}{s x_{10}^2} \right\}}^{z'} \frac{d z''}{z''} \, \int\limits^{\min \left[ x_{10}^2 , x_{21}^2 \frac{z'}{z''} \right]}_\frac{1}{z'' s} \frac{d x^2_{32}}{x^2_{32}} \\ 
& \hspace*{3cm} \times \, \left[ \Gamma^{NS \, \textrm{mag}} (x_{10}^2, x^2_{32}, z'') + 2 \, \Gamma^{NS}_A (x_{10}^2, x^2_{32}, z'') - \Gamma^{NS}_B (x_{10}^2, x^2_{32}, z'') + 3 \, \Gamma^{NS}_C (x_{10}^2, x^2_{32}, z'') \right] . \notag 
\end{tcolorbox}
\end{subequations} 

We have a closed system of eight equations \eqref{FFFF} and \eqref{Gamma4} with eight unknown functions. Given the inhomogeneous terms, these equations can be solved numerically or, perhaps, analytically. The solution of equations \eqref{FFFF} and \eqref{Gamma4} should give us the sub-eikonal asymptotics of the flavor non-singlet Sivers function \eqref{F_S&NS2b}. Below we will solve these equations numerically, leaving an attempt at an analytic solution for future work. Let us stress here that in equations \eqref{FFFF} and \eqref{Gamma4} we assume that $\Lambda$ is a scale characterizing the target which, for the sake of generality, is treated as a perturbative scale: if need be, the integration limits in these equations can be easily modified to make $\Lambda$ an IR cutoff \cite{Kovchegov:2016weo,Kovchegov:2020hgb,Adamiak:2021ppq,Cougoulic:2022gbk}.


\subsubsection{Evolution for $V^{[2]}$ }

Before solving Eqs.~\eqref{FFFF} and \eqref{Gamma4}, we note that there is another dipole amplitude, $F^{[2]}_{10} (z)$, entering the expression \eqref{F_S&NS2b} for the flavor non-singlet Sivers function. Above, in Eqs.~\eqref{FFFF} and \eqref{Gamma4}, we have shown explicitly that $F^{[2]}_{10} (z)$ does not enter/mix with the evolution of the other relevant amplitudes, $F_A, F_B, F_C$ and $F_\textrm{mag}$. In \cite{Kovchegov:2021iyc} it was shown that the initial condition for this amplitude is zero, $F^{[2] (0)}_{10} (z) =0$. (The difference between our $V^{[2]}$ in \eq{ViV2b} and that in Eq.~(77b) of \cite{Kovchegov:2021iyc} does not affect the initial conditions.) The only task left to do is to construct the evolution equation for $F^{[2]}_{10} (z)$ to check whether it contains the amplitudes $F_A, F_B, F_C$ and $F_\textrm{mag}$: if no such mixing is observed and, hence, the evolution of $F^{[2]}_{10} (z)$ involves only this amplitude alone, given zero initial conditions, $F^{[2] (0)}_{10} (z) =0$, we would be able to conclude that $F^{[2]}_{10} (z) =0$. Below, instead of building the full evolution equation for $F^{[2]}_{10} (z)$, we will only complete the steps showing that the evolution equation for $F^{[2]}_{10} (z)$ is closed, and does not mix with the other amplitudes. Combined with $F^{[2] (0)}_{10} (z) =0$, this will allow us to conclude that this dipole amplitude is zero.

One can simplify the gluon part of the operator $V^{[2]}$ from \eq{ViV2b} as
\begin{align}\label{V2_full}
& V_{\un{x}; \un{k}, \un{k}_1}^{[2]} = \frac{i \, p_1^+}{8 \, s}  \int\limits_{-\infty}^{\infty} d{z}^- \ V_{\un{x}} [ \infty, z^-] \, \left[ (\vec{D}^i_x -  \cev{D}_x^i)^2 - (k_1^i -  k^i)^2 \right]  \, V_{\un{x}} [ z^-, -\infty] = \frac{i \, p_1^+}{8 \, s} \, L \, \left[ {\un \pd}_x^2 - (k_1^i -  k^i)^2 \right] \, V_{\un x}  \notag \\ & 
+ \frac{g \, p_1^+}{2 \, s} \int\limits_{-\infty}^{\infty} d{z}^- \ V_{\un{x}} [ \infty, z^-] \, \left[ \left( \frac{1}{2} \, z^- \, \partial^i A^+ + A^i \right) \, \vec{\pd}^i_x  - \cev{\pd}_x^i \, \left( \frac{1}{2} \, z^- \, \partial^i A^+ + A^i \right)  \right]  \, V_{\un{x}} [ z^-, -\infty] + {\cal O} (A_\perp^2). 
\end{align}
Here $L = +\infty^- - (-\infty^-)$ is a regulator of the $z^-$ integral. The first term on the right of \eqref{V2_full} is proportional to the ``standard" eikonal Wilson line. From the definition of the amplitude $F^{[2]}_{10} (z)$ in Eqs.~\eqref{F2NS_def} and \eqref{decompF2} we see that if this term came in without the factor of $i$ in its prefactor, it would have contributed a sub-eikonal correction to the leading-order spin-dependent odderon from Sec.~\ref{sec:Sivers_eik} by generating an approximately constant (in energy or $z$) term in $F^{[2]}_{10} (z)$. However, the presence of the factor of $i$ in the first term on the right of \eqref{V2_full} makes it give a zero contribution to $F^{[2]}_{10} (z)$. We can, therefore, discard it. The terms quadratic in $A_\perp$ in \eq{V2_full} are sub-sub-eikonal and also do not contribute to the sub-eikonal evolution at hand.

The evolution of the second operator on the right in \eq{V2_full} is driven by the following background-field propagators \cite{Cougoulic:2022gbk}:
\begin{align}\label{Vi5}
& \frac{1}{4} \int\limits_{-\infty}^0 dx_{2'}^- \,  
\int\limits_0^\infty dx_2^- \, \Big[ \frac{1}{2} \, x_{2'}^- {\pd}^i a^{+ \, a} (x_{2'}^- , \ul{x}_1)
\contraction[2ex]
{}
{+}{a^{i \, a}
(x_{2'}^- , \ul{x}_1) \Big] \:}
{a}
\: 
+ a^{i \, a} (x_{2'}^- , \ul{x}_1) \Big] \:
a^{+ \, b} (x_2^- , \ul{x}_0)  \\ & = \frac{1}{2 (2 \pi)^3} \int\limits_0^{p_2^-} d k^- \Bigg\{ \int d^2 x_2 \left[ \frac{\epsilon^{ij} x_{20}^j}{x_{20}^2} - 2 x_{21}^i \frac{{\un x}_{21} \times {\un x}_{20}}{x_{21}^2  x_{20}^2} \right] \left( U_{{\ul 2}}^{\textrm{mag}} \right)^{b a} \!\! - i  \int d^2 x_2 d^2 x_{2'} \left[ \frac{ x_{20}^i}{x_{20}^2} - 2 x_{2'1}^i \frac{{\un x}_{2'1} \cdot {\un x}_{20}}{x_{2'1}^2  x_{20}^2} \right] \left( U_{{\ul 2}, {\un 2}'}^{\textrm{phase}} \right)^{b a} \Bigg\} \notag
\end{align}
and
\begin{align}\label{Vi6}
& \frac{1}{4} \int\limits_{-\infty}^0 dx_{2'}^- \,  
\int\limits_0^\infty dx_2^- \, \Big[ \frac{1}{2} \, x_2^- {\pd}^i a^{+ \, b} (x_2^- , \ul{x}_1)
\contraction[2ex]
{}
{+}{a^{i \, b}
(x_2^- , \ul{x}_1) \Big] \:}
{a}
\: 
+ a^{i \, b} (x_2^- , \ul{x}_1) \Big] \:
a^{+ \, a} (x_{2'}^- , \ul{x}_0) \\ & = \frac{1}{2 (2 \pi)^3} \int\limits_0^{p_2^-} d k^- \Bigg\{ \int d^2 x_2 \left[ \frac{\epsilon^{ij} x_{20}^j}{x_{20}^2} - 2 x_{21}^i \frac{{\un x}_{21} \times {\un x}_{20}}{x_{21}^2 x_{20}^2} \right] \left( U_{{\ul 2}}^{\textrm{mag}} \right)^{b a} \!\! + i  \int d^2 x_2 d^2 x_{2'} \left[ \frac{ x_{2'0}^i}{x_{2'0}^2} - 2 x_{21}^i \frac{{\un x}_{21} \cdot {\un x}_{2'0}}{x_{21}^2 x_{2'0}^2} \right] \left( U_{{\ul 2}, {\un 2}'}^{\textrm{phase}} \right)^{b a} \! \Bigg\}. \notag 
\end{align}
Specifically, we need the difference of these propagators (differentiated with $\pd^i_0$ 
in some of the evolution equation terms), in which the first terms cancel,
\begin{align}\label{aaa}
& \frac{1}{4}  \int\limits_{-\infty}^0 dx_{2'}^- \,  
\int\limits_0^\infty dx_2^- \left\{ \Big[ \frac{1}{2} \, x_{2'}^- {\pd}^i a^{+ \, a} (x_{2'}^- , \ul{x}_1)
\contraction[2ex]
{}
{+}{a^{i \, a}
(x_{2'}^- , \ul{x}_1) \Big] \:}
{a}
\: 
+ a^{i \, a} (x_{2'}^- , \ul{x}_1) \Big] \:
a^{+ \, b} (x_2^- , \ul{x}_0) \right. \notag \\
& \hspace*{3cm} \left. - \Big[ \frac{1}{2} \, x_2^- {\pd}^i a^{+ \, b} (x_2^- , \ul{x}_1)
\contraction[2ex]
{}
{+}{a^{i \, b}
(x_2^- , \ul{x}_1) \Big] \:}
{a}
\: 
+ a^{i \, b} (x_2^- , \ul{x}_1) \Big] \:
a^{+ \, a} (x_{2'}^- , \ul{x}_0) \right\} \notag \\ 
& = \frac{-i}{2 (2 \pi)^3} \int\limits_0^{p_2^-} d k^- \int d^2 x_2 d^2 x_{2'} \left[ \frac{ x_{20}^i}{x_{20}^2} - 2 x_{2'1}^i \frac{{\un x}_{2'1} \cdot {\un x}_{20}}{x_{2'1}^2 \, x_{20}^2} + \frac{ x_{2'0}^i}{x_{2'0}^2} - 2 x_{21}^i \frac{{\un x}_{21} \cdot {\un x}_{2'0}}{x_{21}^2 \, x_{2'0}^2}  \right] \left( U_{{\ul 2}, {\un 2}'}^{\textrm{phase}} \right)^{b a} .
\end{align}
Since $U_{{\ul 2}}^{\textrm{mag}}$ canceled out, we conclude that  $F^{\textrm{mag}}$ does not enter the (large-$N_c$ DLA) evolution of $F^{[2]}$.

Employing the adjoint-representation version of the operator in the first line of \eq{V_subeik_2a} (keeping the gluon sector sub-eikonal operator only), after some algebra we simplify the above expression to 
\begin{align}\label{aaa2}
& \frac{1}{4}  \int\limits_{-\infty}^0 dx_{2'}^- \,  
\int\limits_0^\infty dx_2^- \left\{ \Big[ \frac{1}{2} \, x_{2'}^- {\pd}^i a^{+ \, a} (x_{2'}^- , \ul{x}_1)
\contraction[2ex]
{}
{+}{a^{i \, a}
(x_{2'}^- , \ul{x}_1) \Big] \:}
{a}
\: 
+ a^{i \, a} (x_{2'}^- , \ul{x}_1) \Big] \:
a^{+ \, b} (x_2^- , \ul{x}_0) \right. \notag \\
& \hspace*{3cm} \left. -
\Big[ \frac{1}{2} \, x_2^- {\pd}^i a^{+ \, b} (x_2^- , \ul{x}_1)
\contraction[2ex]
{}
{+}{a^{i \, b}
(x_2^- , \ul{x}_1) \Big] \:}
{a}
\: 
+ a^{i \, b} (x_2^- , \ul{x}_1) \Big] \:
a^{+ \, a} (x_{2'}^- , \ul{x}_0) \right\} \notag \\ 
& = \frac{-i}{(2 \pi)^3} \int\limits_0^{p_2^-} d k^- \int d^2 x_2  \left[ \frac{ x_{20}^i}{x_{20}^2} - 2 x_{21}^i \frac{{\un x}_{21} \cdot {\un x}_{20}}{x_{21}^2 \, x_{20}^2} \right] \left( U_{{\ul 2}}^{[2]} \right)^{b a} ,
\end{align}
where we have defined the adjoint version of the operator in \eq{V2_full} by 
\begin{align}
U_{{\ul 2}}^{[2]} = \frac{g \, p_1^+}{2 \, s} \int\limits_{-\infty}^{\infty} d{z}^- \ U_{\un{2}} [ \infty, z^-] \, \left[ \left( \frac{1}{2} \, z^- \, \partial^i {\cal A}^+ + {\cal A}^i \right) \, \vec{\pd}^i_2  - \cev{\pd}_2^i \, \left( \frac{1}{2} \, z^- \, \partial^i {\cal A}^+ + {\cal A}^i \right)  \right]  \, U_{\un{2}} [ z^-, -\infty]
\end{align}
with ${\cal A}^\mu$ the gluon field in the adjoint representation. Note that the eikonal operators (similar to that in the first term on the right of \eq{V2_full}) appear to cancel in arriving at \eq{aaa2}.

The evolution of $F^{[2]}$ also includes the standard eikonal emissions \cite{Mueller:1994rr,Mueller:1994jq,Mueller:1995gb,Balitsky:1995ub,Balitsky:1998ya,Kovchegov:1999yj,Kovchegov:1999ua} with the eikonal propagator $\contraction{}{a^+ \:}{}{a^+} a^+ \:a^+$: as usual, such emissions only come in with the eikonal (dipole) kernel acting on the same dipole amplitude $F^{[2]}$ \cite{Kovchegov:2021iyc}. We conclude from this observation and from \eq{aaa2} that, at large $N_c$, the amplitudes $F^{\textrm{mag}}$ and $F^{i}$ do not enter the DLA evolution of $F^{[2]}$. Hence, just like in \cite{Kovchegov:2021iyc}, the (linearized) evolution of $F^{[2]}$ is closed and only involves this amplitude, with zero initial conditions at the lowest-order (Born) level. Therefore, without completing the construction of the exact evolution equation for $F^{[2]}_{10} (z)$ we can conclude that $F^{[2]}_{10} (z) =0$. This amplitude does not contribute to the Sivers function at DLA and we will not consider it further. 


\subsection{Numerical Solution of the Evolution Equations and the Small-$x$ Asymptotics of the Flavor Non-Singlet Sivers Function}


We now want to solve equations \eqref{FFFF} and \eqref{Gamma4} for the dipole amplitudes $F_A, F_B, F_C$ and $F_\textrm{mag}$ numerically. To do so, let us introduce the dimensionless variables \cite{Kovchegov:2016weo}
\begin{subequations}
\begin{align}
\eta \equiv \sqrt{\frac{\as N_c}{4 \pi}} \, \ln \frac{zs}{\Lambda^2}, \ \ \ s_{10} \equiv \sqrt{\frac{\as N_c}{4 \pi}} \, \ln \frac{1}{x_{10}^2 \Lambda^2} , \\
\eta' \equiv \sqrt{\frac{\as N_c}{4 \pi}} \, \ln \frac{z's}{\Lambda^2}, \ \ \ s_{21} \equiv \sqrt{\frac{\as N_c}{4 \pi}} \, \ln \frac{1}{x_{12}^2 \Lambda^2} , \\
\eta'' \equiv \sqrt{\frac{\as N_c}{4 \pi}} \, \ln \frac{z''s}{\Lambda^2}, \ \ \ s_{32} \equiv \sqrt{\frac{\as N_c}{4 \pi}} \, \ln \frac{1}{x_{32}^2 \Lambda^2} ,
\end{align}
\end{subequations}
allowing us to rewrite the evolution equations \eqref{FFFF} as
\begin{subequations} \label{dimless_bm}
\begin{align}
    F^{NS}_A (s_{10}, \eta) &= F^{NS (0)}_A (s_{10}, \eta) \\
    &+ \int\limits_0^{\eta} \dd{\eta'} \int\limits_{s_{10} - \eta + \eta'}^{\min \left[ s_{10}, \eta' \right]} \dd{s}_{21} \Bigg[ 6 F_A^{NS} (s_{21}, \eta') - F_B^{NS} (s_{21}, \eta') + F_C^{NS} (s_{21}, \eta') \Bigg] , \notag \\
    F^{NS}_B (s_{10}, \eta) &= F^{NS (0)}_B (s_{10}, \eta) \\
    &+ \int\limits_0^{\eta} \dd{\eta'} \int\limits_{s_{10} - \eta + \eta'}^{\min \left[ s_{10}, \eta' \right]} \dd{s}_{21} \Bigg[ -2 F_A^{NS} (s_{21}, \eta') + 5 F_B^{NS} (s_{21}, \eta') - F_C^{NS} (s_{21}, \eta') \Bigg] , \notag \\
    F^{NS}_C (s_{10}, \eta) &= F^{NS (0)}_C (s_{10}, \eta) \\
    &+ \int\limits_0^{\eta} \dd{\eta'} \int\limits_{s_{10} - \eta + \eta'}^{\min \left[ s_{10}, \eta' \right]} \dd{s}_{21} \Bigg[ 6 F_C^{NS} (s_{21}, \eta') + 2 F_{\textrm{mag}}^{NS} (s_{21}, \eta') \Bigg] , \notag \\
    F^{NS}_{\textrm{mag}} (s_{10}, \eta) &= F^{NS (0)}_{\textrm{mag}} (s_{10}, \eta) + 2 \int\limits_{\max [0, s_{10}]}^{\eta} \dd{\eta'} \int\limits_{s_{10}}^{\eta'} \dd{s}_{21} \Bigg[ 2 \Gamma_A^{NS} (s_{10}, s_{21}, \eta') \\
    &- \Gamma_B^{NS} (s_{10}, s_{21}, \eta') + 3 \Gamma_C^{NS} (s_{10}, s_{21}, \eta') + \Gamma_{\textrm{mag}}^{NS} (s_{10}, s_{21}, \eta') \Bigg] , \notag 
\end{align}
\end{subequations}
and also the neighbor dipole equations \eqref{Gamma4} as
\begin{subequations}\label{dimless_neighbor_bm}
\begin{align}
    \Gamma_A^{NS} (s_{10}, s_{21}, \eta') &= F^{NS (0)}_A (s_{10}, \eta') \\
    &+ \int\limits_0^{\eta' - s_{21} + s_{10}} \dd{\eta''} \int\limits_{s_{21} - \eta' + \eta''}^{\textrm{min}\left[ s_{10}, \eta'' \right]} \dd{s}_{32} \Bigg[ 6 F_A^{NS} (s_{32}, \eta'') - F_B^{NS} (s_{32}, \eta'') + F_C^{NS} (s_{32}, \eta'') \Bigg] , \notag \\
    \Gamma_B^{NS} (s_{10}, s_{21}, \eta') &= F^{NS (0)}_B (s_{10}, \eta') \\
    &+ \int\limits_0^{\eta' - s_{21} + s_{10}} \dd{\eta''} \int\limits_{s_{21} - \eta' + \eta''}^{\textrm{min}\left[ s_{10}, \eta'' \right]} \dd{s}_{32} \Bigg[ -2 F_A^{NS} (s_{32}, \eta'') + 5 F_B^{NS} (s_{32}, \eta'') - F_C^{NS} (s_{32}, \eta'') \Bigg] , \notag \\
    \Gamma_C^{NS} (s_{10}, s_{21}, \eta') &= F^{NS (0)}_C (s_{10}, \eta') \\
    &+ \int\limits_0^{\eta' - s_{21} + s_{10}} \dd{\eta''} \int\limits_{s_{21} - \eta' + \eta''}^{\textrm{min}\left[ s_{10}, \eta'' \right]} \dd{s}_{32} \Bigg[ 6 F_C^{NS} (s_{32}, \eta'') + 2 F_{\textrm{mag}}^{NS} (s_{32}, \eta'') \Bigg] , \notag \\
    \Gamma_{\textrm{mag}}^{NS} (s_{10}, s_{21}, \eta') &= F^{NS (0)}_{\textrm{mag}} (s_{10}, \eta') + 2 \int\limits_{\textrm{max}\left[0,s_{10}\right]}^{\eta'} \dd{\eta''} \int\limits_{\textrm{max} \left[s_{10},s_{21} - \eta' + \eta'' \right]}^{\eta''} \dd{s}_{32} \Bigg[ 2 \Gamma_A^{NS} (s_{10}, s_{32}, \eta'') \\
    &- \Gamma_B^{NS} ( s_{10}, s_{32}, \eta'') + 3 \Gamma_C^{NS} ( s_{10}, s_{32}, \eta'') + \Gamma_{\textrm{mag}}^{NS} (s_{10}, s_{32}, \eta'') \Bigg] . \notag
\end{align}
\end{subequations}
We note again that we are going to solve Eqs.~\eqref{dimless_bm} and \eqref{dimless_neighbor_bm} under the assumption that the scale $\Lambda$ simply characterizes the target, and is not necessarily an IR cutoff (cf. \cite{Kovchegov:2016weo,Kovchegov:2020hgb,Adamiak:2021ppq,Cougoulic:2022gbk}). That is, $\Lambda$ is assumed to be still perturbatively large, such that the dipoles may be larger than $1/\Lambda$. This means that $s_{10}, s_{21}$ and $s_{32}$ can assume negative values. 

We can discretize Eqs.~\eqref{dimless_bm} and \eqref{dimless_neighbor_bm} as in \cite{Kovchegov:2016weo,Kovchegov:2020hgb}, by converting the integrals over $\eta'$ and $s_{21}$ into sums over $i' = s_{21} / \Delta s$ and $j' = \eta' / \Delta \eta$ to yield dipole amplitudes of the form $F_{ij} = F (i \, \Delta s, j \, \Delta \eta) = F (s_{10}, \eta)$ in Eqs.~\eqref{dimless_bm} and by converting the integrals over $\eta''$ and $s_{32}$ into sums over $i'' = s_{32} / \Delta s$ and $j'' = \eta'' / \Delta \eta$ to yield neighbor dipole amplitudes of the form $\Gamma_{ikj} = \Gamma ( i \, \Delta s, k \, \Delta s, j \, \Delta \eta) = \Gamma (s_{10}, s_{21}, \eta')$ in Eqs.~\eqref{dimless_neighbor_bm}. We will take a grid with $-N \le i, i', i'' \le N$ and $0 \le j, j', j'' \le N$, where $N$ ($2 N$) is the number of grid steps in the $\eta$-($s_{10}$-)directions, and with equal step sizes $\Delta s = \Delta \eta = \eta_{\textrm{max}} / N$ determined by the maximum $\eta$-value $\eta_{\textrm{max}}$ \cite{Kovchegov:2016weo}.
This discretization yields the following equations for the dipole amplitudes

\begin{subequations} \label{disc_bm}
\begin{align}
    &F^A_{ij} = F^{A (0)}_{ij} + \Delta s \Delta \eta \sum_{j' = 0}^{j-1} \, \sum_{i' = i - j + j' }^{\textrm{min}\left[ i,j' \right] } \left[ 6 F^A_{i'j'} - F^B_{i'j'} + F^C_{i'j'} \right] , \\
    &F^B_{ij} = F^{B (0)}_{ij} + \Delta s \Delta \eta \sum_{j' = 0}^{j-1} \, \sum_{i' = i - j + j' }^{\textrm{min}\left[ i,j' \right] } \left[ -2 F^A_{i'j'} + 5 F^B_{i'j'} - F^C_{i'j'} \right] , \\
    &F^C_{ij} = F^{C (0)}_{ij} + \Delta s \Delta \eta \sum_{j' = 0}^{j-1} \, \sum_{i' = i - j + j' }^{\textrm{min}\left[ i,j' \right] } \left[ 6 F^C_{i'j'} + 2 F^{\textrm{mag}}_{i'j'} \right] , \\
    &F^{\textrm{mag}}_{ij} = F^{\textrm{mag} (0)}_{ij} + 2 \Delta s \Delta \eta \sum_{j' = \max \left[0, i \right]}^{j-1} \, \sum_{i' = i}^{j' } \left[ 2 \Gamma^A_{i i' j'} - \Gamma^B_{i i' j'} + 3 \Gamma^C_{i i' j'} + \Gamma^{\textrm{mag}}_{i i' j'} \right] ,
\end{align}
\end{subequations}

and for the neighbor dipole amplitudes

\begin{subequations} \label{disc_neighbor_bm}
\begin{align}
    &\Gamma^A_{i i' j'} = F^{A (0)}_{ij'} + \Delta s \Delta \eta \sum_{j'' = 0}^{i + j' - i' - 1} \sum_{i'' = i' + j'' - j' }^{\textrm{min} \left[ i, j'' \right]}  \left[ 6 F^A_{i'' j''} - F^B_{i'' j''} + F^C_{i'' j''} \right] , \\
    &\Gamma^B_{i i' j'} = F^{B (0)}_{ij'} + \Delta s \Delta \eta \sum_{j'' = 0}^{i + j' - i' - 1} \sum_{i'' = i' + j'' - j' }^{\textrm{min} \left[ i, j'' \right]}  \left[ -2 F^A_{i'' j''} + 5 F^B_{i'' j''} - F^C_{i'' j''} \right] , \\
    &\Gamma^C_{i i' j'} = F^{C (0)}_{ij'} + \Delta s \Delta \eta \sum_{j'' = 0}^{i + j' - i' - 1} \sum_{i'' = i' + j'' - j'}^{\textrm{min} \left[ i, j'' \right]}  \left[ 6 F^C_{i'' j''} + 2 F^{\textrm{mag}}_{i'' j''} \right] , \\
    &\Gamma^{\textrm{mag}}_{i i' j'} = F^{\textrm{mag} (0)}_{ij'} + 2 \Delta s \Delta \eta \sum_{j'' = \textrm{max}\left[0,i\right]}^{j'-1} \sum_{i'' = \textrm{max} \left[ i, i' + j'' - j' \right]}^{j''}  \left[ 2 \Gamma^A_{i i'' j''} - \Gamma^B_{i i'' j''} + 3 \Gamma^C_{i i'' j''} + \Gamma^{\textrm{mag}}_{i i'' j''} \right] . 
\end{align}
\end{subequations}

These equations can be directly evaluated numerically, but this will require performing a double summation to calculate the dipole amplitude for every point in the grid. Instead, we will use the approach developed in \cite{Kovchegov:2020hgb}: subtracting Eqs.~\eqref{disc_bm} and \eqref{disc_neighbor_bm} with $j \rightarrow j-1$ from Eqs.~\eqref{disc_bm} and \eqref{disc_neighbor_bm}, one can show \cite{Kovchegov:2020hgb} that the equations can be rewritten recursively in the form $F_{ij} = F_{i (j-1)} + \ldots$, allowing one to perform a single summation for each value of the dipole amplitudes. Assuming, for simplicity, that all the inhomogeneous terms are constants independent of the indices $i,j$, one can show that the recursive equations for the dipole amplitudes are  

\begin{subequations}
\begin{align}
    F^A_{ij} = F^A_{i (j-1)} &+ \Delta s \Delta \eta \Bigg[ 6 F^A_{i, j-1} - F^B_{i, j-1} + F^C_{i, j-1} \\
    &+ \sum_{j' = 0}^{j-1}  \Bigg( 6 F^A_{i + j' - j, j'} - F^B_{i + j' - j, j'} + F^C_{i + j' - j, j'} \Bigg) \Bigg] , \notag \\
    F^B_{ij} = F^B_{i (j-1)} &+ \Delta s \Delta \eta \Bigg[ -2 F^A_{i (j-1)} + 5 F^B_{i (j-1)} - F^C_{i (j-1)} \\
    &+ \sum_{j' = 0}^{j-1}  \Bigg( -2 F^A_{(i + j' - j) j'} + 5 F^B_{(i + j' - j)  j'} - F^C_{(i + j' - j) j'} \Bigg) \Bigg] , \notag \\
    F^C_{ij} = F^C_{i (j-1)} &+ \Delta s \Delta \eta \Bigg[ 6 F^C_{i (j-1)} + 2 F^{\textrm{mag}}_{i (j-1)} + \sum_{j' = 0}^{j-1}  \Bigg( 6 F^C_{(i + j' - j) j'} + 2 F^{\textrm{mag}}_{(i + j' - j) j'} \Bigg) \Bigg] , \\
    F^{\textrm{mag}}_{ij} = F^{\textrm{mag}}_{i (j-1)} &+ 2 \Delta s \Delta \eta \Bigg[ \sum_{i' = i}^{j-1}  \Bigg( 2 \Gamma^A_{i i' (j-1)} - \Gamma^B_{i i' (j-1)} + 3 \Gamma^C_{i i' (j-1)} + \Gamma^{\textrm{mag}}_{i i' (j-1)} \Bigg) \Bigg] , \notag
\end{align}
\end{subequations}

and for the neighbor dipoles

\begin{subequations}
\begin{align}
    \Gamma^A_{i k j} &= \Gamma^A_{i k (j-1)} + \Delta s \Delta \eta \Bigg[ 6 F^A_{i \, \textrm{max}\left[ 0, i + j - k - 1 \right]} - F^B_{i \, \textrm{max}\left[ 0, i + j - k - 1 \right]} + F^C_{i \, \textrm{max}\left[ 0, i + j - k - 1 \right]} \\
    &+ \sum_{j'' = 0}^{i + j - k - 1} \Bigg( 6 F^A_{(k + j'' - j) j''} - F^B_{(k + j'' - j) j''} + F^C_{(k + j'' - j) j''} \Bigg) \Bigg] , \notag \\
    \Gamma^B_{i k j} &= \Gamma^B_{i k (j-1)} + \Delta s \Delta \eta \Bigg[ -2 F^A_{i \, \textrm{max}\left[ 0, i + j - k - 1 \right]} + 5 F^B_{i \, \textrm{max}\left[ 0, i + j - k - 1 \right]} - F^C_{i \, \textrm{max}\left[ 0, i + j - k - 1 \right]} \\
    &+ \sum_{j'' = 0}^{i + j - k - 1} \Bigg( -2 F^A_{(k + j'' - j) j''} + 5 F^B_{(k + j'' - j) j''} - F^C_{(k + j'' - j) j''} \Bigg) \Bigg] , \notag \\
    \Gamma^C_{i k j} &= \Gamma^C_{i k (j-1)} + \Delta s \Delta \eta \Bigg[ 6 F^C_{i \, \textrm{max}\left[ 0, i + j - k - 1 \right]} + 2 F^{\textrm{mag}}_{i \, \textrm{max}\left[ 0, i + j - k - 1 \right]} \\
    &+ \sum_{j'' = 0}^{i + j - k - 1} \Bigg( 6 F^C_{(k + j'' - j) j''} - F^{\textrm{mag}}_{(k + j'' - j) j''} + F^C_{(k + j'' - j) j''} \Bigg) \Bigg] , \notag \\
    \Gamma^{\textrm{mag}}_{i k j} 
    &= \Gamma^{\textrm{mag}}_{i k (j-1)} + 2 \Delta s \Delta \eta \Bigg[ \sum_{j'' = \textrm{max} \left[0, i + j - k \right]}^{j - 1} \Bigg( 2 \Gamma^A_{i (k + j'' - j) j''} - \Gamma^B_{i (k + j'' - j) j''} + 3 \Gamma^C_{i (k + j'' - j) j''} + \Gamma^{\textrm{mag}}_{i (k + j'' - j) j''} \Bigg) \\
    &+ \sum_{i'' = k}^{j-1} \Bigg( 2 \Gamma^A_{i i'' (j-1)} - \Gamma^B_{i i'' (j-1)} + 3 \Gamma^C_{i i'' (j-1)} + \Gamma^{\textrm{mag}}_{i i'' (j-1)} \Bigg) \Bigg] . \notag
\end{align}
\end{subequations}


\begin{figure}[ht]
\begin{subfigure}{0.4\textwidth}
  \centering
  \includegraphics[width=1.0\linewidth]{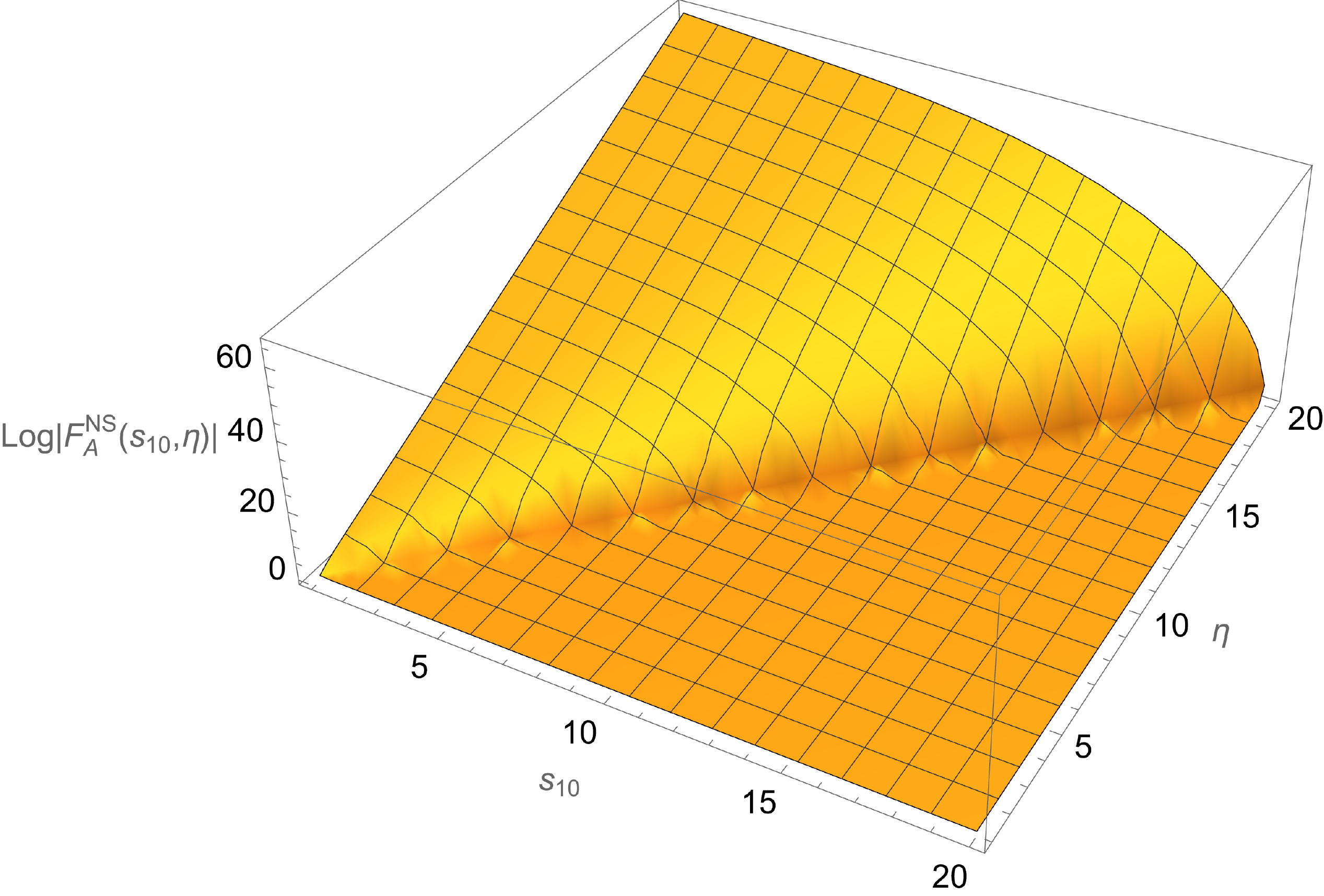}  
  \caption{$F_A^{NS}$}
  \label{FIG:num_dips_A}
\end{subfigure}
\begin{subfigure}{0.4\textwidth}
  \centering
  \includegraphics[width=1.0\linewidth]{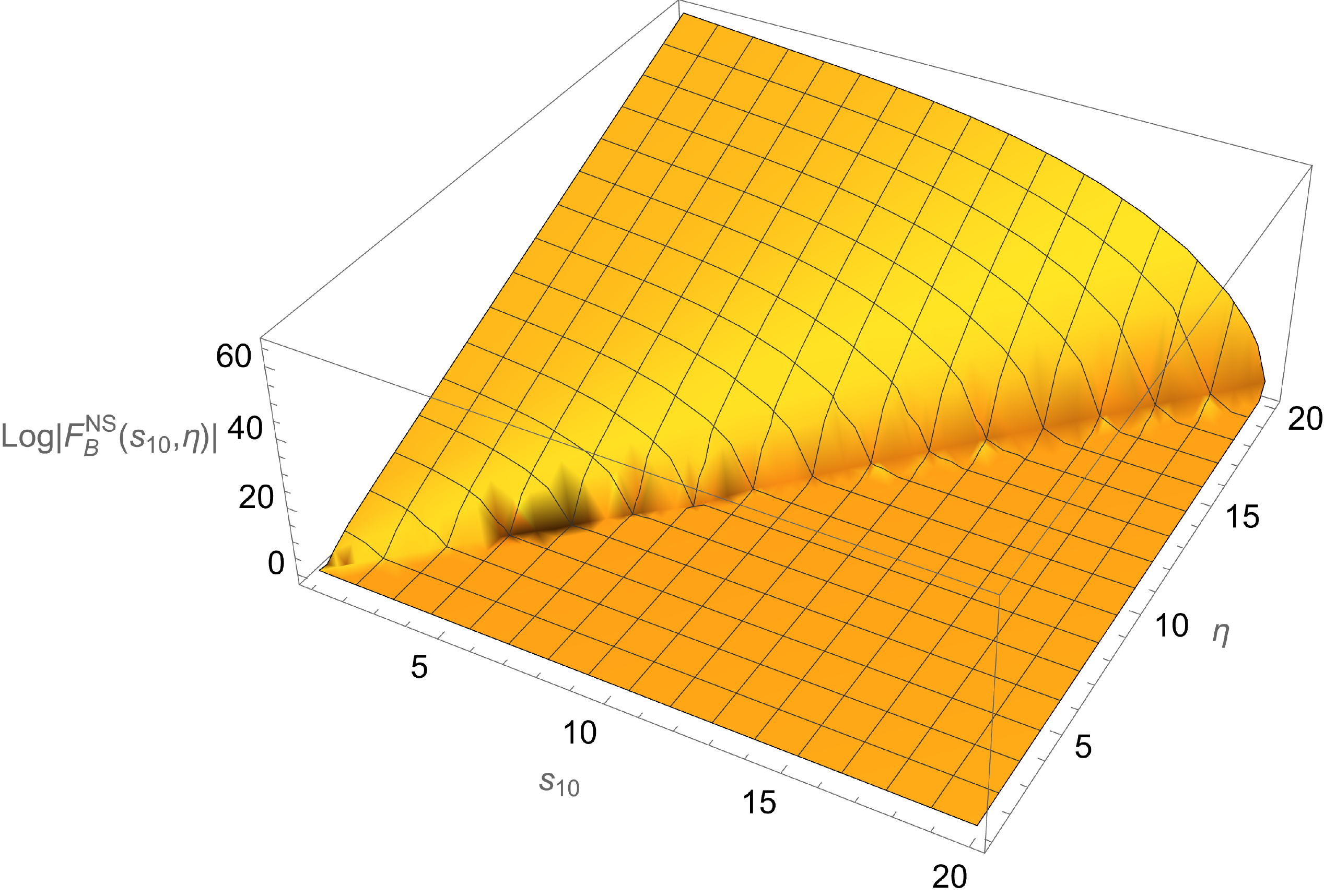}  
  \caption{$F_B^{NS}$}
  \label{FIG:num_dips_B}
\end{subfigure}
\begin{subfigure}{0.4\textwidth}
  \centering
  \includegraphics[width=1.0\linewidth]{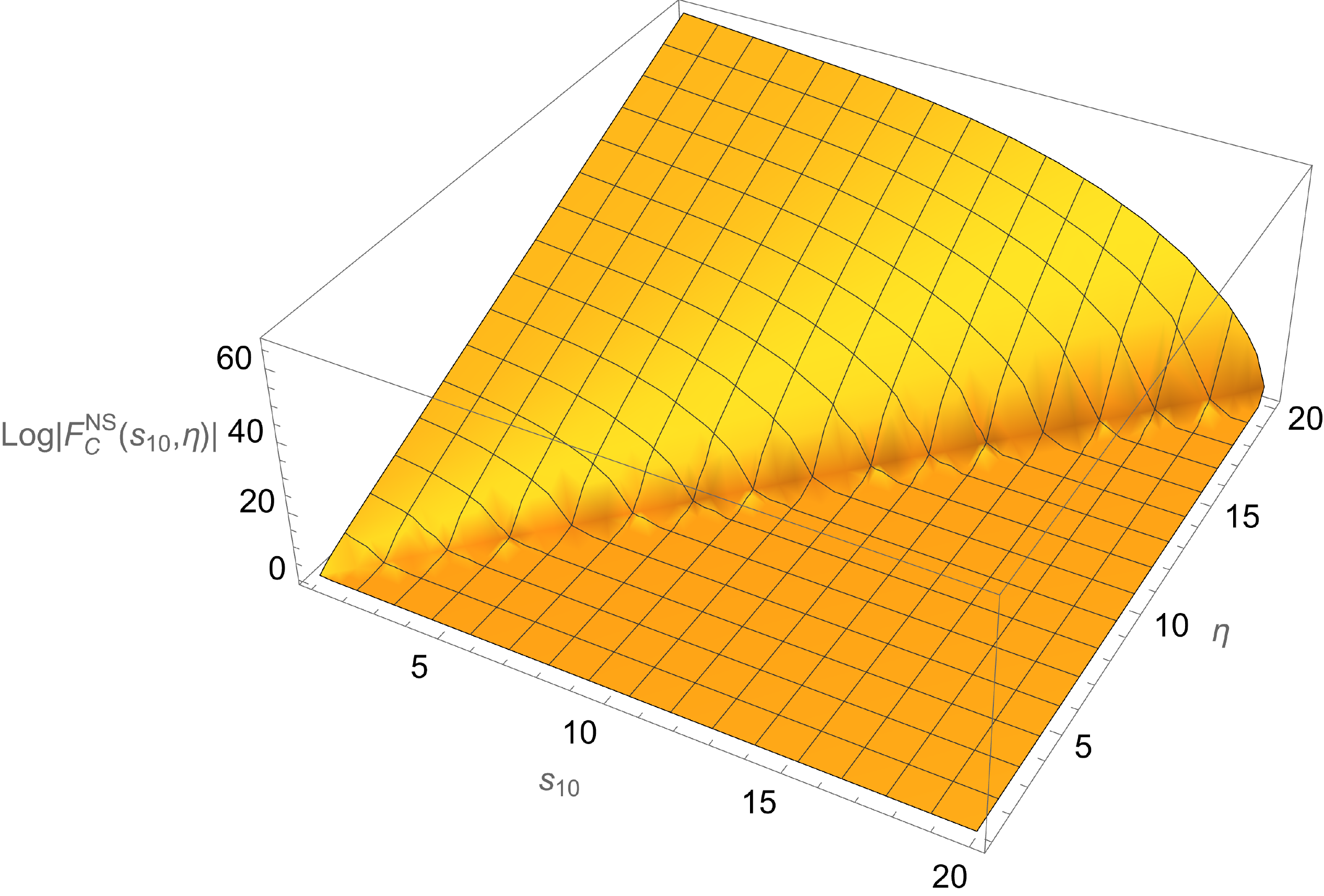}  
  \caption{$F_C^{NS}$}
  \label{FIG:num_dips_C}
\end{subfigure}
\begin{subfigure}{0.4\textwidth}
  \centering
  \includegraphics[width=1.0\linewidth]{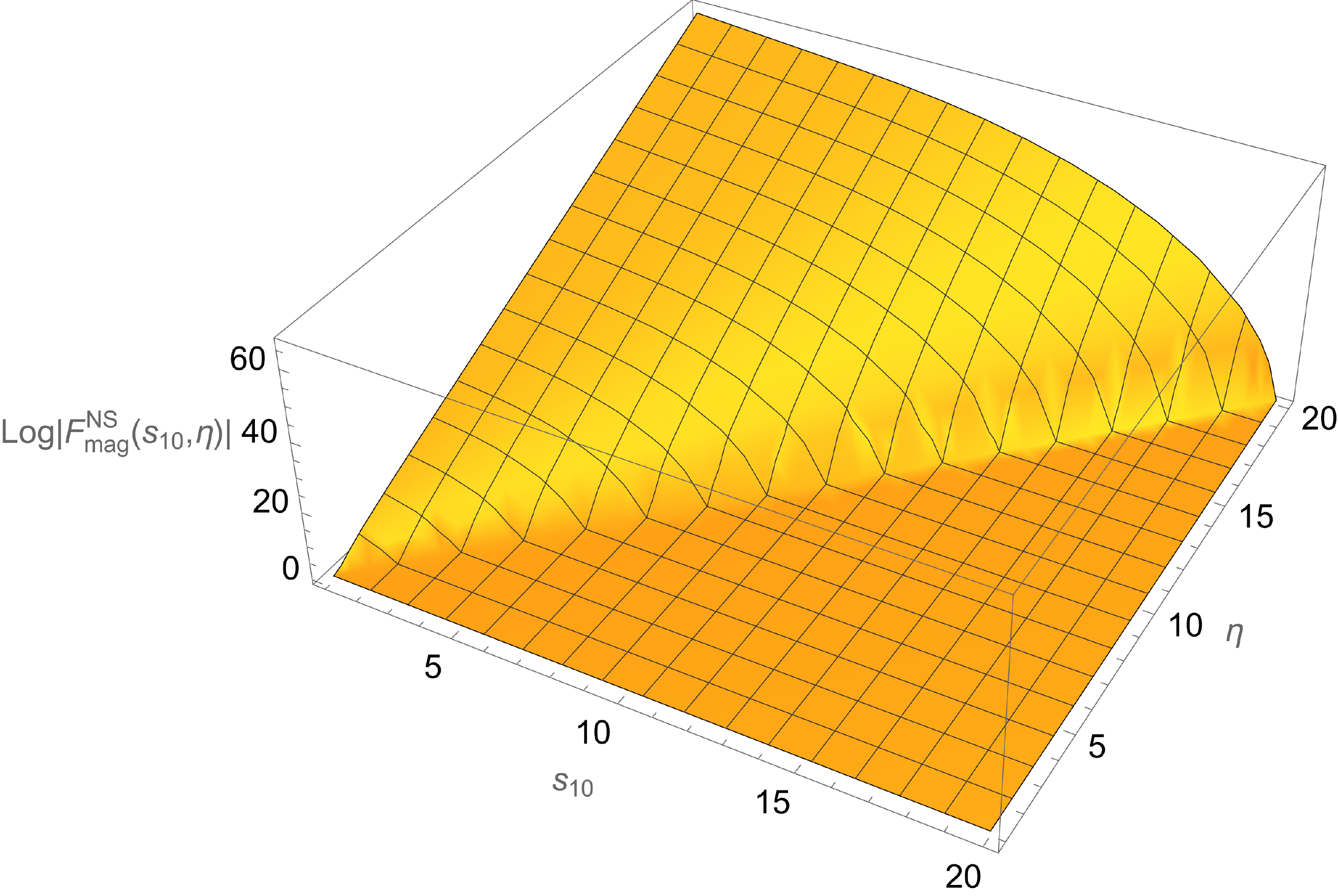}  
  \caption{$F_{\textrm{mag}}^{NS}$}
  \label{FIG:num_dips_m}
\end{subfigure}
\caption{Plots of our numerical solutions for the dipole amplitudes numerically calculated at $\eta_{\textrm{max}} = 20$ and $\Delta \eta = \Delta s = \eta_{\textrm{max}} /N=0.033$. The plots depict the logarithm of the absolute value of each dipole amplitude as a function of variables $\eta$ and $s_{10}$.}
\label{FIG:num_dips}
\end{figure}

\begin{figure}[h]
\begin{subfigure}{0.4\textwidth}
  \centering
  \includegraphics[width=1.0\linewidth]{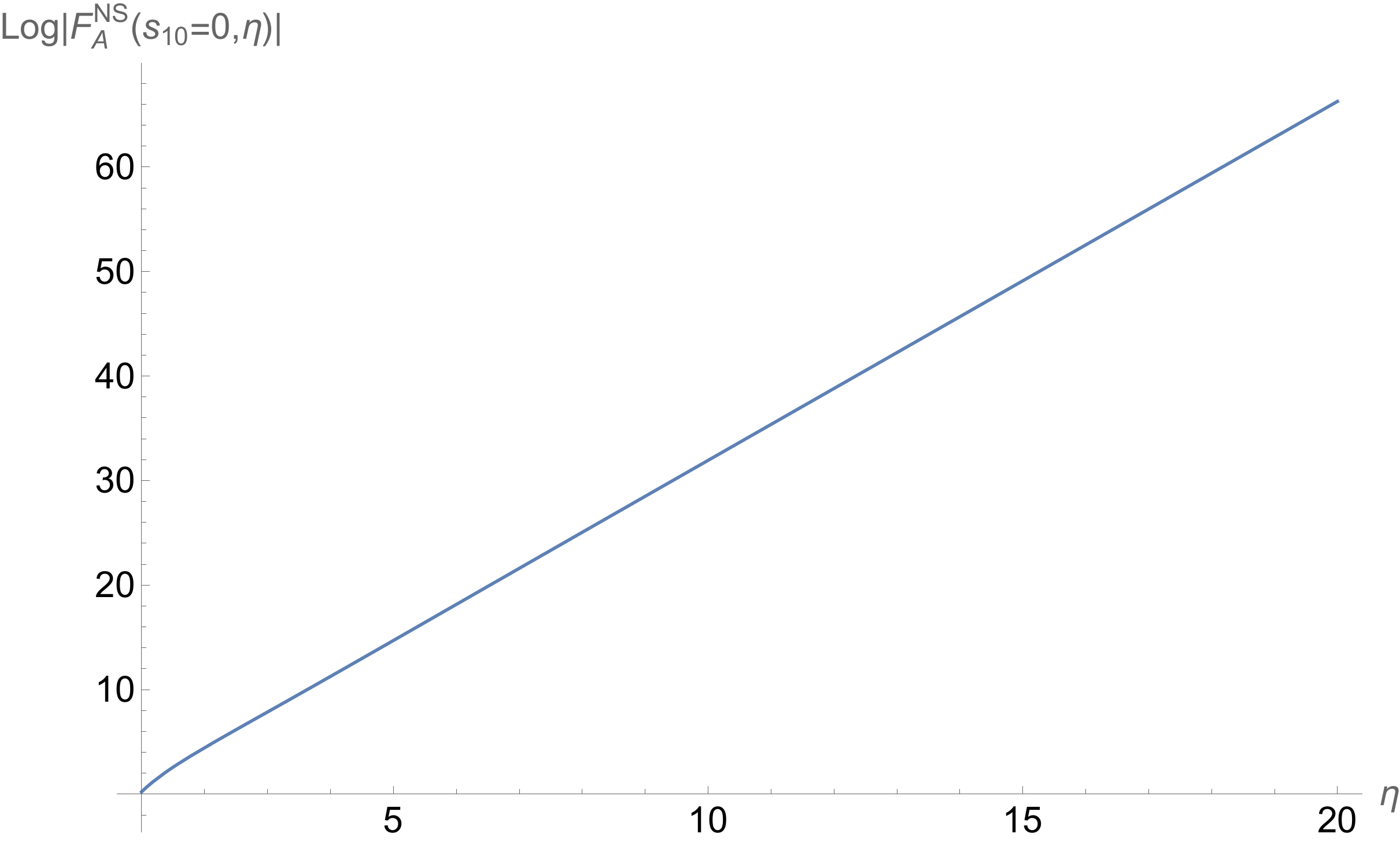}  
  \caption{$F_A^{NS}$}
  \label{FIG:num_dips_A_1d}
\end{subfigure}
\begin{subfigure}{0.4\textwidth}
  \centering
  \includegraphics[width=1.0\linewidth]{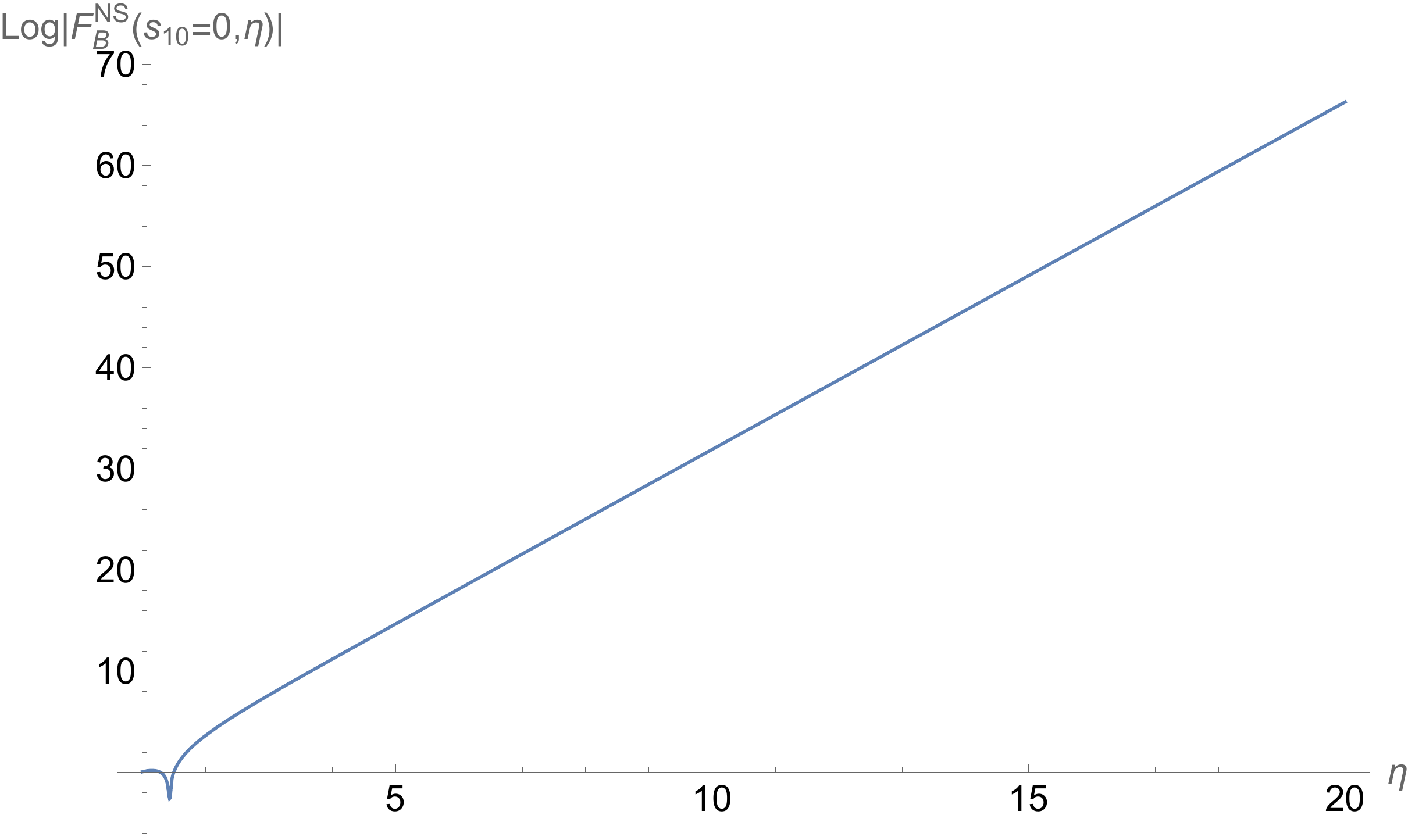}  
  \caption{$F_B^{NS}$}
  \label{FIG:num_dips_B_1d}
\end{subfigure}
\begin{subfigure}{0.4\textwidth}
  \centering
  \includegraphics[width=1.0\linewidth]{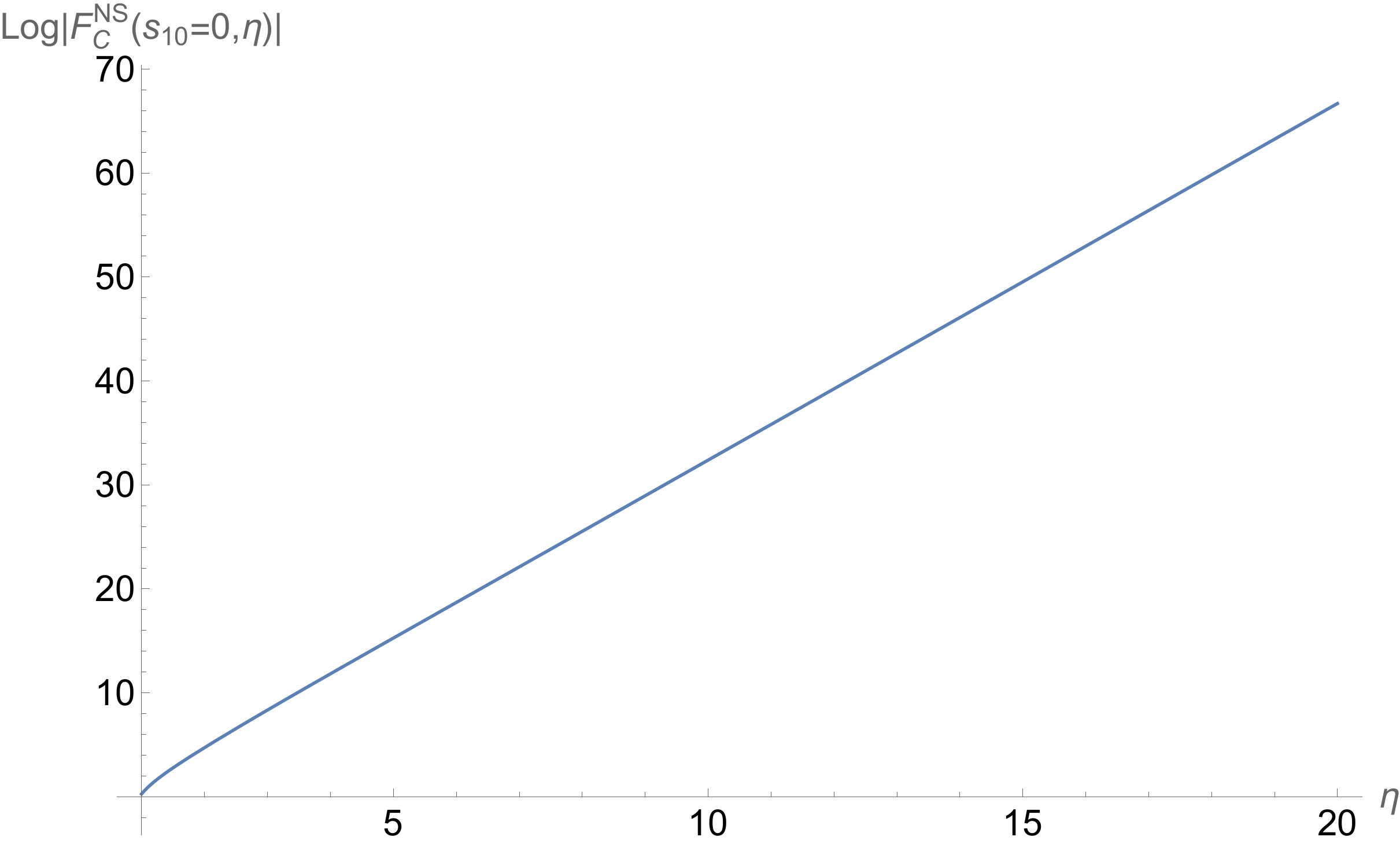}  
  \caption{$F_C^{NS}$}
  \label{FIG:num_dips_C_1d}
\end{subfigure}
\begin{subfigure}{0.4\textwidth}
  \centering
  \includegraphics[width=1.0\linewidth]{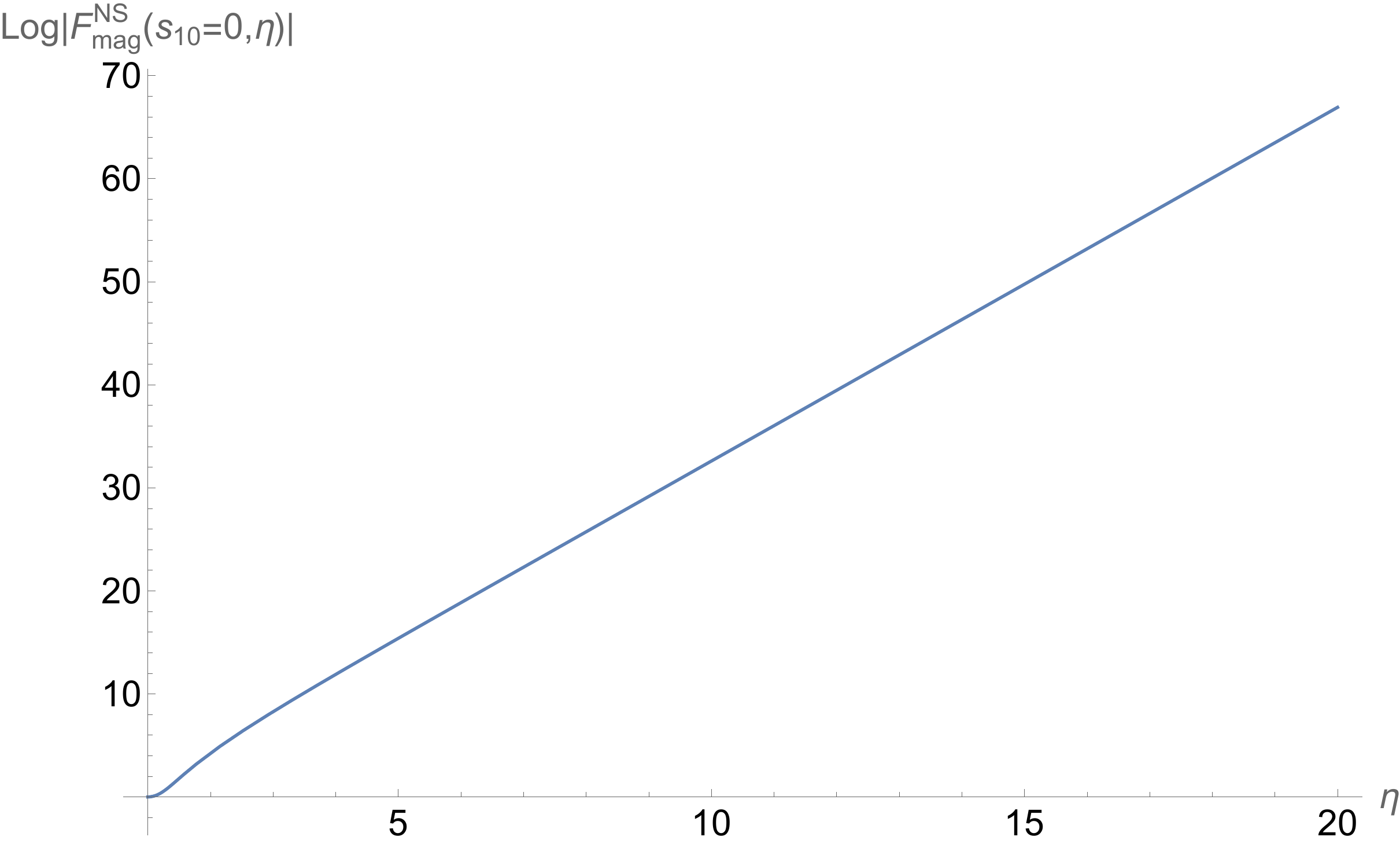}  
  \caption{$F_{\textrm{mag}}^{NS}$}
  \label{FIG:num_dips_m_1d}
\end{subfigure}
\caption{Plots of our numerical solutions for the dipole amplitudes numerically calculated at $\eta_{\textrm{max}} = 20$ and $\Delta \eta = \Delta s = \eta_{\textrm{max}} /N=0.033$. The plots depict the logarithm of the absolute value of each dipole amplitude as a function of the single variable $\eta$ along the line $s_{10}=0$.}
\label{FIG:num_dips_1d}
\end{figure}

We obtain the dipole amplitudes in the physical region $\eta \geq s_{10}$ by performing these sums over the range $j - N \leq i \leq k \leq j$, starting with $j = 1$ and running up to $j = N$ where $\eta = \eta_{\textrm{max}}$. For simplicity in the numerical calculations and extrapolations we take the initial conditions of all the dipole amplitudes to be unity, $F^{A (0)} = F^{B (0)} = F^{C (0)} = F^{\textrm{mag} (0)} = 1$. In \fig{FIG:num_dips} we plot the logarithm of the absolute value of the numerically calculated polarized dipole amplitudes as a function of $\eta$ and $s_{10}$ at $1/\eta_{\textrm{max}} = 0.05$ and $\eta_{\textrm{max}} /N=0.033$. In \fig{FIG:num_dips_1d} we plot the same logarithm of the absolute value, now as a function of $\eta$ only along the line $s_{10}=0$, demonstrating the asymptotic linear behavior. This indicates that the dipole amplitudes grow exponentially with $\eta$: we, therefore, write
\begin{subequations}\label{dip_asympt}
\begin{align}
    &F_A^{NS} (s_{10}=0, \eta) = F_A^{NS (0)} e^{\alpha_A \, \eta} , \\
    &F_B^{NS} (s_{10}=0, \eta)= F_B^{NS (0)} e^{\alpha_B \, \eta } , \\
    &F_C^{NS} (s_{10}=0, \eta)= F_C^{NS (0)} e^{\alpha_C \, \eta} , \\
    &F_{\textrm{mag}}^{NS} (s_{10}=0, \eta) = F_{\textrm{mag}}^{NS (0)} e^{\alpha_{\textrm{mag}} \, \eta } ,
\end{align}
\end{subequations}
such that each dipole amplitude contributes a factor proportional to $(1/x)^{\alpha \sqrt{\alpha_s N_c / 4 \pi}}$ to the flavor non-singlet Sivers function, with the rescaled intercept $\alpha$.

Our numerical solution follows the steps performed in \cite{Kovchegov:2016weo,Kovchegov:2020hgb,Cougoulic:2022gbk} for helicity evolution. Once we obtain the numerical data for the amplitudes, we extract the intercept by taking a linear regression of the logarithm of the absolute value of the dipole amplitudes for $s_{10}=0$ and $0.75 \ \eta_{\textrm{max}} \leq \eta \leq \eta_{\textrm{max}}$, keeping the standard error of the regression as our uncertainty. We calculate this intercept for values of $\eta_{\textrm{max}}$ ranging from $10$ up to $40$ and grid spacing $\eta_{\textrm{max}}/N$ ranging from $0.1$ down to $0.025$. We find that the uncertainties from the linear regression decrease rapidly as we increase $\eta_{\textrm{max}}$ while keeping $\eta_{\textrm{max}}/N$ fixed. The decrease of the error is exponentially fast, falling from $\mathcal{O}(10^{-4})$ for $\eta_{\textrm{max}} = 10$ to $\mathcal{O}(10^{-8})$ for $\eta_{\textrm{max}} = 40$. If we perform the regression over the same $\eta$ range using an amplitude calculated with two different $\eta_{\textrm{max}}$ values, say $\eta_{\textrm{Ma}}$ and $\eta_{\textrm{Mb}}$ with
\begin{align}
\eta_{\textrm{Ma}} > \eta_{\textrm{Mb}},    
\end{align}
and taking the $\eta$ range to be from $0.75 \eta_{\textrm{Mb}}$ to $\eta_{\textrm{Mb}}$, both amplitudes will yield identical results including the uncertainty from the statistical error of the linear regression. This demonstrates that the uncertainties are not due to numerical artifacts and that there is some physical `pre-asymptotic' behavior which hinders the precise numerical extraction of the intercept. 
We find that the following modified ansatz (cf. Eqs.~\eqref{dip_asympt}) captures the `pre-asymptotic' behavior rather well:
\begin{subequations}\label{dip_asympt_adj}
\begin{align}
    &F_A^{NS} (s_{10}=0, \eta) = F_A^{NS (0)} e^{\alpha_A \, \eta \, + \, \beta_A \, e^{-\gamma_A \, \eta}} , \\
    &F_B^{NS} (s_{10}=0, \eta)= F_B^{NS (0)} e^{\alpha_B \, \eta \, + \, \beta_B \, e^{-\gamma_B \, \eta}} , \\
    &F_C^{NS} (s_{10}=0, \eta)= F_C^{NS (0)} e^{\alpha_C \, \eta \, + \, \beta_C \, e^{-\gamma_C \, \eta}} , \\
    &F_{\textrm{mag}}^{NS} (s_{10}=0, \eta) = F_{\textrm{mag}}^{NS (0)} e^{\alpha_{\textrm{mag}} \, \eta \, + \, \beta_{\textrm{mag}} \, e^{-\gamma_{\textrm{mag}} \, \eta}} .
\end{align}
\end{subequations}
The statistical errors in the linear regressions are at most $\mathcal{O} (10^{-4})$, so these exponentially suppressed `pre-asymptotic' $\beta \, e^{-\gamma \, \eta}$ terms added into the exponent are very small in the $0.75 \ \eta_{\textrm{max}} \leq \eta \leq \eta_{\textrm{max}}$ range we use to extract the intercept $\alpha$. Thus, the parameters $\beta_A, \gamma_A, \beta_B, \gamma_B,$ etc. are difficult to extract numerically, but unimportant for extracting the intercepts $\alpha_A, \alpha_B, \alpha_C$ and $\alpha_{\textrm{mag}}$ to a precision $\mathcal{O} (10^{-3})$. Our estimates of the `pre-asymptotic' parameters are
\begin{subequations}
\begin{align}
    &\beta_A \approx -1.2, \ \gamma_A \approx 0.4 , \\
    &\beta_B \approx -1.2, \ \gamma_B \approx 0.4 , \\
    &\beta_C \approx 0.7, \ \gamma_C \approx 0.4 , \\
    &\beta_{\textrm{mag}} \approx 0.04, \ \gamma_{\textrm{mag}} \approx 0.3 .
\end{align}
\end{subequations}

One may worry that the ``exponential of an exponential" structure in Eqs.~\eqref{dip_asympt_adj} may appear somewhat unusual in small-$x$ QCD. However, since at small $x$ we have $\eta \sim \ln (1/x)$, one can see that 
\begin{align}\label{inf_series}
    e^{\alpha \, \eta \, + \, \beta \, e^{-\gamma \, \eta}} \to \left( \frac{1}{x} \right)^{\alpha \, \sqrt{\frac{\as \, N_c}{4 \pi}}} \, e^{\beta \, x^{\gamma \, \sqrt{\frac{\as \, N_c}{4 \pi}}}} = \left( \frac{1}{x} \right)^{\alpha \, \sqrt{\frac{\as \, N_c}{4 \pi}}} + \beta \, \left( \frac{1}{x} \right)^{(\alpha - \gamma) \, \sqrt{\frac{\as \, N_c}{4 \pi}}} + \frac{\beta^2}{2} \, \left( \frac{1}{x} \right)^{(\alpha - 2 \gamma) \, \sqrt{\frac{\as \, N_c}{4 \pi}}} + \ldots . 
\end{align}
We see that the pre-asymptotic $\beta \, e^{-\gamma \, \eta}$ terms may arise from summing an infinite series of  power-of-$x$ terms which are sub-leading at small $x$. While the infinite series in \eq{inf_series} is still somewhat unusual and its origin is unclear to us at the moment, the individual power-of-$x$ terms in it are quite common in small-$x$ physics. 


We want to extrapolate our numerical results to the continuum solution by taking the limit where we send both $1/ \eta_{\textrm{max}} \rightarrow 0$ and $\eta_{\textrm{max}}/N = \Delta \rightarrow 0$. To do so, we fit multiple polynomial models for the four intercepts as functions of $1/\eta_{\textrm{max}}$ and $\Delta$ using the the results of our linear regressions for data in the $(1/\eta_{\textrm{max}}, \Delta)$ plane. As the errors due to the `pre-asymptotic' behavior are smaller than the precision we are looking for, we fit the data points without any weighting by their uncertainties. Weighting by the uncertainties would give a better continuum extrapolation in principle, but for our data the uncertainties vary by several orders of magnitude with changing $\eta_{\textrm{max}}$, so the weights for points with $\eta_{\textrm{max}} \sim 40$ which are well within the asymptotic behavior regime will wash out the data points with $\eta_{\textrm{max}} \sim 10$ and the fitting routine will not accurately capture the approach to the continuum limit with increasing $\eta_{\textrm{max}}$. Our polynomial models range from just a constant term up to a third-order polynomial in both variables including all possible interaction terms, so for a given intercept $\alpha$ we have (cf. \cite{Kovchegov:2016weo,Kovchegov:2020hgb,Cougoulic:2022gbk})
\begin{subequations}
\begin{align}
    &\alpha^{\textrm{const}} = a , \\
    &\alpha^{\textrm{lin}} = a + b_1 \ \frac{1}{\eta_{\textrm{max}}} + b_2 \ \Delta ,\\
    &\alpha^{\textrm{quad}} = a + b_1 \ \frac{1}{\eta_{\textrm{max}}} + b_2 \ \Delta + c_1 \ \frac{1}{\eta^2_{\textrm{max}}} + c_2 \ \Delta^2 + c_3 \ \Delta \ \frac{1}{\eta_{\textrm{max}}} ,\\ 
    &\alpha^{\textrm{cub}} = a + b_1 \ \frac{1}{\eta_{\textrm{max}}} + b_2 \ \Delta + c_1 \ \frac{1}{\eta^2_{\textrm{max}}} + c_2 \ \Delta^2 + c_3 \ \Delta \ \frac{1}{\eta_{\textrm{max}}} + d_1 \ \frac{1}{\eta^3_{\textrm{max}}} + d_2 \ \Delta^3 \\
    & \hspace{1cm} + d_3 \ \Delta \ \frac{1}{\eta^2_{\textrm{max}}} + d_4 \ \Delta^2 \ \frac{1}{\eta_{\textrm{max}}}. \notag 
\end{align}
\end{subequations}
We fit these polynomials to our data for the four dipole intercepts and then calculate the Akaike Information Criterion (AIC) \cite{Akaike:1974} for each model. The AIC increases going from the constant term model up to the quadratic model, then decreases moving to the cubic model, and the new parameters introduced for the cubic model beyond those in the quadratic model all have insignificant t-statistics. Thus, we take the intercepts from the best fit quadratic polynomials
\begin{figure}[ht]
\begin{subfigure}{0.4\textwidth}
  \centering
  \includegraphics[width=1.0\linewidth]{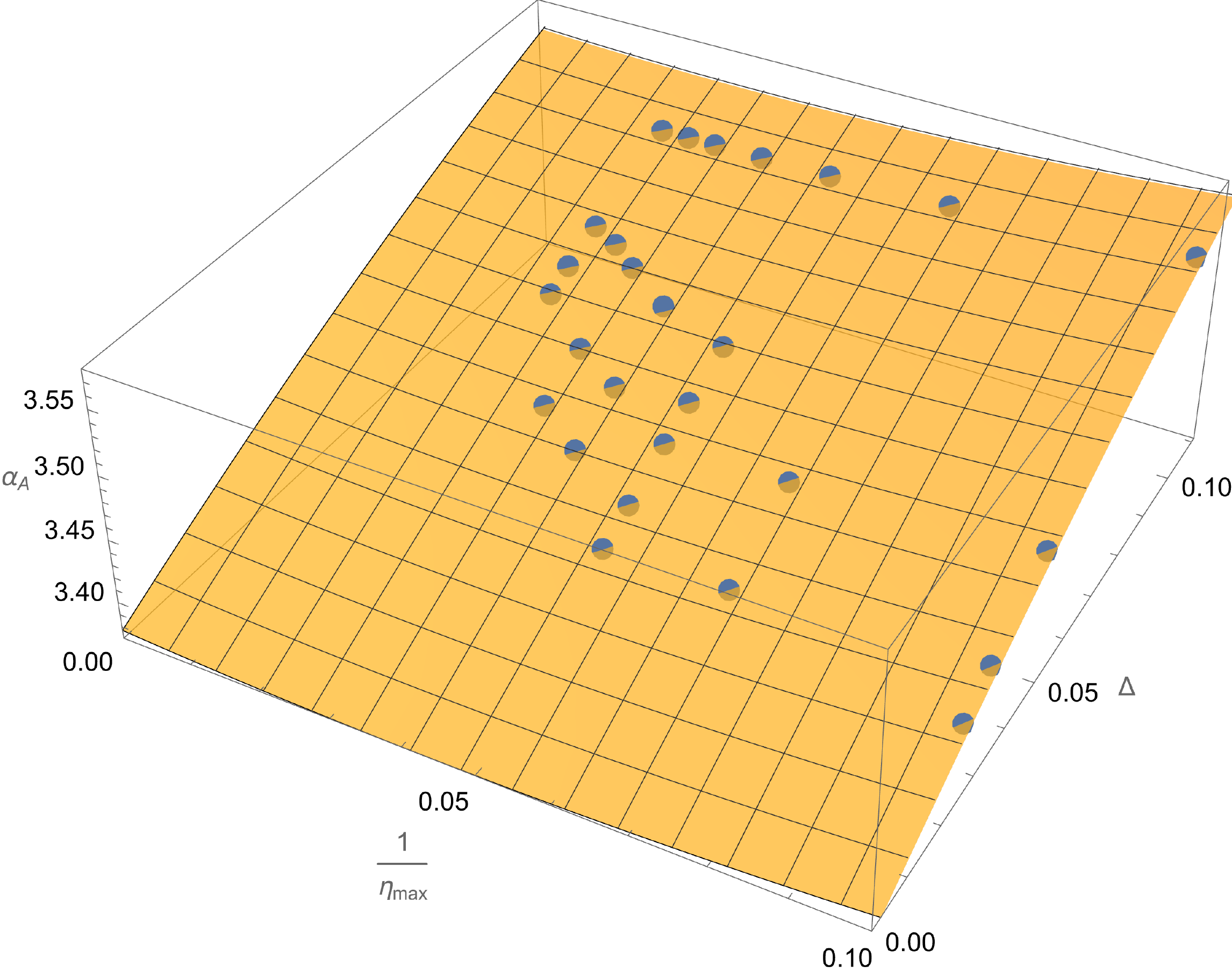}  
  \caption{$\alpha_A$}
  \label{FIG:best_fit_A}
\end{subfigure}
\begin{subfigure}{0.4\textwidth}
  \centering
  \includegraphics[width=1.0\linewidth]{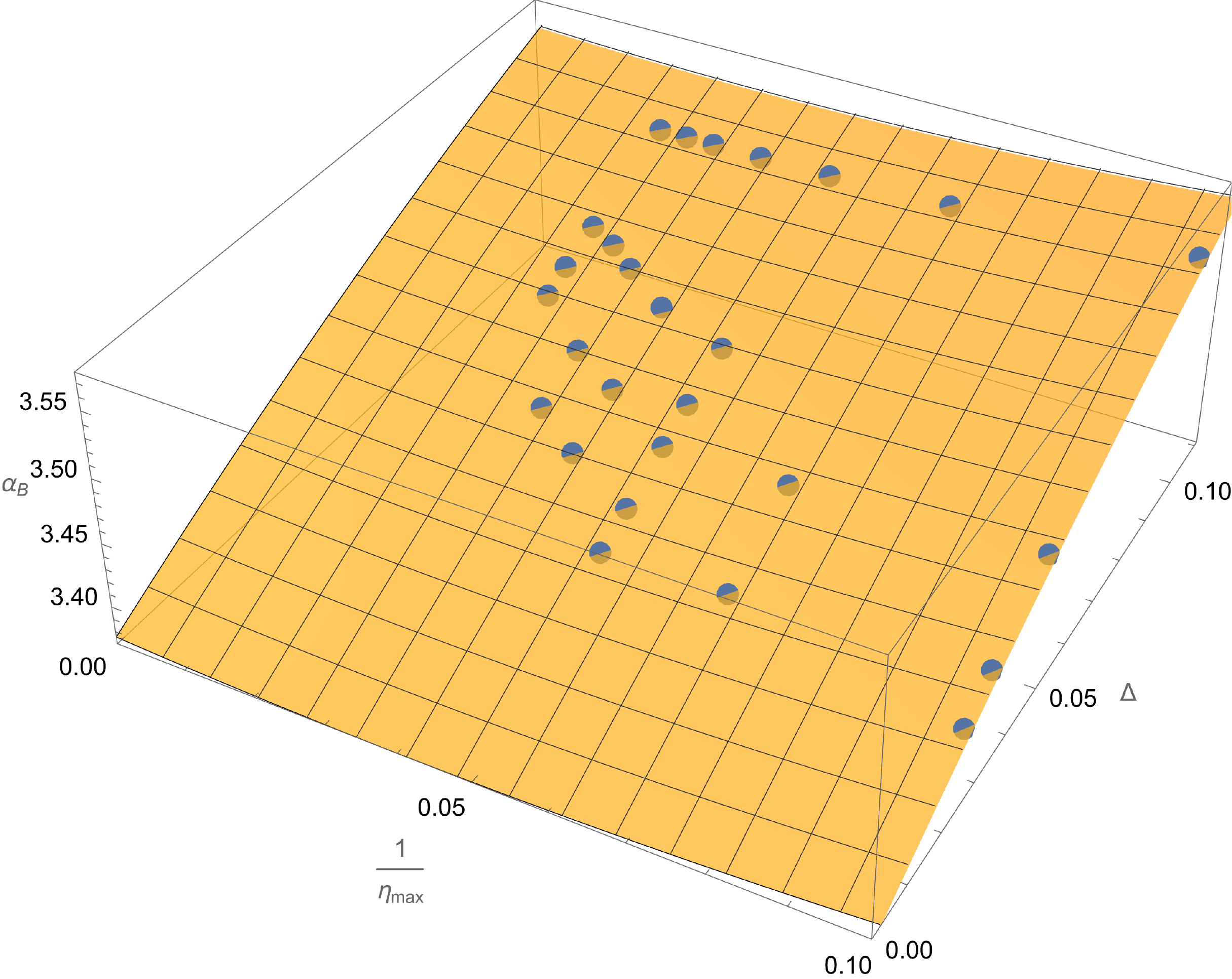}  
  \caption{$\alpha_B$}
  \label{FIG:best_fit_B}
\end{subfigure}
\begin{subfigure}{0.4\textwidth}
  \centering
  \includegraphics[width=1.0\linewidth]{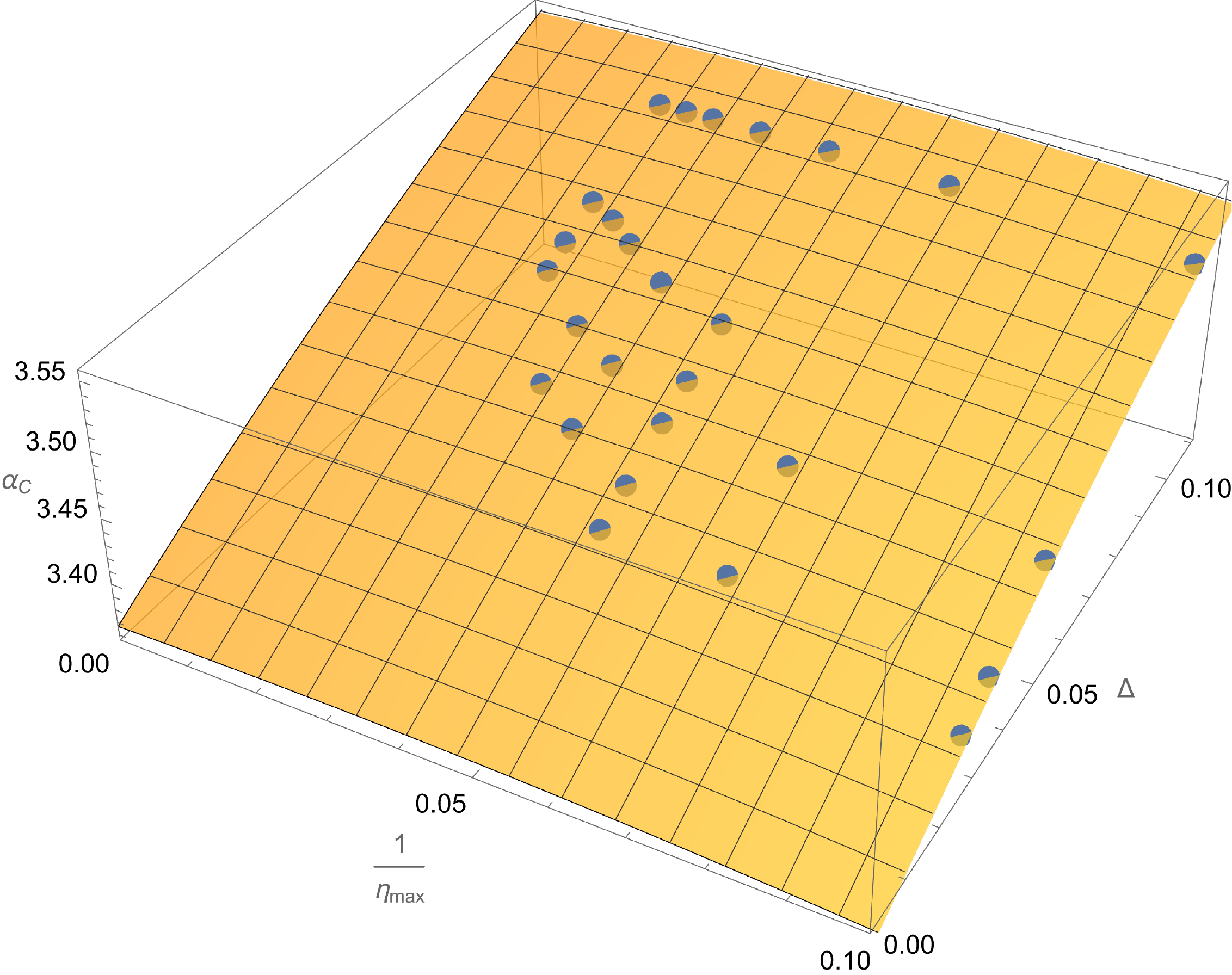}  
  \caption{$\alpha_C$}
  \label{FIG:best_fit_C}
\end{subfigure}
\begin{subfigure}{0.4\textwidth}
  \centering
  \includegraphics[width=1.0\linewidth]{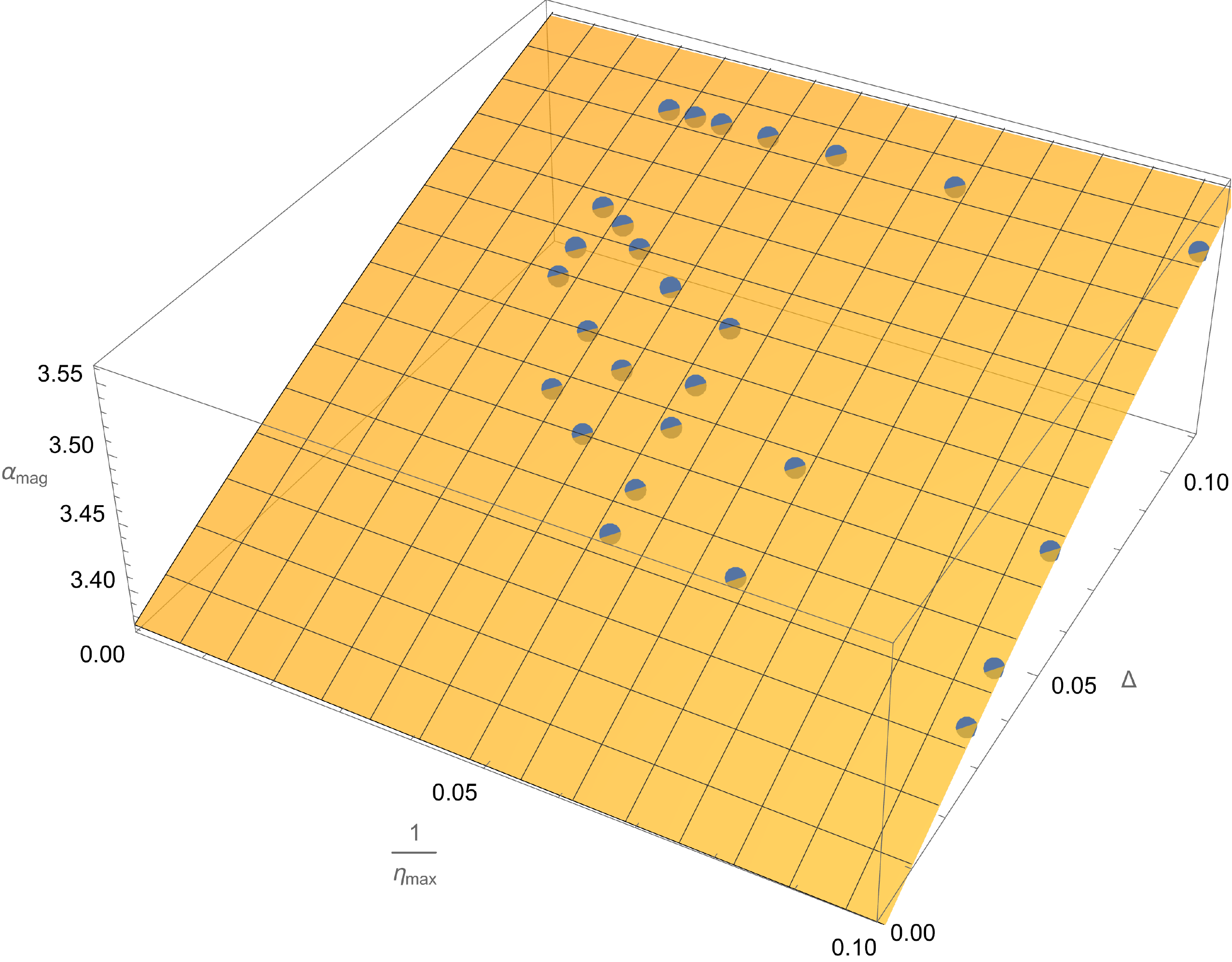}  
  \caption{$\alpha_{\textrm{mag}}$}
  \label{FIG:best_fit_m}
\end{subfigure}
\caption{Plots of the approximate intercepts (blue dots) for the numerically calculated dipoles at each point in the $(1/\eta_{\textrm{max}}, \Delta)$ plane as well as the best fit quadratic surface for each amplitude. The continuum limit $1/\eta_{\textrm{max}}\rightarrow 0, \Delta \rightarrow 0$ corresponds to the lower left corner.}
\label{FIG:best_fit}
\end{figure}
\begin{subequations}\label{intercepts_quad}
\begin{align}
    & \alpha_A = 3.377 - 0.317 \frac{1}{\eta_{\textrm{max}}} + 3.815 \frac{1}{\eta^2_{\textrm{max}}} + 1.986 \Delta - 3.729 \Delta^2 + 0.501 \Delta \frac{1}{\eta_{\textrm{max}}} , \\
    & \alpha_B = 3.376 - 0.282 \frac{1}{\eta_{\textrm{max}}} + 3.616 \frac{1}{\eta^2_{\textrm{max}}} + 2.006 \Delta - 3.800 \Delta^2 + 0.277 \Delta \frac{1}{\eta_{\textrm{max}}} , \\
    & \alpha_C = 3.376 + 0.168 \frac{1}{\eta_{\textrm{max}}} - 2.413 \frac{1}{\eta^2_{\textrm{max}}} + 2.027 \Delta - 3.886 \Delta^2 + 0.155 \Delta \frac{1}{\eta_{\textrm{max}}} , \\
    & \alpha_{\textrm{mag}} = 3.373 - 0.071 \frac{1}{\eta_{\textrm{max}}} + 0.165 \frac{1}{\eta^2_{\textrm{max}}} + 1.982 \Delta - 3.778 \Delta^2 + 0.852 \Delta \frac{1}{\eta_{\textrm{max}}} .
\end{align}
\end{subequations}
In \fig{FIG:best_fit} we plot the best fit quadratic surface for each dipole amplitude's intercepts along with the data points from the linear regression in the $(1/\eta_{\textrm{max}},\Delta)$ plane. Taking the $1/ \eta_{\textrm{max}} \rightarrow 0$ and $\eta_{\textrm{max}}/N\rightarrow 0$ limits yields the continuum extrapolation of the small-$x$ asymptotics as 
\begin{subequations}
\begin{align}
    &F_A^{NS} \sim \left( \frac{1}{x} \right)^{(3.377 \pm 0.003) \, \sqrt{\frac{\alpha_s N_c}{4 \pi}}} , \\
    &F_B^{NS} \sim \left( \frac{1}{x} \right)^{(3.376 \pm 0.004) \, \sqrt{\frac{\alpha_s N_c}{4 \pi}}} , \\
    &F_C^{NS} \sim \left( \frac{1}{x} \right)^{(3.367 \pm 0.003) \, \sqrt{\frac{\alpha_s N_c}{4 \pi}}} , \\
    &F_{\textrm{mag}}^{NS} \sim \left( \frac{1}{x} \right)^{(3.373 \pm 0.003) \, \sqrt{\frac{\alpha_s N_c}{4 \pi}}} ,
\end{align}
\end{subequations}
where the uncertainties come from the standard error of the quadratic polynomial fits. Plugging these into \eq{F_S&NS2b} and rounding up the intercepts values to the first decimal point yields
\begin{align}\label{Sivers_sub_eik8}
f_{1 \: T}^{\perp \: NS} (x \ll 1, {k}_T^2) \Big|_\textrm{sub-eikonal} \sim \left( \frac{1}{x} \right)^{3.4 \, \sqrt{\frac{\alpha_s N_c}{4 \pi}}}.
\end{align}


\subsection{Small-$x$ Asymptotics of the Flavor Non-Singlet Sivers Function: a Summary}

We conclude this Section by summarizing the results of our calculations. The flavor non-singlet Sivers function at small $x$ receives contributions at the eikonal and sub-eikonal order. The eikonal contribution is coming from the spin-dependent odderon (see \eq{siv}) and leads to $f_{1 \: T}^{\perp \: NS} \sim 1/x$ asymptotics with an almost non-perturbative accuracy \cite{Bartels:1999yt,Kovchegov:2003dm,Kovchegov:2012rz,Caron-Huot:2013fea,Brower:2008cy,Avsar:2009hc,Brower:2014wha}, in agreement with \cite{Dong:2018wsp,Boer:2015pni}. The sub-eikonal contribution to the flavor non-singlet Sivers function is calculated in this Section. At large $N_c$ and in DLA it is given by \eq{Sivers_sub_eik8}. We, therefore, conclude that one can describe the small-$x$ asymptotics of the quark Sivers function as
\begin{align}\label{Sivers_small_x}
f_{1 \: T}^{\perp \: NS} (x, k_T^2)  = C_O (x, k_T^2) \, \frac{1}{x} + C_1 (x, k_T^2) \, \left( \frac{1}{x} \right)^{3.4 \, \sqrt{\frac{\alpha_s N_c}{4 \pi}}} + \ldots
\end{align}
with some functions $C_O (x, k_T^2)$ and $C_1 (x, k_T^2)$, which can be numerically obtained from the above results. These functions also depend on $x$, but in a much slower way than the powers explicitly shown in \eq{Sivers_small_x} (see, e.g., Eq.~(26) in \cite{Contreras:2020lrh}). The ellipsis in \eq{Sivers_small_x} denote the further energy-suppressed sub-sub-eikonal corrections.



\section{The Boer-Mulders Function}
\label{sec:evolution_BM}


\subsection{The Boer-Mulders Function at Small $x$}

Our next goal is to study the other leading-twist T-odd quark TMD, the Boer-Mulders function \cite{Boer:1997nt}, at small values of $x$. The LCOT analysis of \cite{Kovchegov:2018znm,Kovchegov:2018zeq} applies to this TMD as well, such that we will need only the contribution of the diagram B in the classification of \cite{Kovchegov:2018znm,Kovchegov:2018zeq}, shown above in \fig{FIG:diagbdet}. The Boer-Mulders function, unlike the Sivers function, receives no eikonal-level contribution (as will become apparent shortly): hence, the diagram C from  \cite{Kovchegov:2018zeq} need not be considered. 

Employing the standard definition \cite{Meissner:2007rx} for the quark Boer-Mulders function $h_1^{\perp \, q} (x, k_T^2)$ we write, taking only the diagram B and it complex conjugate into account,
\begin{align}\label{h1q1}
- \frac{\epsilon^{ij} k^i}{M_P} \, h_1^{\perp \, q} (x, k_T^2) & \subset \int \frac{d r^- \, d^2 r_\perp}{2 (2\pi)^3} \, e^{i k \cdot r} \bra{P} \bar{\psi}(0) \mathcal{U}[0,r] \frac{ i \, \sigma^{j+} \, \gamma^5}{2} \psi(r) \ket{P}  \\ 
& = \int \frac{d r^- \, d^2 r_\perp}{2 (2\pi)^3} \, e^{i k \cdot r} \bra{P} \bar{\psi}(0) \mathcal{U}[0,r] \frac{ \gamma^5 \gamma^+ \gamma^j}{2} \psi(r) \ket{P} \notag \\ 
& = \frac{2 p_1^+}{2 (2 \pi)^3} \sum_X \int d{\xi^-} d^2 {\xi_{\perp}} d{\zeta^-} d^2 {\zeta_{\perp}} e^{i k \cdot (\zeta - \xi)}  \Big[\frac{\gamma^5 \gamma^+ \gamma^j}{2} \Big]_{\alpha \beta} \Big{\langle} \bar{\psi}_{\alpha} (\xi) V_{\underline{\xi}} [\xi^-,\infty] | X \rangle \langle X | V_{\underline{\zeta}} [\infty, \zeta^-] \psi_{\beta} (\zeta) \Big{\rangle} \notag \\
& =  -\frac{2 p_1^+}{2 (2 \pi)^3} \int d^2 {\zeta_{\perp}}  d^2 {w_{\perp}} \frac{ d^2{k_{1 \perp}} d{k_1^-}}{(2\pi)^3} e^{i (\underline{k}_1 + \underline{k}) \cdot (\un{w} - \un{\zeta})} \theta (k_1^-) \, \frac{1}{(x p_1^+ k_1^- + \underline{k}_1^2 ) (x p_1^+ k_1^- + \underline{k}^2)}  \notag \\
\times & \sum_{\chi_1 , \chi_2} \bar{v}_{\chi_2} (k_2) \frac{\gamma^5 \gamma^+ \gamma^j}{2} v_{\chi_1}(k_1) \, \Big{\langle} \tord V_{\underline{\zeta}}^{mn} \, \bar{v}_{\chi_1} (k_1) \left( \hat{V}_{{\un w}}^\dagger \right)^{nm}  v_{\chi_2} (k_2) \Big{\rangle} \Bigg|_{k_2^- = k_1^-, k_1^2 =0, k_2^2 =0, \un{k}_2 = - \un{k}} + \mbox{c.c.} \notag
\end{align}
with the future-pointing (semi-inclusive deep inelastic scattering or SIDIS) fundamental Wilson-line staple $\mathcal{U}[0,r]$, fundamental color indices $m, n$, and the $\pm$-reversed BL spinors from \eq{chi_def}.

Choosing, for definiteness, the transverse index $j=1$ in \eq{h1q1} yields
\begin{align}\label{h1q2}
\frac{k^y}{M_P} \, h_1^{\perp \, q} (x, k_T^2) & \subset   -\frac{2 p_1^+}{2 (2 \pi)^3} \int d^2 {\zeta_{\perp}}  d^2 {w_{\perp}} \frac{ d^2{k_{1 \perp}} d{k_1^-}}{(2\pi)^3} e^{i (\underline{k}_1 + \underline{k}) \cdot (\un{w} - \un{\zeta})} \theta (k_1^-) \, \frac{1}{(x p_1^+ k_1^- + \underline{k}_1^2 ) (x p_1^+ k_1^- + \underline{k}^2)}  \notag \\
\times & \sum_{\chi_1 , \chi_2} \bar{v}_{\chi_2} (k_2) \frac{\gamma^5 \gamma^+ \gamma^1}{2} v_{\chi_1}(k_1) \, \Big{\langle} \tord V_{\underline{\zeta}}^{mn} \, \bar{v}_{\chi_1} (k_1) \left( \hat{V}_{{\un w}}^\dagger \right)^{nm}  v_{\chi_2} (k_2) \Big{\rangle} \Bigg|_{k_2^- = k_1^-, k_1^2 =0, k_2^2 =0, \un{k}_2 = - \un{k}} + \mbox{c.c.} .
\end{align}
Employing the $\pm$-reversed BL spinors \eqref{chi_def} we find for massless quarks \cite{Kovchegov:2018zeq}
\begin{align}\label{mel}
\bar{v}_{\chi_2} (k_2) \frac{\gamma^5 \gamma^+ \gamma^1}{2} v_{\chi_1}(k_1)  = \frac{1}{2 \, \sqrt{k_1^- \, k_2^-}} \, \Big[ & \, - \chi_1 \, \delta_{\chi_1 \chi_2} \, (2 \, {\un S} \cdot {\un k}_1 \, {\un S} \cdot {\un k}_2 - {\un k}_1 \cdot {\un k}_2 )  \\ & - i \, \chi_1 \, \delta_{\chi_1, \, - \chi_2} \, ({\un S} \times {\un k}_1 \, {\un S} \cdot {\un k}_2 + {\un S} \cdot {\un k}_1 \, {\un S} \times {\un k}_2 )  \Big]  \notag
\end{align}
where ${\un S} = {\hat x}$ is the unit vector in the direction of the spin quantization axis for the spinors \eqref{chi_def}.

The replacement \eqref{V_repl},
\begin{align}\label{V_repl2}
\bar{v}_{\chi_1} (k_1) \left( \hat{V}_{{\un w}}^\dagger \right)^{nm}  v_{\chi_2} (k_2)  \to 2 \sqrt{k_1^- \, k_2^-}  \, \int d^2 z_\perp \, \left( V^{\textrm{pol} \, \dagger}_{{\un z}, {\un w}; \chi_2 , \chi_1} \right)^{nm} ,
\end{align}
along with $e^{i \un{k} \cdot (\un{w} - \un{\zeta})} \to e^{i \un{k} \cdot (\un{z} - \un{\zeta})}$ yields
\begin{align}\label{BM2}
& \frac{k^y}{M_P} \, h_1^{\perp \, q} (x, k_T^2) \subset   -\frac{2 p_1^+}{2 (2 \pi)^3} \int d^2 {\zeta_{\perp}}  d^2 {w_{\perp}} d^2 z_\perp \, \frac{ d^2{k_{1 \perp}} d{k_1^-}}{(2\pi)^3} e^{i \underline{k}_1 \cdot (\un{w} - \un{\zeta}) + i \underline{k} \cdot (\un{z} - \un{\zeta})} \theta (k_1^-) \, \frac{1}{(x p_1^+ k_1^- + \underline{k}_1^2 ) (x p_1^+ k_1^- + \underline{k}^2)}  \\
\times & \sum_{\chi_1 , \chi_2} \Big[  \chi_1 \, \delta_{\chi_1 \chi_2} \, (2 \, {\un S} \cdot {\un k}_1 \, {\un S} \cdot {\un k} - {\un k}_1 \cdot {\un k} )  + i \, \chi_1 \, \delta_{\chi_1, \, - \chi_2} \, ({\un S} \times {\un k}_1 \, {\un S} \cdot {\un k} + {\un S} \cdot {\un k}_1 \, {\un S} \times {\un k} )  \Big]   \Big{\langle} \tord \tr \left[ V_{\underline{\zeta}} \, V^{\textrm{pol} \, \dagger}_{{\un z}, {\un w}; \chi_2 , \chi_1} \right] \Big{\rangle}  + \mbox{c.c.} . \notag 
\end{align}

We see that only the $\chi_1 \, \delta_{\chi_1 \chi_2}$ and $\chi_1 \, \delta_{\chi_1, \, - \chi_2}$ structures in $V^{\textrm{pol} \, \dagger}_{{\un z}, {\un w}; \chi_2 , \chi_1}$ survive in \eq{BM2}. As promised above, this eliminates the eikonal contribution to $V^{\textrm{pol} \, \dagger}_{{\un z}, {\un w}; \chi_2 , \chi_1}$, which comes in with $\delta_{\chi_1 \chi_2}$ only. (The same is true for the eikonal contribution to the diagram C of \cite{Kovchegov:2018znm}.) Moreover, at the sub-eikonal order, the relevant polarization structures can be found from \eq{Vphase&mag}: these are $\delta_{\chi_1 \chi_2}$ and $\delta_{\chi_1, - \chi_2}$ \cite{Kovchegov:2021iyc}. These structures in $V^{\textrm{pol} \, \dagger}_{{\un z}, {\un w}; \chi_2 , \chi_1}$ also do not contribute to \eq{BM2}, due to the projections on $\chi_1 \, \delta_{\chi_1 \chi_2}$ and $\chi_1 \, \delta_{\chi_1, \, - \chi_2}$ in it. We conclude that only the sub-sub-eikonal part of $V^{\textrm{pol} \, \dagger}_{{\un z}, {\un w}; \chi_2 , \chi_1}$ can contribute to \eq{BM2}.

The sub-sub-eikonal contribution to the quark $S$-matrix has to involve a quark-sector operator \cite{Kovchegov:2021iyc}. Only the terms quadratic in the quark or gluon fields may contribute to the DLA evolution: the sub-sub-eikonal operators which are at most quadratic in the non-eikonal field and come in with the $\chi_1 \, \delta_{\chi_1 \chi_2}$ and $\chi_1 \, \delta_{\chi_1, \, - \chi_2}$ polarization structures exist only in the quark sector \cite{Kovchegov:2021iyc}. Sub-sub-eikonal operators containing the gluon fields either come in proportional to the quark mass, which we neglect, or give zero projections onto $\chi_1 \, \delta_{\chi_1 \chi_2}$ and $\chi_1 \, \delta_{\chi_1, \, - \chi_2}$, or enter through terms containing more than two sub-eikonal fields, which do not contribute to DLA evolution. Keeping only the sub-sub-eikonal quark terms, we write \cite{Kovchegov:2021iyc} 
\begin{align}\label{Vpol_qq}
V^{\textrm{pol q}\overline{\textrm{q}}}_{\un{x}, \un{y}; \chi', \chi} = \int\limits_{-\infty}^{\infty} d{z}_1^- d^2 z_1 \int\limits_{z_1^-}^\infty d z_2^- d^2 z_2 \ V_{\un{x}} [ \infty, z_2^-] \, \delta^2 (\un{x} - \un{z}_2) \, \mathcal{O}^{\textrm{pol q}\overline{\textrm{q}}}_{\chi', \chi} (z_2^-, z_1^-;  \un{z}_2, \un{z}_1)  \, V_{\un{y}} [ z_1^-, -\infty] \, \delta^2 (\un{y} - \un{z}_1) + \ldots
\end{align}
with 
\begin{align}
& \mathcal{O}^{\textrm{pol q}\overline{\textrm{q}}}_{\chi', \chi} (z_2^-, z_1^-;  \un{z}_2, \un{z}_1) \supset \frac{g^2 \, (p_1^+)^2}{8 \, s^2} \, t^b \, \psi_{\beta} (z_2^-,\un{z}_2) \, U_{\un{z}_2}^{ba} [z_2^-,z_1^-] \, \delta^2 (\un{z}_2 - \un{z}_1)  \\ 
&  \Bigg\{ \chi \delta_{\chi, \chi'}  \Bigg[ \left[ i \gamma^5 \underline{S} \cdot \cev{\underline{D}}_{z_2} - \underline{S} \times \cev{\underline{D}}_{z_2} \right] (1 - i \gamma^5 \gamma^1 \gamma^2) +  \left[ i \gamma^5 \underline{S} \cdot \underline{D}_{z_1}  - \underline{S} \times \underline{D}_{z_1} \right] (1 + i \gamma^5 \gamma^1 \gamma^2) \Bigg] \notag \\ 
 & - \chi \delta_{\chi, - \chi'}  \Bigg[ \left[ i \underline{S} \cdot \cev{\underline{D}}_{z_2} - \gamma^5 \underline{S} \times \cev{\underline{D}}_{z_2} \right] (1 - i \gamma^5 \gamma^1 \gamma^2) +  \left[ i \underline{S} \cdot \underline{D}_{z_1}  - \gamma^5 \underline{S} \times \underline{D}_{z_1} \right] (1 + i \gamma^5 \gamma^1 \gamma^2) \Bigg] \Bigg{\}}_{\alpha \beta} \!\!\!\!\!  \bar{\psi}_\alpha (z_1^-,\un{z}_1) \, t^a . \notag
\end{align}
Here the covariant derivatives act on the spinors only. Only the polarization structures $\chi \delta_{\chi, \chi'}$ and $\chi \delta_{\chi, - \chi'}$ that survive in \eq{BM2} are shown explicitly. The ellipsis in \eq{Vpol_qq} denote the operators containing either multiple insertions of $\mathcal{O}^{\textrm{pol q}\overline{\textrm{q}}}_{\chi', \chi}$ or a four-quark field operator or additional powers of sub-eikonal gluon field components \cite{Kovchegov:2021iyc}: none of these are relevant for the DLA evolution we are about to construct.  

We define
\begin{subequations}\label{VTVT}
\begin{align}
V_{\un{x}}^\textrm{T} \equiv & \, \frac{g^2 \, (p_1^+)^2}{16 \, s^2} \, \int\limits_{-\infty}^{\infty} d{z}_1^- \int\limits_{z_1^-}^\infty d z_2^-  \ V_{\un{x}} [ \infty, z_2^-] \, t^b \, \psi_{\beta} (z_2^-,\un{x}) \, U_{\un{x}}^{ba} [z_2^-,z_1^-] \, \Bigg[ \left[ i \gamma^5 \underline{S} \cdot \cev{\underline{D}}_{x} - \underline{S} \times \cev{\underline{D}}_{x} \right] \, \gamma^+ \gamma^- \\ 
& +  \left[ i \gamma^5 \underline{S} \cdot \underline{D}_{x}  - \underline{S} \times \underline{D}_{x} \right] \gamma^- \gamma^+ \Bigg]_{\alpha \beta} \bar{\psi}_\alpha (z_1^-,\un{x}) \, t^a \, V_{\un{x}} [ z_1^-, -\infty]  , \notag \\
V_{\un{x}}^{\textrm{T} \, \perp} \equiv & \, - \frac{g^2 \, (p_1^+)^2}{16 \, s^2} \, \int\limits_{-\infty}^{\infty} d{z}_1^- \int\limits_{z_1^-}^\infty d z_2^-  \ V_{\un{x}} [ \infty, z_2^-] \, t^b \, \psi_{\beta} (z_2^-,\un{x}) \, U_{\un{x}}^{ba} [z_2^-,z_1^-] \,  \Bigg[ \left[ i \underline{S} \cdot \cev{\underline{D}}_{x} - \gamma^5 \underline{S} \times \cev{\underline{D}}_{x} \right]  \, \gamma^+ \gamma^-  \\ 
& +  \left[ i \underline{S} \cdot \underline{D}_{x}  - \gamma^5 \underline{S} \times \underline{D}_{x} \right] \, \gamma^- \gamma^+ \Bigg]_{\alpha \beta}   \bar{\psi}_\alpha (z_1^-,\un{z}_1) \, t^a \, V_{\un{x}} [ z_1^-, -\infty]  . \notag
\end{align}
\end{subequations}
Note that the color matrix in $\cev{\underline{D}}_{x}$ is inserted between $t^b$ and $\psi_{\beta}$, while the color matrix of $\underline{D}_{x}$ is inserted between $\bar{\psi}_\alpha$ and $t^a$ \cite{Kovchegov:2021iyc}. 

With this notation we have
\begin{align}
V^{\textrm{pol q}\overline{\textrm{q}}}_{\un{x}, \un{y}; \chi', \chi} \supset \left[ \chi \, \delta_{\chi, \chi'} \,  V_{\un{x}}^\textrm{T}  + \chi \, \delta_{\chi, - \chi'} \, V_{\un{x}}^{\textrm{T} \, \perp} \right] \, \delta^2 ({\un x} - {\un y}),
\end{align}
such that \eq{BM2} becomes
\begin{align}\label{BM3}
& \frac{k^y}{M_P} \, h_1^{\perp \, q} (x, k_T^2) \subset  -\frac{4 p_1^+}{2 (2 \pi)^3} \int d^2 {\zeta_{\perp}}  d^2 {w_{\perp}} \, \frac{ d^2{k_{1 \perp}} d{k_1^-}}{(2\pi)^3} e^{i (\underline{k}_1 + \underline{k} ) \cdot (\un{w} - \un{\zeta})  } \theta (k_1^-) \, \frac{1}{(x p_1^+ k_1^- + \underline{k}_1^2 ) (x p_1^+ k_1^- + \underline{k}^2)}   \\
& \hspace*{4cm} \times \Big[  (2 \, {\un S} \cdot {\un k}_1 \, {\un S} \cdot {\un k} - {\un k}_1 \cdot {\un k} )  \Big{\langle} \tord \tr \left[ V_{\underline{\zeta}} \, V^{\textrm{T} \, \dagger}_{{\un w}} \right] + \atord \tr \left[ V_{\underline{w}}^\dagger \, V^{\textrm{T} }_{{\un \zeta}} \right] \Big{\rangle} \notag \\ 
& \hspace*{4cm} + i \,  ({\un S} \times {\un k}_1 \, {\un S} \cdot {\un k} + {\un S} \cdot {\un k}_1 \, {\un S} \times {\un k} )  \Big{\langle} \tord \tr \left[ V_{\underline{\zeta}} \, V^{\textrm{T} \, \perp \, \dagger}_{{\un w}} \right] - \atord \tr \left[ V_{\underline{w}}^\dagger \, V^{\textrm{T} \, \perp}_{{\un \zeta}} \right] \Big{\rangle} \Big]   . \notag
\end{align}

Finally, expanding the denominators in the powers of $x$, 
\begin{align}
\frac{1}{(x p_1^+ k_1^- + \underline{k}_1^2 ) (x p_1^+ k_1^- + \underline{k}^2)} = \frac{1}{\underline{k}_1^2 \, \underline{k}^2} - \frac{x p_1^+ k_1^-}{\underline{k}_1^2 \, \underline{k}^2} \left(  \frac{1}{\underline{k}_1^2} +  \frac{1}{\underline{k}^2}  \right) + \ldots
\end{align}
while remembering that for sub-sub-eikonal quantities the leading term of such expansion is canceled by the instantaneous term in the quark propagator of diagram B 
\cite{Kovchegov:2018znm}, we arrive at the expression for the quark Boer-Mulders function,
\begin{align}\label{BM4}
& \frac{k^y}{M_P} \, h_1^{\perp \, q} (x, k_T^2) \subset  \frac{4 x (p_1^+)^2}{2 (2 \pi)^3} \int d^2 {\zeta_{\perp}}  d^2 {w_{\perp}} \, \frac{ d^2{k_{1 \perp}} d{k_1^-}}{(2\pi)^3} e^{i (\underline{k}_1 + \underline{k} ) \cdot (\un{w} - \un{\zeta})  } \theta (k_1^-) \, \frac{k_1^-}{\underline{k}_1^2 \, \underline{k}^2} \left(  \frac{1}{\underline{k}_1^2} +  \frac{1}{\underline{k}^2}  \right)   \\
& \hspace*{4cm} \times \Big[  (2 \, {\un S} \cdot {\un k}_1 \, {\un S} \cdot {\un k} - {\un k}_1 \cdot {\un k} )  \Big{\langle} \tord \tr \left[ V_{\underline{\zeta}} \, V^{\textrm{T} \, \dagger}_{{\un w}} \right] + \atord \tr \left[ V_{\underline{w}}^\dagger \, V^{\textrm{T} }_{{\un \zeta}} \right] \Big{\rangle} \notag \\ 
& \hspace*{4cm} + i \,  ({\un S} \times {\un k}_1 \, {\un S} \cdot {\un k} + {\un S} \cdot {\un k}_1 \, {\un S} \times {\un k} )  \Big{\langle} \tord \tr \left[ V_{\underline{\zeta}} \, V^{\textrm{T} \, \perp \, \dagger}_{{\un w}} \right] - \atord \tr \left[ V_{\underline{w}}^\dagger \, V^{\textrm{T} \, \perp}_{{\un \zeta}} \right] \Big{\rangle} \Big]   . \notag
\end{align}

Repeating the above steps for the anti-quark Boer-Mulders distribution yields
\begin{align}\label{BM5}
& \frac{k^y}{M_P} \, h_1^{\perp \, {\bar q}} (x, k_T^2) \subset  \frac{4 x (p_1^+)^2}{2 (2 \pi)^3} \int d^2 {\zeta_{\perp}}  d^2 {w_{\perp}} \, \frac{ d^2{k_{1 \perp}} d{k_1^-}}{(2\pi)^3} e^{i (\underline{k}_1 + \underline{k} ) \cdot (\un{w} - \un{\zeta})  } \theta (k_1^-) \, \frac{k_1^-}{\underline{k}_1^2 \, \underline{k}^2} \left(  \frac{1}{\underline{k}_1^2} +  \frac{1}{\underline{k}^2}  \right)   \\
& \hspace*{4cm} \times \Big[  (2 \, {\un S} \cdot {\un k}_1 \, {\un S} \cdot {\un k} - {\un k}_1 \cdot {\un k} )  \Big{\langle} \tord \tr \left[ V_{\underline{\zeta}}^\dagger \, V^{\textrm{T} }_{{\un w}} \right] + \atord \tr \left[ V_{\underline{w}} \, V^{\textrm{T} \, \dagger}_{{\un \zeta}} \right] \Big{\rangle} \notag \\
& \hspace*{4cm} + i \,  ({\un S} \times {\un k}_1 \, {\un S} \cdot {\un k} + {\un S} \cdot {\un k}_1 \, {\un S} \times {\un k} )  \Big{\langle} \tord \tr \left[ V_{\underline{\zeta}}^\dagger \, V^{\textrm{T} \, \perp}_{{\un w}} \right] - \atord \tr \left[ V_{\underline{w}} \, V^{\textrm{T} \, \perp \, \dagger}_{{\un \zeta}} \right] \Big{\rangle} \Big]   . \notag
\end{align}

The flavor-singlet and non-singlet Boer-Mulders distributions are 
\begin{subequations}
\begin{align}
\label{BM_S}
& \frac{k^y}{M_P} \, h_1^{\perp \, \textrm{S}} (x, k_T^2) \subset  \frac{4 x (p_1^+)^2}{2 (2 \pi)^3} \sum_f \int d^2 {\zeta_{\perp}}  d^2 {w_{\perp}} \, \frac{ d^2{k_{1 \perp}} d{k_1^-}}{(2\pi)^3} e^{i (\underline{k}_1 + \underline{k} ) \cdot (\un{w} - \un{\zeta})  } \theta (k_1^-) \, \frac{k_1^-}{\underline{k}_1^2 \, \underline{k}^2} \left(  \frac{1}{\underline{k}_1^2} +  \frac{1}{\underline{k}^2}  \right)   \\
\times & \Big[  (2 \, {\un S} \cdot {\un k}_1 \, {\un S} \cdot {\un k} - {\un k}_1 \cdot {\un k} )  \Big{\langle} \tord \tr \left[ V_{\underline{\zeta}} \, V^{\textrm{T} \, \dagger}_{{\un w}} \right] + \atord \tr \left[ V_{\underline{w}}^\dagger \, V^{\textrm{T} }_{{\un \zeta}} \right] +  \tord \tr \left[ V_{\underline{\zeta}}^\dagger \, V^{\textrm{T} }_{{\un w}} \right] + \atord \tr \left[ V_{\underline{w}} \, V^{\textrm{T} \, \dagger}_{{\un \zeta}} \right] \Big{\rangle} \notag \\ 
& + i \,  ({\un S} \times {\un k}_1 \, {\un S} \cdot {\un k} + {\un S} \cdot {\un k}_1 \, {\un S} \times {\un k} )  \Big{\langle} \tord \tr \left[ V_{\underline{\zeta}} \, V^{\textrm{T} \, \perp \, \dagger}_{{\un w}} \right] - \atord \tr \left[ V_{\underline{w}}^\dagger \, V^{\textrm{T} \, \perp}_{{\un \zeta}} \right] + \tord \tr \left[ V_{\underline{\zeta}}^\dagger \, V^{\textrm{T} \, \perp}_{{\un w}} \right] - \atord \tr \left[ V_{\underline{w}} \, V^{\textrm{T} \, \perp \, \dagger}_{{\un \zeta}} \right]\Big{\rangle} \Big]   , \notag \\
& \frac{k^y}{M_P} \, h_1^{\perp \, \textrm{NS}} (x, k_T^2) \subset  \frac{4 x (p_1^+)^2 }{2 (2 \pi)^3} \int d^2 {\zeta_{\perp}}  d^2 {w_{\perp}} \, \frac{ d^2{k_{1 \perp}} d{k_1^-}}{(2\pi)^3} e^{i (\underline{k}_1 + \underline{k} ) \cdot (\un{w} - \un{\zeta})  } \theta (k_1^-) \, \frac{k_1^-}{\underline{k}_1^2 \, \underline{k}^2} \left(  \frac{1}{\underline{k}_1^2} +  \frac{1}{\underline{k}^2}  \right)   \\
\times & \Big[  (2 \, {\un S} \cdot {\un k}_1 \, {\un S} \cdot {\un k} - {\un k}_1 \cdot {\un k} )  \Big{\langle} \tord \tr \left[ V_{\underline{\zeta}} \, V^{\textrm{T} \, \dagger}_{{\un w}} \right] + \atord \tr \left[ V_{\underline{w}}^\dagger \, V^{\textrm{T} }_{{\un \zeta}} \right] -  \tord \tr \left[ V_{\underline{\zeta}}^\dagger \, V^{\textrm{T} }_{{\un w}} \right] - \atord \tr \left[ V_{\underline{w}} \, V^{\textrm{T} \, \dagger}_{{\un \zeta}} \right] \Big{\rangle} \notag \\ 
& + i \,  ({\un S} \times {\un k}_1 \, {\un S} \cdot {\un k} + {\un S} \cdot {\un k}_1 \, {\un S} \times {\un k} )  \Big{\langle} \tord \tr \left[ V_{\underline{\zeta}} \, V^{\textrm{T} \, \perp \, \dagger}_{{\un w}} \right] - \atord \tr \left[ V_{\underline{w}}^\dagger \, V^{\textrm{T} \, \perp}_{{\un \zeta}} \right] - \tord \tr \left[ V_{\underline{\zeta}}^\dagger \, V^{\textrm{T} \, \perp}_{{\un w}} \right] + \atord \tr \left[ V_{\underline{w}} \, V^{\textrm{T} \, \perp \, \dagger}_{{\un \zeta}} \right]\Big{\rangle} \Big] . \notag
\end{align}
\end{subequations}

Defining
\begin{subequations}\label{HS_defs}
\begin{align}
& {\hat H}^{1}_{ {\un w}, {\un \zeta}} (z) \equiv \frac{1}{2 N_c} \, \llangle \tord \tr \left[ V_{\underline{\zeta}} \, V^{\textrm{T} \, \dagger}_{{\un w}} \right] -  \tord \tr \left[ V_{\underline{\zeta}}^\dagger \, V^{\textrm{T} }_{{\un w}} \right] \rrangle, \\
& {\hat H}^{2}_{ {\un w}, {\un \zeta}} (z) \equiv \frac{1}{2 N_c} \,  \llangle \tord \tr \left[ V_{\underline{\zeta}} \, V^{\textrm{T} \, \perp \, \dagger}_{{\un w}} \right] - \tord \tr \left[ V_{\underline{\zeta}}^\dagger \, V^{\textrm{T} \, \perp}_{{\un w}} \right]  \rrangle , 
\end{align}
\end{subequations}
where, now and for the rest of this Section,
\begin{align}\label{sub_sub_eik_braket}
\llangle \ldots \rrangle \equiv (z s)^2 \, \Big\langle \ldots \Big\rangle = (p_1^+ k_1^-)^2  \, \Big\langle \ldots \Big\rangle
\end{align}
(cf. \eq{sub_eik_braket}), we rewrite the non-singlet distribution as
\begin{align}\label{BM_NS_1}
& \frac{k^y}{M_P} \, h_1^{\perp \, \textrm{NS}} (x, k_T^2) \subset  \frac{8 x N_c}{2 (2 \pi)^3} \int d^2 {\zeta_{\perp}}  d^2 {w_{\perp}} \, \frac{ d^2{k_{1 \perp}} }{(2\pi)^3} \, e^{i (\underline{k}_1 + \underline{k} ) \cdot (\un{w} - \un{\zeta})  } \, \frac{1}{\underline{k}_1^2 \, \underline{k}^2} \left(  \frac{1}{\underline{k}_1^2} +  \frac{1}{\underline{k}^2}  \right)   \int\limits_\frac{\Lambda^2}{s}^1 \frac{dz}{z} \\
\times & \Big\{  (2 \, {\un S} \cdot {\un k}_1 \, {\un S} \cdot {\un k} - {\un k}_1 \cdot {\un k} )  \left[ {\hat H}^{1}_{ {\un w}, {\un \zeta}} (z) + {\hat H}^{1 \, *}_{ {\un \zeta}, {\un w}} (z) \right] + i \,  ({\un S} \times {\un k}_1 \, {\un S} \cdot {\un k} + {\un S} \cdot {\un k}_1 \, {\un S} \times {\un k} )  \left[ {\hat H}^{2}_{ {\un w}, {\un \zeta}} (z) - {\hat H}^{2 \, *}_{ {\un \zeta}, {\un w}} (z) \right] \Big\} , \notag
\end{align}
where, as before, $z=k_1^-/p_2^-$.

The operators \eqref{VTVT} are linear in $\un S$. Hence, the amplitude ${\hat H}^1$ and ${\hat H}^2$ should be linear in $\un S$ too. In addition, to contribute to the Boer-Mulders function, ${\hat H}^{1}$ should contain $\epsilon^{ij}$, while ${\hat H}^{2}$ should not. We thus write (while changing the notation from ${\un w}, {\un \zeta}$ to ${\un x}_1, {\un x}_0$)
\begin{subequations}\label{HS_defs2}
\begin{align}
& \int d^2 b_\perp \, {\hat H}^{1}_{ 10} (z) = {\un x}_{10} \times {\un S} \ {\hat H}^{1} (x_{10}^2, z ) , \\
& \int d^2 b_\perp \, {\hat H}^{2}_{ 10} (z) = {\un x}_{10} \cdot {\un S} \ {\hat H}^{2} (x_{10}^2, z ) .
\end{align}
\end{subequations}
Employing these results in \eq{BM_NS_1} we get
\begin{align}\label{BM_NS_2}
& \frac{k^y}{M_P} \, h_1^{\perp \, \textrm{NS}} (x, k_T^2) =  \frac{8 i x N_c}{(2 \pi)^3} \int d^2 {x_{10}} \, \frac{ d^2{k_{1 \perp}} }{(2\pi)^3} \, e^{i (\underline{k}_1 + \underline{k} ) \cdot {\un x}_{10}  } \, \frac{1}{\underline{k}_1^2 \, \underline{k}^2} \left(  \frac{1}{\underline{k}_1^2} +  \frac{1}{\underline{k}^2}  \right)   \int\limits_\frac{\Lambda^2}{s}^1 \frac{dz}{z} \\
\times & \Big\{  (2 \, {\un S} \cdot {\un k}_1 \, {\un S} \cdot {\un k} - {\un k}_1 \cdot {\un k} ) \, ({\un x}_{10} \times {\un S}) \, \mbox{Im} \, {\hat H}^{1} (x_{10}^2, z )  + ({\un S} \times {\un k}_1 \, {\un S} \cdot {\un k} + {\un S} \cdot {\un k}_1 \, {\un S} \times {\un k} )  \, ({\un x}_{10} \cdot {\un S}) \,\mbox{Re} \, {\hat H}^{2} (x_{10}^2, z ) \Big\} . \notag
\end{align}
Finally, defining 
\begin{subequations}\label{HS_defs3}
\begin{align}
& H^{1}_{10} (z) \equiv \frac{1}{2 N_c} \, \mbox{Im} \,  \llangle \tord \tr \left[ V_{\underline{0}} \, V^{\textrm{T} \, \dagger}_{{\un 1}} \right] - \tord \tr \left[ V_{\underline{0}}^\dagger \, V^{\textrm{T} }_{{\un 1}} \right] \rrangle, \\
& H^{2}_{10} (z) \equiv \frac{1}{2 N_c} \, \mbox{Re} \,  \llangle \tord \tr \left[ V_{\underline{0}} \, V^{\textrm{T} \, \perp \, \dagger}_{{\un 1}} \right] - \tord \tr \left[ V_{\underline{0}}^\dagger \, V^{\textrm{T} \, \perp}_{{\un 1}} \right]  \rrangle , 
\end{align}
\end{subequations}
along with 
\begin{subequations}\label{HS_defs4}
\begin{align}
& \int d^2 b_\perp \, H^{1}_{10} (z) = {\un x}_{10} \times {\un S} \, H^{1} (x_{10}^2, z ) ,  \label{HS_defs4a} \\
& \int d^2 b_\perp \, H^{2}_{10} (z) = {\un x}_{10} \cdot {\un S} \, H^{2} (x_{10}^2, z ) , \label{HS_defs4b} 
\end{align}
\end{subequations}
we obtain our final expression for the flavor non-singlet Boer-Mulders function at small $x$,
\begin{align}\label{BM_NS_2}
& \frac{k^y}{M_P} \, h_1^{\perp \, \textrm{NS}} (x, k_T^2) =  \frac{i x N_c}{\pi^3} \int d^2 {x_{10}} \, \frac{ d^2{k_{1 \perp}} }{(2\pi)^3} \, e^{i (\underline{k}_1 + \underline{k} ) \cdot {\un x}_{10}  } \, \frac{1}{\underline{k}_1^2 \, \underline{k}^2} \left(  \frac{1}{\underline{k}_1^2} +  \frac{1}{\underline{k}^2}  \right)   \int\limits_\frac{\Lambda^2}{s}^1 \frac{dz}{z} \\
\times & \Big\{  (2 \, {\un S} \cdot {\un k}_1 \, {\un S} \cdot {\un k} - {\un k}_1 \cdot {\un k} ) \, ({\un x}_{10} \times {\un S}) \, H^{1} (x_{10}^2, z )  + ({\un S} \times {\un k}_1 \, {\un S} \cdot {\un k} + {\un S} \cdot {\un k}_1 \, {\un S} \times {\un k} ) \, ({\un x}_{10} \cdot {\un S}) \, H^{2} (x_{10}^2, z ) \Big\} . \notag
\end{align}


\subsection{Evolution of the Flavor Non-Singlet Boer-Mulders Function at Small $x$}

We are now ready to construct small-$x$ evolution equations determining the small-$x$ asymptotics for the Boer-Mulders function. We want to build the DLA evolution of 
\begin{subequations}\label{HS_defs5}
\begin{align}
& H^{1}_{10} (z) = \frac{1}{2 N_c} \, \mbox{Im} \,  \llangle \tord \tr \left[ V_{\underline{0}} \, V^{\textrm{T} \, \dagger}_{{\un 1}} \right] - \tord \tr \left[ V_{\underline{0}}^\dagger \, V^{\textrm{T} }_{{\un 1}} \right] \rrangle, \label{H10_1} \\
& H^{2}_{10} (z) = \frac{1}{2 N_c} \, \mbox{Re} \,  \llangle \tord \tr \left[ V_{\underline{0}} \, V^{\textrm{T} \, \perp \, \dagger}_{{\un 1}} \right] - \tord \tr \left[ V_{\underline{0}}^\dagger \, V^{\textrm{T} \, \perp}_{{\un 1}} \right]  \rrangle . \label{H10_2}
\end{align}
\end{subequations}

Since we need only the operators quadratic in non-eikonal fields for the DLA evolution at hand, we first simplify the operators \eqref{VTVT} by replacing ${\un D} \to {\un \pd}$, 
\begin{subequations}\label{VTVT2}
\begin{align}
V_{\un{x}}^\textrm{T} = & \, \frac{g^2 \, (p_1^+)^2}{16 \, s^2} \, \int\limits_{-\infty}^{\infty} d{z}_1^- \int\limits_{z_1^-}^\infty d z_2^-  \ V_{\un{x}} [ \infty, z_2^-] \, t^b \, \psi_{\beta} (z_2^-,\un{x}) \, U_{\un{x}}^{ba} [z_2^-,z_1^-] \, \Bigg[ \left[ i \gamma^5 \underline{S} \cdot \cev{\underline{\pd}}_{x} - \underline{S} \times \cev{\underline{\pd}}_{x} \right] \, \gamma^+ \gamma^- \\ 
& +  \left[ i \gamma^5 \underline{S} \cdot \underline{\pd}_{x}  - \underline{S} \times \underline{\pd}_{x} \right] \gamma^- \gamma^+ \Bigg]_{\alpha \beta} \bar{\psi}_\alpha (z_1^-,\un{x}) \, t^a \, V_{\un{x}} [ z_1^-, -\infty]  , \notag \\
V_{\un{x}}^{\textrm{T} \, \perp} = & \, - \frac{g^2 \, (p_1^+)^2}{16 \, s^2} \, \int\limits_{-\infty}^{\infty} d{z}_1^- \int\limits_{z_1^-}^\infty d z_2^-  \ V_{\un{x}} [ \infty, z_2^-] \, t^b \, \psi_{\beta} (z_2^-,\un{x}) \, U_{\un{x}}^{ba} [z_2^-,z_1^-] \,  \Bigg[ \left[ i \underline{S} \cdot \cev{\underline{\pd}}_{x} - \gamma^5 \underline{S} \times \cev{\underline{\pd}}_{x} \right]  \, \gamma^+ \gamma^-  \\ 
& +  \left[ i \underline{S} \cdot \underline{\pd}_{x}  - \gamma^5 \underline{S} \times \underline{\pd}_{x} \right] \, \gamma^- \gamma^+ \Bigg]_{\alpha \beta}   \bar{\psi}_\alpha (z_1^-,\un{z}_1) \, t^a \, V_{\un{x}} [ z_1^-, -\infty]  . \notag
\end{align}
\end{subequations}

\begin{figure}[ht]
\centering
\includegraphics[width= 0.8 \linewidth]{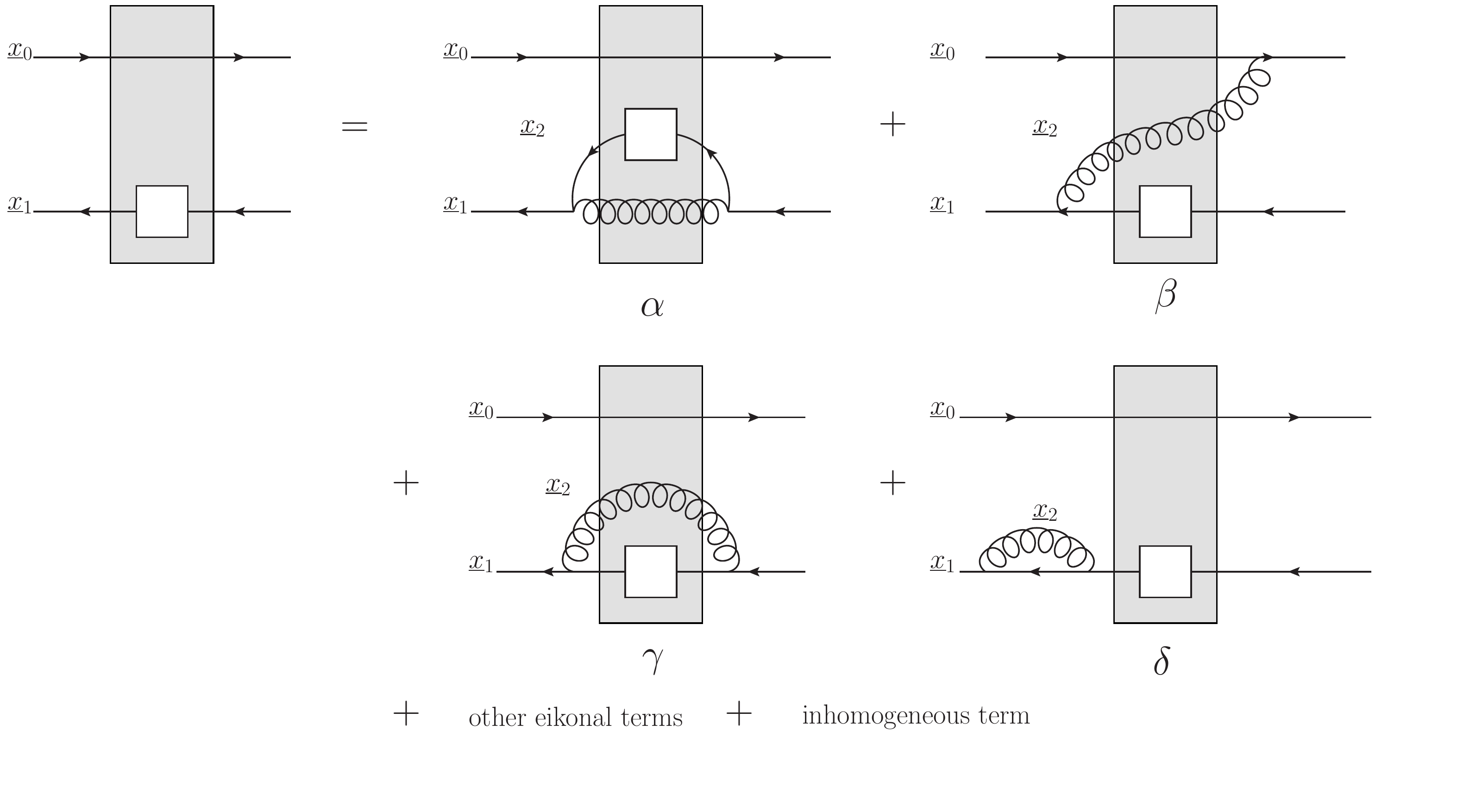}
\caption{The main types of diagrams contributing to the DLA evolution of the amplitudes $H^{1}_{10} (z)$ and $H^{2}_{10} (z)$ from \eq{HS_defs5}. The diagram $\alpha$ on the right contains a sub-sub-eikonal soft-quark emission, while the remaining diagrams $\beta, \gamma, \delta$, etc. are a sample of the eikonal emission diagrams \cite{Mueller:1994rr,Mueller:1994jq,Mueller:1995gb,Balitsky:1995ub,Balitsky:1998ya,Kovchegov:1999yj,Kovchegov:1999ua}. }
\label{FIG:BM_evolution}
\end{figure}

The diagrams contributing to the DLA evolution of the amplitudes $H^{1}_{10} (z)$ and $H^{2}_{10} (z)$ from \eq{HS_defs5} are shown in \fig{FIG:BM_evolution}. To evaluate the diagram $\alpha$ we employ the quark propagator \cite{Kovchegov:2018zeq,Kovchegov:2018znm,Cougoulic:2022gbk}
\begin{align}\label{q_propagator2}
& \int\limits_{-\infty}^0 dx_{2'}^- \, 
\int\limits_0^\infty dx_2^- \, 
\contraction
{}
{\bar\psi^i_\alpha}
{(x_2^- , {\un x}_1) \:}
{psi^j_\beta}
\bar\psi^i_\alpha (x_2^- , {\un x}_1) \: 
\psi^j_\beta (x_{2'}^- , {\un x}_1) 
= - \frac{1}{\pi} \sum_{\chi, \chi'}  \int d k^- \, k^-  \int d^2 x_2 \, d^2 x_{2'} \, \left[ \int
 \frac{d^2 k_{2'}}{(2\pi)^2} \, 
e^{i \ul{k}_{2'} \cdot \ul{x}_{2'1}} \,  \frac{1}{{\un k}_{2'}^2 }  \, \left( v_{\chi'} (k_{2'}) \right)_\beta \right] \notag
\\ & \times
\left( V_{{\ul 2}, {\un 2}'; \chi, \chi'}^{\textrm{pol} \, \dagger} \right)^{ji} \, \left[ \int \frac{d^2 k_2}{(2\pi)^2} \, e^{- i \ul{k}_2 \cdot \ul{x}_{21} }  \, \frac{1}{{\un k}_2^2} \, \left( {\bar v}_\chi (k_2) \right)_\alpha \right]
\end{align}
with the matrix element for the transverse spinors \eqref{chi_def} (and for massless quarks)
\begin{align}
    {\bar v}_\chi (k_2) \left[ (\gamma^5 \, k^x_{2'} + i k^y_{2'}) \gamma^+ \gamma^- - (\gamma^5 \, k^x_{2} + i k^y_{2}) \gamma^- \gamma^+ \right] v_{\chi'} (k_{2'}) = 4 i \,  \frac{k_2^- + k_{2'}^-}{\sqrt{k_2^- \, k_{2'}^-}} \, \left[ i \chi \, \delta_{\chi, \chi'} \, {\un k}_2 \cdot {\un k}_{2'} - \chi \, \delta_{\chi, -\chi'} \,  {\un k}_2 \times {\un k}_{2'} \right] .
\end{align}
The contribution of the diagram $\alpha$ to the evolution of $H^{1}_{ 10} (z)$ is
\begin{align}\label{H1alpha}
\delta_\alpha  H^{1}_{ 10} (z) = - \frac{\as }{2 \pi^2 \, N_c} \,  \int\limits_\frac{\Lambda^2}{s}^z \frac{dz'}{z'} \int \frac{d^2 x_2}{x_{21}^2} \, \textrm{Im} \llangle \tord \tr \left[ t^a V_{\un 0}^\dagger t^b V_{\un 2}^\textrm{T} \right] \, U_{\un 1}^{ba} - \tord \tr \left[ t^b V_{\un 0} t^a V_{\un 2}^{\textrm{T} \, \dagger} \right] \, U_{\un 1}^{ba} \rrangle (z') . 
\end{align}
The contribution of the $\alpha$-diagram to $H^{2}_{ 10} (z)$ can be calculated similarly and is 
\begin{align}\label{H2alpha}
\delta_\alpha  H^{2}_{ 10} (z) = - \frac{\as}{2 \pi^2\, N_c} \,  \int\limits_\frac{\Lambda^2}{s}^z \frac{dz'}{z'} \int \frac{d^2 x_2}{x_{21}^2} \, \textrm{Re} \llangle \tord \tr \left[ t^a V_{\un 0}^\dagger t^b V_{\un 2}^{\textrm{T} \, \perp} \right] \, U_{\un 1}^{ba} - \tord \tr \left[ t^b V_{\un 0} t^a V_{\un 2}^{\textrm{T} \, \perp \, \dagger} \right] \, U_{\un 1}^{ba} \rrangle (z') . 
\end{align}
Adding the eikonal contributions from the diagrams $\beta, \gamma, \delta$, etc., using the standard technique \cite{Mueller:1994rr,Mueller:1994jq,Mueller:1995gb,Balitsky:1995ub,Balitsky:1998ya,Kovchegov:1999yj,Kovchegov:1999ua} we arrive at the evolution equations for the amplitudes $H^{1}_{10} (z)$ and $H^{2}_{10} (z)$ 
\begin{subequations}\label{H10_gen}
\begin{align}
    & H_{10}^1 (z) = H^{1 \, (0)}_{10}(z) + \frac{\alpha_s N_c}{2\pi^2} \int\limits_{\frac{\Lambda^2}{s}}^z \frac{\dd{z}'}{z'} \int \dd[2]{x}_2 \frac{x_{10}^2}{x_{21}^2 x_{20}^2} \frac{1}{2} \, \mbox{Im} \, \llangle \frac{1}{N_c^2} \,  \tord \tr \left[ V_{\un{0}} t^a V_{\un{1}}^{\textrm{T} \dagger} t^b \right] \, \left( U_{\un{2}} \right)^{ba} \label{H1_ev} \\
    &- \frac{C_F}{N_c^2} \,  \tord \tr \left[ V_{\un{0}} \,  V_{\un{1}}^{\textrm{T} \dagger} \right] - \frac{1}{N_c^2} \,  \tord \tr \left[ V_{\un{0}}^{\dagger} t^b V_{\un{1}}^{\textrm{T}} t^a \right] \, \left( U_{\un{2}} \right)^{ba} + \frac{C_F}{N_c^2} \,  \tord \tr \left[ V_{\un{0}}^{\dagger} \,  V_{\un{1}}^{\textrm{T}} \right] \rrangle (z')  \nonumber \\
    &+ \frac{\alpha_s N_c}{2\pi^2} \int\limits_{\frac{\Lambda^2}{s}}^z \frac{\dd{z}'}{z'} \int  \frac{\dd[2]{x}_2}{x_{21}^2} \, \textrm{Im} \, \llangle \frac{1}{N_c^2}   \tord \tr \Big[ t^b V_{\un{0}} t^a V_{\un{2}}^{\textrm{T} \dagger} \Big] \, U_{\un{1}}^{ba} - \frac{1}{N_c^2}  \tord \tr \Big[  t^a V_{\un{0}}^{\dagger} t^b V_{\un{2}}^{\textrm{T}} \Big] \, U_{\un{1}}^{ab} \rrangle  (z')    , \nonumber \\
    & H_{10}^2 (z) = H^{2 \, (0)}_{10}(z) + \frac{\alpha_s N_c}{2\pi^2} \int\limits_{\frac{\Lambda^2}{s}}^z \frac{\dd{z}'}{z'} \int \dd[2]{x}_2 \frac{x_{10}^2}{x_{21}^2 x_{20}^2} \frac{1}{2} \mbox{Re} \, \llangle \frac{1}{N_c^2} \,  \tord \tr \left[ V_{\un{0}} t^a V_{\un{1}}^{\textrm{T} \perp \dagger} t^b \right] \, \left( U_{\un{2}} \right)^{ba} \label{H2_ev} \\
    &- \frac{C_F}{N_c^2} \,  \tord \tr \left[ V_{\un{0}} \,  V_{\un{1}}^{\textrm{T} \perp \dagger} \right] - \frac{1}{N_c^2} \,  \tord \tr \left[ V_{\un{0}}^{\dagger} t^b V_{\un{1}}^{\textrm{T} \perp} t^a \right] \, \left( U_{\un{2}} \right)^{ba} + \frac{C_F}{N_c^2} \,  \tord \tr \left[ V_{\un{0}}^{\dagger} \,  V_{\un{1}}^{\textrm{T} \perp} \right] \rrangle (z')  \nonumber \\
    &+ \frac{\alpha_s N_c}{2\pi^2} \int\limits_{\frac{\Lambda^2}{s}}^z \frac{\dd{z}'}{z'} \int \frac{\dd[2]{x}_2}{x_{21}^2} \textrm{Re} \llangle \frac{1}{N_c^2}  \tord \tr \Big[ t^b V_{\un{0}} t^a V_{\un{2}}^{\textrm{T} \perp \dagger} \Big] U_{\un{1}}^{ba} - \frac{1}{N_c^2}   \tord \tr \Big[  t^a V_{\un{0}}^{\dagger} t^b V_{\un{2}}^{\textrm{T} \perp} \Big] \, U_{\un{1}}^{ab} \rrangle  (z')   . \nonumber
\end{align}
\end{subequations}
These are the general equations for the small-$x$ evolution of the flavor non-singlet dipole amplitudes for the Boer-Mulders TMD. Let us note that the two equations are decoupled from each other: the evolution of $H^{1}_{10} (z)$ only depends on $V^\textrm{T}$, which, in turn, enters the definition \eqref{H10_1} of $H^{1}_{10} (z)$. Similarly, the evolution of $H^{2}_{10} (z)$ only depends on $V^{\textrm{T} \perp}$ entering \eqref{H10_2}.


\subsection{Large-$N_c$ Evolution}

The equations \eqref{H10_gen} we derived above are a general result for the amplitudes $H^{1}_{10} (z)$ and $H^{2}_{10} (z)$, but they are not a closed system of equations. The dipole amplitudes on the left-hand side of \eq{H1_ev} and \eq{H2_ev} are not present on the right-hand side, just as we saw with the Sivers function. Again we need to take the large-$N_c$ limit in order to obtain closed equations.

Employing the Fierz identity and  \eq{Ui+mag_Fierzc}, the contribution of the diagram $\alpha$ to \eq{H1_ev} at large $N_c$ (the last line of \eq{H1_ev}) can be readily shown to be
\begin{align}\label{H1alpha2}
\delta_\alpha  H^{1}_{ 10} (z) = \frac{\as \, N_c}{2 \pi^2} \,  \int\limits_\frac{\Lambda^2}{s}^z \frac{dz'}{z'} \int \frac{d^2 x_2}{x_{21}^2} \, H^{1}_{21} (z') . 
\end{align}
Integrating over impact parameters while employing \eq{HS_defs4a} yields zero, which results from the integration over the angles of ${\un x}_{21}$. In arriving at \eq{H1alpha2} we have also assumed that we are outside the saturation region, and linearized the expression on its right-hand side by putting $S=1$: the result may be non-zero inside the saturation region. 

Similarly, the linearized ($S=1$) contribution of the diagram $\alpha$ to \eq{H2_ev} is
\begin{align}\label{H2alpha2}
\delta_\alpha  H^{2}_{ 10} (z) = \frac{\as \, N_c}{2 \pi^2} \,  \int\limits_\frac{\Lambda^2}{s}^z \frac{dz'}{z'} \int \frac{d^2 x_2}{x_{21}^2} \, H^{2}_{21} (z') . 
\end{align}
This is also zero after integration over impact parameters, this time due to \eq{HS_defs4b}.

Therefore, only the eikonal evolution contributes to both $H^{1}_{ 10} (z)$ and $H^{2}_{10} (z)$. After applying the Fierz identity and  \eq{Ui+mag_Fierzc}, the linearized evolution equations become 
\begin{subequations}\label{HS_evol}
\begin{align}
& H^{1}_{ 10} (z) = H^{1 \, (0)}_{ 10} (z) + \frac{\as \, N_c}{2 \pi^2} \,  \int\limits_\frac{\Lambda^2}{s}^z \frac{dz'}{z'} \int d^2 x_2 \, \frac{x_{10}^2}{x_{21}^2 \, x_{20}^2} \, \left[ H^{1}_{12} (z') - \Gamma^{1}_{20, 21} (z') \right], \label{HS_evol_a} \\
& H^{2}_{ 10} (z) = H^{2 \, (0)}_{ 10} (z) + \frac{\as \, N_c}{2 \pi^2} \,  \int\limits_\frac{\Lambda^2}{s}^z \frac{dz'}{z'} \int d^2 x_2 \, \frac{x_{10}^2}{x_{21}^2 \, x_{20}^2} \, \left[ H^{2}_{12} (z') - \Gamma^{2}_{20, 21} (z') \right]  . 
\end{align}
\end{subequations}
Here $\Gamma^{1}_{20, 21} (z')$ and $\Gamma^{2}_{20, 21} (z')$ are the neighbor dipole amplitudes for $H^{1}_{10} (z)$ and $H^{2}_{10} (z)$, respectively \cite{Kovchegov:2015pbl,Kovchegov:2016zex}, which we wrote in anticipation that only the $x_{21} \ll x_{10}$ region would contribute in DLA. This is indeed the case. We assume, like for the Sivers function analysis, that all amplitudes do not contain integer powers of the dipole size, and may contain only $\sim \sqrt{\as}$ or $\sim \as$ powers or logarithms of the dipole sizes. In the $\un{x}_2 \rightarrow \un{x}_0$ limit the two terms in the brackets of Eqs.~\eqref{HS_evol} cancel each other, and the eikonal kernel itself has no IR logarithms, so we only have the $\un{x}_2 \rightarrow \un{x}_1$ ultraviolet (UV) logarithms to consider. Keeping only this UV logarithmic contribution in Eqs.~\eqref{HS_evol}  and integrating the resulting equations over all impact parameters while employing Eqs.~\eqref{HS_defs4} along with
\begin{subequations}\label{HS_defs40}
\begin{align}
& \int d^2 b_\perp \, \Gamma^{1}_{10, 21} (z) = {\un x}_{10} \times {\un S} \ \Gamma^{1} (x_{10}^2, x_{21}^2, z ) ,  \label{HS_defs4a} \\
& \int d^2 b_\perp \, \Gamma^{2}_{10, 21} (z) = {\un x}_{10} \cdot {\un S} \ \Gamma^{2} (x_{10}^2, x_{21}^2,  z) , \label{HS_defs4b} 
\end{align}
\end{subequations}
we arrive at 
\begin{subequations}\label{HS_evol2}
\begin{align}
& {\un x}_{10} \times {\un S} \, H^{1} (x_{10}^2, z) = {\un x}_{10} \times {\un S} \, H^{1 \, (0)} (x_{10}^2, z) + \frac{\as \, N_c}{2 \pi} \,  \int\limits_\frac{\Lambda^2}{s}^z \frac{dz'}{z'} \int\limits^{x_{10}^2}_\frac{1}{z' s} \frac{d x_{21}^2}{x_{21}^2} \, \left[ {\un x}_{12} \times {\un S} \, H^{1} (x^2_{12} , z') \right. \\ 
& \hspace*{10cm} \left.  - {\un x}_{10} \times {\un S} \, \Gamma^{1} (x_{10}^2, x_{21}^2 , z') \right], \notag \\
& {\un x}_{10} \cdot {\un S}  \, H^{2} (x_{10}^2, z) = {\un x}_{10} \cdot {\un S}  \, H^{2 \, (0)} (x_{10}^2, z) + \frac{\as \, N_c}{2 \pi^2} \,  \int\limits_\frac{\Lambda^2}{s}^z \frac{dz'}{z'} \int\limits^{x_{10}^2}_\frac{1}{z' s} \frac{d x_{21}^2}{x_{21}^2} \, \left[ {\un x}_{12} \cdot {\un S} \, H^{2} (x^2_{12} , z') \right. \\ 
& \hspace*{10cm} \left. - {\un x}_{10} \cdot {\un S} \, \Gamma^{2} (x^2_{20} , x^2_{21} , z') \right]  . \notag
\end{align}
\end{subequations}

Just like with the DLA contribution of the diagram $\alpha$, the first term in the square brackets of each equation vanishes after the integral over the ${\un x}_{21}$ angles. We are left with 
\begin{subequations}\label{HS_evol3}
\begin{align}
& H^{1} (x_{10}^2, z) = H^{1 \, (0)} (x_{10}^2, z) - \frac{\as \, N_c}{2 \pi} \,  \int\limits_\frac{\Lambda^2}{s}^z \frac{dz'}{z'} \int\limits^{x_{10}^2}_\frac{1}{z' s} \frac{d x_{21}^2}{x_{21}^2} \, \Gamma^{1} (x_{10}^2, x_{21}^2 , z') , \\
& H^{2} (x_{10}^2, z) = H^{2 \, (0)} (x_{10}^2, z) - \frac{\as \, N_c}{2 \pi^2} \,  \int\limits_\frac{\Lambda^2}{s}^z \frac{dz'}{z'} \int\limits^{x_{10}^2}_\frac{1}{z' s} \frac{d x_{21}^2}{x_{21}^2} \, \Gamma^{2} (x^2_{20} , x^2_{21} , z') . 
\end{align}
\end{subequations}

Since the equations \eqref{HS_evol3} are identical, we will solve only one of them, with the solution applicable to the other equation. To close the system, we need an equation for the neighbor dipole amplitude, which can be readily derived following \cite{Kovchegov:2015pbl,Kovchegov:2016zex}. We thus have
\begin{subequations}\label{HS_evol4}
\begin{align}
    &H (x_{10}^2, z) = H^{(0)} (x_{10}^2, z) - \frac{\alpha_s N_c}{2 \pi} \int\limits_{\frac{\Lambda^2}{s}}^{z} \frac{\dd{z}'}{z'} \int\limits_\frac{1}{z' s}^{x_{10}^2} \frac{\dd{x}_{21}^2}{x_{21}^2} \Gamma (x_{10}^2, x_{21}^2, z') , \label{h_eq} \\
    &\Gamma (x_{10}^2, x_{21}^2, z') = H^{(0)} (x_{10}^2, z) - \frac{\alpha_s N_c}{2 \pi} \int\limits_{\frac{\Lambda^2}{s}}^{z'} \frac{\dd{z}''}{z''} \int\limits_\frac{1}{z'' s}^{\textrm{min}\{ x_{10}^2, x_{21}^2 \frac{z'}{z''} \}} \frac{\dd{x}_{32}^2}{x_{32}^2} \Gamma (x_{10}^2, x_{32}^2, z') \label{h_gam_eq} .
\end{align}
\end{subequations}

We can rewrite these equations in terms of the dimensionless variables similar to those we used for the Sivers function dipole amplitudes
\begin{subequations}\label{HS_evol55}
\begin{align}
\eta \equiv \sqrt{\frac{\as N_c}{2 \pi}} \, \ln \frac{zs}{\Lambda^2}, \ \ \ s_{10} \equiv \sqrt{\frac{\as N_c}{2 \pi}} \, \ln \frac{1}{x_{10}^2 \Lambda^2} , \\
\eta' \equiv \sqrt{\frac{\as N_c}{2 \pi}} \, \ln \frac{z's}{\Lambda^2}, \ \ \ s_{21} \equiv \sqrt{\frac{\as N_c}{2 \pi}} \, \ln \frac{1}{x_{12}^2 \Lambda^2} , \\
\eta'' \equiv \sqrt{\frac{\as N_c}{2 \pi}} \, \ln \frac{z''s}{\Lambda^2}, \ \ \ s_{32} \equiv \sqrt{\frac{\as N_c}{2 \pi}} \, \ln \frac{1}{x_{32}^2 \Lambda^2} .
\end{align}
\end{subequations}
Performing this substitution and putting $H^{(0)} (x_{10}^2, z) = 1$ for simplicity, we have
\begin{subequations}\label{HS_evol66}
\begin{align}
    &H (s_{10}, \eta) = 1 - \int\limits_{s_{10}}^{\eta} \dd{\eta}' \int\limits_{s_{10}}^{\eta'} \dd{s}_{21} \Gamma (s_{10}, s_{21} ,\eta' ) , \label{h_ev_dimless} \\
    &\Gamma ( s_{10}, s_{21}, \eta' ) = 1 - \int\limits_{s_{10}}^{\eta'} \dd{\eta}' \int\limits_{\textrm{max}\{s_{10}, s_{21} + \eta'' - \eta'\}}^{\eta''} \dd{s}_{32} \Gamma (s_{10}, s_{32}, \eta'') \label{h_gam_ev_dimless} .
\end{align}
\end{subequations}

Equations \eqref{HS_evol66} are similar to those solved in \cite{Kovchegov:2017jxc}. Their kernel structure admits the scaling ansatz for the dipole amplitudes, 
\begin{align} \label{h_dimless_ansatz}
    &H (s_{10}, \eta) = H (\eta - s_{10}) = H (\zeta) \\
    &\Gamma (s_{10}, s_{21}, \eta') = \Gamma (\eta' - s_{10}, \eta' - s_{21}) = \Gamma ( \zeta, \zeta'). \notag
\end{align}
This ansatz converts Eqs.~\eqref{HS_evol66} into the following evolution equations 
\begin{subequations}\label{HS_evol77}
\begin{align}
    &H (\zeta) = 1 - \int\limits_{0}^{\zeta} \dd{\xi} \int\limits_{0}^{\xi} \dd{\xi}' \Gamma ( \xi, \xi') \label{h_scale_ev} \\
    &\Gamma ( \zeta, \zeta') = H (\zeta') - \int\limits_{\zeta'}^{\zeta} \dd{\xi} \int\limits_{0}^{\zeta'} \dd{\xi}' \Gamma ( \xi, \xi') . \label{h_gam_scale_ev}
\end{align}
\end{subequations}
Note that  $H (\zeta) = \Gamma (\zeta, \zeta)$.

We solve Eqs.~\eqref{HS_evol77} in Appendix~\ref{sec:app_sol}, obtaining
\begin{align} \label{Hsolved}
    H (\zeta) &= \frac{1}{\zeta} J_1 (2 \zeta). 
\end{align}
Plugging this solution back into the expression \eqref{BM_NS_2} for the flavor non-singlet Boer-Mulders function (both for $H^{1} (x_{10}^2, z)$ and $H^{2} (x_{10}^2, z)$) we find
\begin{align}
    \frac{k_T^y}{M_P} h_1^{\perp \, NS} (x, k_T^2) \propto x \, \sqrt{\frac{2 \pi}{\alpha_s N_c}} \int\limits_{\sqrt{\frac{\alpha_s N_c}{2\pi}} \ln (\Lambda^2 x_{10}^2)}^{\sqrt{\frac{\alpha_s N_c}{2\pi}} \ln (s x_{10}^2)} \frac{d \zeta}{\zeta} \, J_1  (2 \zeta) . 
\end{align}
Since 
\begin{align}\label{Bessel_asympt}
    \frac{1}{\zeta} J_1 (2 \zeta) \Bigg|_{\zeta \gg 1} \approx \sqrt{\frac{1}{\pi}} \frac{1}{\zeta^{3/2}} \cos \left( 2 \zeta - \frac{3 \pi}{4}  \right)
\end{align}
we see that the $\zeta$ integral is convergent at high energies, giving us an $s$-independent constant in the $s \to \infty$ limit. This means that DLA evolution leaves the small-$x$ flavor non-singlet Boer-Mulders function with its exact naive sub-sub-eikonal scaling, 
\begin{align} \label{bm_evolved}
    h_1^{\perp NS} (x \ll 1, k_T^2) \sim  \left( \frac{1}{x} \right)^{-1} .
\end{align}
Note that the exact solution of the above evolution equations for $H^{1} (x_{10}^2, z)$ and $H^{2} (x_{10}^2, z)$, when employed in \eq{BM_NS_2} would allow us to generate a more detailed expression for the Boer-Mulders function, which would include its $k_T$-dependence (potentially mixing with the $x$-dependence) as well. In \eq{bm_evolved} we are only showing the dominant power of $x$ for this TMD. 

One may notice that the lowest-order calculation of the quark Boer-Mulders function in the quark--scalar diquark model of the proton results in an additive sub-eikonal term proportional to the light quark mass: see Eq.~(A11) in \cite{Meissner:2007rx}. This may motivate one to explore the possibility that keeping the (light) quark mass-dependent terms in \eq{mel} may result in a different (perhaps sub-eikonal) small-$x$ asymptotics for the Boer-Mulders function. In Appendix~\ref{sec:app_mass} we explore the mass corrections to our result \eqref{bm_evolved}. We find that such sub-eikonal correction is possible: however, in DLA the evolution of the corresponding dipole amplitudes $H^{NS} (x_{10}^2, z)$ and $H^{NS \, [2]} (x_{10}^2, z)$ is trivial, the amplitudes do not evolve, as we also show in Appendix~\ref{sec:app_mass}. Hence, the light quark mass correction to \eq{bm_evolved} comes in with the amplitudes given entirely by their initial conditions, which, therefore, may simply be zero. If, similar to \cite{Meissner:2007rx}, either one or both of the initial condition amplitudes $H^{NS \, (0)} (x_{10}^2, z)$ and $H^{NS \, [2] \, (0)} (x_{10}^2, z)$ from Eqs.~\eqref{H_NS_2} and \eqref{HS_NS_5} in Appendix~\ref{sec:app_mass} are not zero, then \eq{bm_evolved} would receive an additive $\sim x^0 (m/k_T)$ correction to its right-hand side. Whether this correction dominates over the sub-sub-eikonal term given in \eq{bm_evolved} would depend on the $(x, k_T^2)$ region of interest and on the value of the light quark mass $m$.


\section{Conclusions}
\label{sec:conc}

In this paper we studied the small-$x$ asymptotics of T-odd leading-twist quark TMDs, the Sivers and Boer-Mulders functions. We have revised the sub-eikonal DLA small-$x$ evolution for the flavor non-singlet Sivers function by including the terms omitted in our previous paper on the subject \cite{Kovchegov:2021iyc}. The final expression for the sub-eikonal contribution to the flavor non-singlet Sivers function is given in \eq{F_S&NS2b}. The DLA large-$N_c$ evolution of the corresponding dipole amplitudes $F_A^{NS} (x_{10}^2, z)$, $F_B^{NS} (x_{10}^2, z)$, $F_C^{NS} (x_{10}^2, z)$, and $F^{NS \, \textrm{mag}} (x_{10}^2, z)$ is given in Eqs.~\eqref{FFFF} and \eqref{Gamma4}, with the remaining amplitude $F^{NS \, [2]} (x_{10}^2, z) =0$ as shown above as well. The solution of Eqs.~\eqref{FFFF} and \eqref{Gamma4}, when used in \eq{F_S&NS2b} would give us a complete expression for the flavor non-singlet Sivers function at small $x$, including all the $x$ and $k_T$ dependence of this function. Leaving the search for an analytic solution of Eqs.~\eqref{FFFF} and \eqref{Gamma4} for the future, we have solved these equations numerically with the result shown in \fig{FIG:num_dips}. While the corresponding numerical construction of $f_{1 \: T}^{\perp \: NS} (x, k_T^2)$ can readily be accomplished using the amplitudes from \fig{FIG:num_dips}, we have instead concentrated on extracting the {\sl leading} $x$-dependence of these amplitudes. This resulted in the following dominant small-$x$ asymptotics of the flavor non-singlet Sivers function,
\begin{align}\label{NS_Sivers}
f_{1 \: T}^{\perp \: NS} (x \ll 1 ,k_T^2) =  C_O (x, k_T^2) \, \frac{1}{x} + C_1 (x, k_T^2) \, \left( \frac{1}{x} \right)^{3.4 \, \sqrt{\frac{\as \, N_c}{4 \pi}}} + \ldots,
\end{align}
which also includes the (eikonal) spin-dependent odderon \cite{Boer:2015pni,Dong:2018wsp} term (the first term on the right), with our sub-eikonal result given by the second term on the right. The conclusion \eqref{NS_Sivers}, given by the sum of the odderon and DLA terms, appears to be qualitatively similar to that found in \cite{Kirschner:1996jj} for the transverse structure function. The intercept (power of $1/x$) of our sub-eikonal correction appears to be numerically large, $3.4 \, \sqrt{\tfrac{\as \, N_c}{4 \pi}} \approx 0.91$ for $\as = 0.3$, very close to the intercept of $1$ for the odderon. If this DLA intercept remains large when higher-order corrections in $\as$ are included, our result may indicate a difficulty in experimentally searching for the spin-dependent odderon caused by the ``background" sub-eikonal correction having the $x$-dependence very similar to that of the odderon. 

In the second part of the paper we have found the leading contribution to the flavor non-singlet Boer-Mulders function at small-$x$ given by \eq{BM_NS_2}, which depends on the sub-sub-eikonal dipole amplitudes $H^1_{10}$ and $H^2_{10}$ defined in Eqs.~\eqref{HS_defs3}. We obtained nonlinear evolution equations for these dipole amplitudes in the large-$N_c$ limit, and solved the linearized large-$N_c$ equations \eqref{HS_evol} analytically. It is interesting to note that the equations \eqref{HS_evol} are fully decoupled from each other, although the nonlinear versions of those equations couple to the unpolarized eikonal dipole $S$-matrices which obey the Balitsky-Kovchegov (BK) equations \cite{Balitsky:1995ub,Balitsky:1998ya,Kovchegov:1999yj,Kovchegov:1999ua}. In addition, the equations \eqref{HS_evol} for $H^1_{10}$ and $H^2_{10}$ are actually identical. Both of these features appear to be in contrast to the complex coupled system of equations we found for the dipole amplitudes defining the Sivers function. 

We solved the linearized DLA equations \eqref{HS_evol} for the Boer-Mulders dipole amplitudes analytically, obtaining the solution \eqref{Hsolved} which oscillates with decreasing oscillation amplitude as $\ln(1/x)$ increases. Plugging the evolved dipole amplitudes back into the expression \eqref{BM_NS_2} for this TMD led to the oscillations being practically ``washed out", and we found that the leading sub-sub-eikonal contribution to the Boer-Mulders function receives no power corrections to its scaling with $x$, that is 
\begin{align}\label{NS_Boer_Mulders}
    h_1^{\perp NS} (x \ll 1, k_T^2) = C (x, k_T^2) \left( \frac{1}{x} \right)^{-1} + \ldots .
\end{align}
Here the ellipsis denote corrections further suppressed by powers of $x$ while $C (x, k_T^2)$ is a slowly-varying function of $x$ which can be exactly determined by our formalism developed above. In addition, we discussed that the light quark mass corrections may bring in the sub-eikonal $\sim x^0 \, m/k_T$ additive term in \eq{NS_Boer_Mulders}.

The situation with small-$x$ asymptotics of the Boer-Mulders function is very similar to the case of the spin-dependent odderon which gave the leading eikonal contribution to the quark Sivers function \cite{Boer:2015pni,Szymanowski:2016mbq,Dong:2018wsp} (cf. \eq{siv}). As we briefly mentioned above, the odderon is known to have a small-$x$ intercept of one \cite{Bartels:1999yt,Kovchegov:2003dm,Kovchegov:2012rz,Caron-Huot:2013fea,Brower:2008cy,Avsar:2009hc,Brower:2014wha} (at leading and sub-leading logarithmic order, and to all orders in $\as \, N_c$), which translates into the eikonal scaling  $\sim 1/x$ for the Sivers function. While the asymptotics \eqref{NS_Boer_Mulders} for the Boer-Mulders function is not known with such a high precision, we observe that with the accuracy of our DLA calculation, the intercept receives no correction to the lowest-order sub-sub-eikonal $x$-scaling. Additionally, a number of T-odd gluon TMDs with the dipole Wilson line staple studied in \cite{Boer:2015pni} have their small-$x$ asymptotics dominated by the spin-dependent odderon, with the intercept, again, receiving no corrections proportional to any power of $\as \, N_c$. 

At this point one may speculate that the leading small-$x$ asymptotics of T-odd TMDs does not receive any power-of-$\as$ corrections to its leading intercept (leading power of $1/x$). It seems that we can conjecture that the T-odd TMDs have some protection from $\as$-corrections to their leading power of $1/x$ under small-$x$ evolution, possibly due to some hidden symmetry which has not yet been identified. It would be very interesting to further investigate this apparent protection, possibly by studying the higher-order in $\as$ corrections to our result in \eq{NS_Boer_Mulders}. We already know from the sub-eikonal contribution to the quark Sivers function shown in \eq{NS_Sivers} that this protection does not hold when one goes beyond the leading-eikonality (leading power of $1/x$) small-$x$ contribution: while the protection holds for the Sivers function at the eikonal order, it is clearly broken in the sub-eikonal (the second on the right-hand side) term in \eq{NS_Sivers}. Still it may be interesting to see whether this protection holds in the leading term of \eq{NS_Boer_Mulders} beyond the zero DLA-order correction ($\sim \sqrt{\as}$) found here.


\section*{Acknowledgments}

The author are indebted to Yossathorn (Josh) Tawabutr for his insights and advice on the numerical solution of the evolution equations for the Sivers function. 

This material is based upon work supported by the U.S. Department of Energy, Office of Science, Office of Nuclear Physics under Award Number DE-SC0004286 (YK and MGS) and under Contract Number DE-SC0020682 (MGS), and by the Center for Nuclear Femtography, Southeastern Universities Research Association, Washington, D.C. (MGS). \\



\appendix
\section{Extraction of the DLA Evolution Equations for the Sivers Function}
\label{sec:app_DLA}

In this Appendix we extract the DLA parts of Eqs.~\eqref{F_NS_2} and \eqref{F_mag_eq_2}, beginning with the former equation.

We start with the first term under the integral in \eq{F_NS_2}, 
\begin{align}\label{term1}
\int d^2 x_2 \, 2 \, \left[ \frac{\epsilon^{ij} x_{21}^j}{x_{21}^2} - \frac{\epsilon^{ij} x_{20}^j}{x_{20}^2} + 2 x_{21}^i \frac{{\un x}_{21} \times {\un x}_{20}}{x_{21}^2 \, x_{20}^2} \right] \, \left[ - {\un x}_{21} \cdot {\un S}_P \, F^{NS \, \textrm{mag}} (x_{21}^2, z')  +  {\un x}_{20} \cdot {\un S}_P \, F^{NS \, \textrm{mag}} (x_{20}^2, z')   \right] .
\end{align}
It clearly has no ultraviolet (UV) logarithms, neither when $2 \to 1$ nor when $2 \to 0$. It does have an IR logarithm (coming from the last term in the first square brackets), which gives
\begin{align}\label{term1_DLA}
2 \pi \,  \epsilon^{ij} \, x_{10}^j \, {\un x}_{10} \cdot {\un S}_P \, \int\limits_{x_{10}^2}^{\frac{z}{z'} x_{10}^2} \frac{d x_{21}^2}{x_{21}^2} \, F^{NS \, \textrm{mag}} (x_{21}^2, z') .
\end{align}

Next we analyze the $F_A^{NS}$ contribution to the second term under the integral in \eq{F_NS_2}, which we rewrite as 
\begin{align}\label{term2A}
& \epsilon^{jk} \, S_P^k \,  \int d^2 x_2 \left[ \delta^{ij} \left( \frac{2}{x_{21}^2} + \frac{x_{10}^2}{x_{21}^2 \, x_{20}^2} - \frac{2}{x_{20}^2} \right)  - 2 \frac{x_{21}^i \, x_{20}^j}{x_{21}^2 \, x_{20}^2} \left( 2 \frac{{\un x}_{20} \cdot {\un x}_{21}}{x_{20}^2} + 1 \right) + 2 \frac{x_{21}^i \, x_{21}^j}{x_{21}^2 \, x_{20}^2} \left( 2 \frac{{\un x}_{20} \cdot {\un x}_{21}}{x_{21}^2} + 1 \right) + 2 \frac{x_{20}^i \, x_{20}^j}{x_{20}^4} \right. \notag \\ 
& \left. - 2 \frac{x_{21}^i \, x_{21}^j}{x_{21}^4}   \right] \, \left[ - x_{21}^2 \, F_A^{NS} (x_{21}^2, z') + x_{20}^2 \, F_A^{NS} (x_{20}^2, z') \right]. 
\end{align}
One can readily verify that there is no UV logarithm as $2 \to 0$. There is a UV logarithm at $2 \to 1$, which gives
\begin{align}\label{term2A_DLA_UV}
2 \pi \, \epsilon^{ij} \, S_P^j \, x_{10}^2 \,  \int\limits^{x_{10}^2}_\frac{1}{z' s} \frac{d x^2_{21}}{x^2_{21}} \, \Gamma_A^{NS} (x_{10}^2, x^2_{21}, z').
\end{align}
Here we have employed the impact-parameter integrated ``neighbor" dipole amplitude $\Gamma$ \cite{Kovchegov:2015pbl,Kovchegov:2016zex,Kovchegov:2018znm}, defined for any of the amplitudes $F_A, F_B, F_C$ and $F_\textrm{mag}$ as
\begin{align}
    \Gamma  (x_{10}^2, x^2_{21}, z) = \int d^2 b_\perp \, \Gamma_{10,21} (z),
\end{align}
where the neighbor dipole amplitude $\Gamma_{10,21} (z)$ is the amplitude $F_{10} (z)$, but with the lifetime cutoff on its evolution dependent on another ``neighbor" dipole size, $x_{21}$. (See \eq{neighbor_b} in the main text and the discussion above it.)

The IR logarithm in \eq{term2A} is 
\begin{align}\label{term2A_DLA_IR}
2 \pi \, ( 2 \,  \epsilon^{ij} \, S_P^j \, x_{10}^2 - x_{10}^i \, {\un x}_{10} \times {\un S}_P ) \,  \int\limits_{x_{10}^2}^{\frac{z}{z'} \, x_{10}^2} \frac{d x^2_{21}}{x^2_{21}} \, F_A^{NS} (x^2_{21}, z').
\end{align}

Next we move on to the $F_B^{NS}$ contribution to the second term under the integral in \eq{F_NS_2}
\begin{align}\label{term2B}
& \int d^2 x_2 \left[ \delta^{ij} \left( \frac{2}{x_{21}^2} + \frac{x_{10}^2}{x_{21}^2 \, x_{20}^2} - \frac{2}{x_{20}^2} \right)  - 2 \frac{x_{21}^i \, x_{20}^j}{x_{21}^2 \, x_{20}^2} \left( 2 \frac{{\un x}_{20} \cdot {\un x}_{21}}{x_{20}^2} + 1 \right) + 2 \frac{x_{21}^i \, x_{21}^j}{x_{21}^2 \, x_{20}^2} \left( 2 \frac{{\un x}_{20} \cdot {\un x}_{21}}{x_{21}^2} + 1 \right) + 2 \frac{x_{20}^i \, x_{20}^j}{x_{20}^4} \right. \notag \\ 
& \left. - 2 \frac{x_{21}^i \, x_{21}^j}{x_{21}^4}   \right] \, \left[ - x_{21}^j \, {\un x}_{21} \times {\un S}_P \, F^{NS}_B (x_{21}^2, z') + x_{20}^j \, {\un x}_{20} \times {\un S}_P \, F^{NS}_B (x_{20}^2, z') \right]. 
\end{align}
Again, there is no UV logarithm as $2 \to 0$. From the $2 \to 1$ UV limit we get
\begin{align}\label{term2B_DLA_UV}
2 \pi \, x_{10}^i  \, {\un x}_{10} \times {\un S}_P \,  \int\limits^{x_{10}^2}_\frac{1}{z' s} \frac{d x^2_{21}}{x^2_{21}} \, \Gamma_B^{NS} (x_{10}^2, x^2_{21}, z').
\end{align}
In the IR we have
\begin{align}\label{term2B_DLA_IR}
\pi \, ( 5\,  x_{10}^i \, {\un x}_{10} \times {\un S}_P - 2 \, \epsilon^{ij} \, S_P^j \, x_{10}^2 ) \,  \int\limits_{x_{10}^2}^{\frac{z}{z'} \, x_{10}^2} \frac{d x^2_{21}}{x^2_{21}} \, F_B^{NS} (x^2_{21}, z').
\end{align}

Moving on to the $F_C^{NS}$ contribution to the second term under the integral in \eq{F_NS_2}
\begin{align}\label{term2C}
& \int d^2 x_2 \left[ \delta^{ij} \left( \frac{2}{x_{21}^2} + \frac{x_{10}^2}{x_{21}^2 \, x_{20}^2} - \frac{2}{x_{20}^2} \right)  - 2 \frac{x_{21}^i \, x_{20}^j}{x_{21}^2 \, x_{20}^2} \left( 2 \frac{{\un x}_{20} \cdot {\un x}_{21}}{x_{20}^2} + 1 \right) + 2 \frac{x_{21}^i \, x_{21}^j}{x_{21}^2 \, x_{20}^2} \left( 2 \frac{{\un x}_{20} \cdot {\un x}_{21}}{x_{21}^2} + 1 \right) + 2 \frac{x_{20}^i \, x_{20}^j}{x_{20}^4} \right. \notag \\ 
& \left. - 2 \frac{x_{21}^i \, x_{21}^j}{x_{21}^4}   \right] \, \left[ -  \epsilon^{jk} \, x_{21}^k \, {\un x}_{21} \cdot {\un S}_P \, F^{NS}_C (x_{21}^2, z')  +  \epsilon^{jk} \, x_{20}^k \, {\un x}_{20} \cdot {\un S}_P \, F^{NS}_C (x_{20}^2, z')  \right] 
\end{align}
we again notice the lack of the UV logarithm as $2 \to 0$. The $2 \to 1$ UV logarithm is
\begin{align}\label{term2C_DLA_UV}
2 \pi \, \epsilon^{ij} \, x_{10}^j  \, {\un x}_{10} \cdot {\un S}_P \,  \int\limits^{x_{10}^2}_\frac{1}{z' s} \frac{d x^2_{21}}{x^2_{21}} \, \Gamma_C^{NS} (x_{10}^2, x^2_{21}, z').
\end{align}
In the IR we have
\begin{align}\label{term2C_DLA_IR}
\pi \, ( 6 \, \epsilon^{ij} \, x_{10}^j \, {\un x}_{10} \cdot {\un S}_P -  x_{10}^i \, {\un x}_{10} \times {\un S}_P  ) \,  \int\limits_{x_{10}^2}^{\frac{z}{z'} \, x_{10}^2} \frac{d x^2_{21}}{x^2_{21}} \, F_C^{NS} (x^2_{21}, z').
\end{align}

Finally we consider the eikonal evolution term, 
\begin{align}\label{term_eik_DLA}
& \int d^2 x_2 \, \frac{x_{10}^2}{x_{21}^2 \, x_{20}^2} \, \Bigg\{ \epsilon^{ij} \, S_P^j \, x_{21}^2 \, F_A^{NS} (x_{21}^2, z') + x_{21}^i \, {\un x}_{21} \times {\un S}_P \, F^{NS}_B (x_{21}^2, z') +  \epsilon^{ij} \, x_{21}^j \, {\un x}_{21} \cdot {\un S}_P \, F^{NS}_C (x_{21}^2, z') \notag \\ 
&  - \epsilon^{ij} \, S_P^j \, x_{10}^2 \, \Gamma_A^{NS} (x_{10}^2, x_{21}^2, z') - x_{10}^i \, {\un x}_{10} \times {\un S}_P \, \Gamma^{NS}_B (x_{10}^2, x_{21}^2, z') -  \epsilon^{ij} \, x_{10}^j \, {\un x}_{10} \cdot {\un S}_P \, \Gamma^{NS}_C (x_{10}^2, x_{21}^2, z')  \Bigg\} .
\end{align}
Again, there is no UV logarithm as $2 \to 0$. In the $2 \to 1$ UV limit only the virtual terms contribute, yielding
\begin{align}\label{term_eik_DLA_UV}
- \pi \, \int\limits^{x_{10}^2}_\frac{1}{z' s} \frac{d x^2_{21}}{x^2_{21}} \, \left[ \epsilon^{ij} \, S_P^j \, x_{10}^2 \, \Gamma_A^{NS} (x_{10}^2, x_{21}^2, z') + x_{10}^i \, {\un x}_{10} \times {\un S}_P \, \Gamma^{NS}_B (x_{10}^2, x_{21}^2, z') +  \epsilon^{ij} \, x_{10}^j \, {\un x}_{10} \cdot {\un S}_P \, \Gamma^{NS}_C (x_{10}^2, x_{21}^2, z') \right] .
\end{align}
In the IR limit, only the real terms remain, giving 
\begin{align}\label{term_eik_DLA_IR}
\frac{\pi}{2} \, \epsilon^{ij} \, S_P^j \, x_{10}^2 \, \int\limits_{x_{10}^2}^{\frac{z}{z'} \, x_{10}^2} \frac{d x^2_{21}}{x^2_{21}} \, \left[ 2 \, F_A^{NS} (x^2_{21}, z')  + F_B^{NS} (x^2_{21}, z')  + F_C^{NS} (x^2_{21}, z') \right] .
\end{align}

We thus arrive at the following evolution equation:
\begin{align}\label{F_NS_DLA}
& \epsilon^{ij} \, S_P^j \, x_{10}^2 \, F_A^{NS} (x_{10}^2, z) + x_{10}^i \, {\un x}_{10} \times {\un S}_P \, F^{NS}_B (x_{10}^2, z) +  \epsilon^{ij} \, x_{10}^j \, {\un x}_{10} \cdot {\un S}_P \, F^{NS}_C (x_{10}^2, z) \notag \\ 
& = \epsilon^{ij} \, S_P^j \, x_{10}^2 \, F_A^{NS\, (0)} (x_{10}^2, z) + x_{10}^i \, {\un x}_{10} \times {\un S}_P \, F^{NS\, (0)}_B (x_{10}^2, z) +  \epsilon^{ij} \, x_{10}^j \, {\un x}_{10} \cdot {\un S}_P \, F^{NS\, (0)}_C (x_{10}^2, z)  \notag \\ 
& + \frac{\as \, N_c}{2 \pi} \, \epsilon^{ij} \, x_{10}^j \, {\un x}_{10} \cdot {\un S}_P \, \int\limits_{\frac{\Lambda^2}{s}}^z \frac{d z'}{z'} \,  \int\limits_{x_{10}^2}^{\frac{z}{z'} x_{10}^2} \frac{d x_{21}^2}{x_{21}^2} \, F^{NS \, \textrm{mag}} (x_{21}^2, z') + \frac{\as \, N_c}{2 \pi} \, \epsilon^{ij} \, S_P^j \, x_{10}^2 \,  \int\limits_{\frac{1}{s \, x_{10}^2}}^z \frac{d z'}{z'} \ \int\limits^{x_{10}^2}_\frac{1}{z' s} \frac{d x^2_{21}}{x^2_{21}} \, \Gamma_A^{NS} (x_{10}^2, x^2_{21}, z') \notag \\ 
& + \frac{\as \, N_c}{2 \pi} \,  ( 2 \, \epsilon^{ij} \, S_P^j \, x_{10}^2 - x_{10}^i \, {\un x}_{10} \times {\un S}_P ) \, \int\limits_{\frac{\Lambda^2}{s}}^z \frac{d z'}{z'} \,  \int\limits_{x_{10}^2}^{\frac{z}{z'} \, x_{10}^2} \frac{d x^2_{21}}{x^2_{21}} \, F_A^{NS} (x^2_{21}, z') \notag \\ 
& +  \frac{\as \, N_c}{2 \pi} \, x_{10}^i  \, {\un x}_{10} \times {\un S}_P \, \int\limits_{\frac{1}{s \, x_{10}^2}}^z \frac{d z'}{z'} \   \int\limits^{x_{10}^2}_\frac{1}{z' s} \frac{d x^2_{21}}{x^2_{21}} \, \Gamma_B^{NS} (x_{10}^2, x^2_{21}, z')  \notag  \\ 
& + \frac{\as \, N_c}{4 \pi} \, ( 5\,  x_{10}^i \, {\un x}_{10} \times {\un S}_P - 2 \, \epsilon^{ij} \, S_P^j \, x_{10}^2 ) \, \int\limits_{\frac{\Lambda^2}{s}}^z \frac{d z'}{z'} \, \int\limits_{x_{10}^2}^{\frac{z}{z'} \, x_{10}^2} \frac{d x^2_{21}}{x^2_{21}} \, F_B^{NS} (x^2_{21}, z') \notag \\
& +  \frac{\as \, N_c}{2 \pi} \, \epsilon^{ij} \, x_{10}^j  \, {\un x}_{10} \cdot {\un S}_P \, \int\limits_{\frac{1}{s \, x_{10}^2}}^z \frac{d z'}{z'} \   \int\limits^{x_{10}^2}_\frac{1}{z' s} \frac{d x^2_{21}}{x^2_{21}} \, \Gamma_C^{NS} (x_{10}^2, x^2_{21}, z')  \notag  \\ 
& + \frac{\as \, N_c}{4 \pi} \, ( 6 \, \epsilon^{ij} \, x_{10}^j \, {\un x}_{10} \cdot {\un S}_P -  x_{10}^i \, {\un x}_{10} \times {\un S}_P  ) \, \int\limits_{\frac{\Lambda^2}{s}}^z \frac{d z'}{z'} \, \int\limits_{x_{10}^2}^{\frac{z}{z'} \, x_{10}^2} \frac{d x^2_{21}}{x^2_{21}} \, F_C^{NS} (x^2_{21}, z') \notag \\
& - \frac{\as \, N_c}{2 \pi} \int\limits_{\frac{1}{s \, x_{10}^2}}^z \frac{d z'}{z'}  \int\limits^{x_{10}^2}_\frac{1}{z' s} \frac{d x^2_{21}}{x^2_{21}}  \left[ \epsilon^{ij} \, S_P^j \, x_{10}^2 \, \Gamma_A^{NS} (x_{10}^2, x_{21}^2, z') + x_{10}^i \, {\un x}_{10} \times {\un S}_P \, \Gamma^{NS}_B (x_{10}^2, x_{21}^2, z') \right. \notag \\ 
& \hspace*{10cm} \left. +  \epsilon^{ij} \, x_{10}^j \, {\un x}_{10} \cdot {\un S}_P \, \Gamma^{NS}_C (x_{10}^2, x_{21}^2, z') \right] \notag \\
& +  \frac{\as \, N_c}{4 \pi} \epsilon^{ij} \, S_P^j \, x_{10}^2 \, \int\limits_{\frac{\Lambda^2}{s}}^z \frac{d z'}{z'} \,  \int\limits_{x_{10}^2}^{\frac{z}{z'} \, x_{10}^2} \frac{d x^2_{21}}{x^2_{21}} \, \left[ 2 \, F_A^{NS} (x^2_{21}, z')  + F_B^{NS} (x^2_{21}, z')  + F_C^{NS} (x^2_{21}, z') \right] . 
\end{align}
Equating the coefficients of the three tensor structures we arrive at the equations \eqref{FABC} in the main text. 

Next, we will obtain the DLA version of \eq{F_mag_eq_2}. To extract the DLA evolution from \eq{F_mag_eq_2}, we begin with the first term on its right, 
\begin{align}\label{term1mag}
\int d^2 x_2 \, 2 \, \left[ \frac{1}{x_{21}^2} -  \frac{{\un x}_{21}}{x_{21}^2} \cdot \frac{{\un x}_{20}}{x_{20}^2} \right] \, \left[ - {\un x}_{21} \cdot {\un S}_P \, F^{NS \, \textrm{mag}} (x_{21}^2, z') +  {\un x}_{20} \cdot {\un S}_P \, F^{NS \, \textrm{mag}} (x_{20}^2, z')  \right]. 
\end{align}
There is no UV logarithm when $2 \to 0$. There is no IR logarithm either. When $2 \to 1$ we get
\begin{align}\label{term1_DLA_UV}
2 \pi \, {\un x}_{10} \cdot {\un S}_P \,  \int\limits^{x_{10}^2}_\frac{1}{z' s} \frac{d x^2_{21}}{x^2_{21}} \, \Gamma^{NS \, \textrm{mag}} (x_{10}^2, x^2_{21}, z').
\end{align}

Next we look at the $F_A^{NS}$ contribution to the second term in \eq{F_mag_eq_2}, which, in DLA, is 
\begin{align}\label{term2Amag}
& \epsilon^{ik} \, S_P^k \, \int d^2 x_2  \left[ 2 \frac{\epsilon^{ij} \, x_{21}^j}{x_{21}^4} - \frac{\epsilon^{ij} \, (x_{20}^j + x_{21}^j)}{x_{20}^2 \, x_{21}^2}  - \frac{2 \, {\un x}_{20} \times {\un x}_{21}}{x_{20}^2 \, x_{21}^2} \left( \frac{x_{21}^i}{x_{21}^2} - \frac{x_{20}^i}{x_{20}^2}\right) \right] \left[ -  x_{21}^2 \, F_A^{NS} (x_{21}^2, z')  + x_{20}^2 \, F_A^{NS} (x_{20}^2, z') \right] \notag \\
& \approx 2 \pi \, {\un x}_{10} \cdot {\un S}_P \,  \int\limits^{x_{10}^2}_\frac{1}{z' s} \frac{d x^2_{21}}{x^2_{21}} \, \Gamma^{NS}_A (x_{10}^2, x^2_{21}, z').
\end{align}
There is neither IR nor UV $2 \to 0$ logarithms, and the DLA contribution only comes from the $2 \to 1$ region, originating from the first term in the square brackets.

Similarly, for the $F_B^{NS}$ contribution to the second term in \eq{F_mag_eq_2} we write
\begin{align}\label{term2Bmag}
&  \int d^2 x_2  \left[ 2 \frac{\epsilon^{ij} \, x_{21}^j}{x_{21}^4} - \frac{\epsilon^{ij} \, (x_{20}^j + x_{21}^j)}{x_{20}^2 \, x_{21}^2}  - \frac{2 \, {\un x}_{20} \times {\un x}_{21}}{x_{20}^2 \, x_{21}^2} \left( \frac{x_{21}^i}{x_{21}^2} - \frac{x_{20}^i}{x_{20}^2}\right) \right] \left[ - x_{21}^i \, {\un x}_{21} \times {\un S}_P \, F^{NS}_B (x_{21}^2, z') \right. \notag \\
& \left. + x_{20}^i \, {\un x}_{20} \times {\un S}_P \, F^{NS}_B (x_{20}^2, z') \right] \approx - \pi \, {\un x}_{10} \cdot {\un S}_P \,  \int\limits^{x_{10}^2}_\frac{1}{z' s} \frac{d x^2_{21}}{x^2_{21}} \, \Gamma^{NS}_B (x_{10}^2, x^2_{21}, z').
\end{align}

The contribution of $F_C^{NS}$ to the second term in \eq{F_mag_eq_2} is
\begin{align}\label{term2Cmag}
&  \int d^2 x_2  \left[ 2 \frac{\epsilon^{ij} \, x_{21}^j}{x_{21}^4} - \frac{\epsilon^{ij} \, (x_{20}^j + x_{21}^j)}{x_{20}^2 \, x_{21}^2}  - \frac{2 \, {\un x}_{20} \times {\un x}_{21}}{x_{20}^2 \, x_{21}^2} \left( \frac{x_{21}^i}{x_{21}^2} - \frac{x_{20}^i}{x_{20}^2}\right) \right] \left[ -  \epsilon^{ik} \, x_{21}^k \, {\un x}_{21} \cdot {\un S}_P \, F^{NS}_C (x_{21}^2, z') \right. \notag \\
& \left. +  \epsilon^{ik} \, x_{20}^k \, {\un x}_{20} \cdot {\un S}_P \, F^{NS}_C (x_{20}^2, z') \right]  \approx 3 \pi \, {\un x}_{10} \cdot {\un S}_P \,  \int\limits^{x_{10}^2}_\frac{1}{z' s} \frac{d x^2_{21}}{x^2_{21}} \, \Gamma^{NS}_C (x_{10}^2, x^2_{21}, z').
\end{align}

Finally, the eikonal term on the right of \eq{F_mag_eq_2} contributes only the UV logarithm at $2 \to 1$ coming from the virtual term,
\begin{align}
& \frac{\as \, N_c}{2 \pi^2} \, \int\limits_{\frac{\Lambda^2}{s}}^z \frac{d z'}{z'} \, \int d^2 x_2 \, \frac{x_{10}^2}{x_{21}^2 \, x_{20}^2} \,  \Bigg\{ - {\un x}_{21} \cdot {\un S}_P \, F^{NS \, \textrm{mag}} (x_{21}^2, z')    - {\un x}_{10} \cdot {\un S}_P \, \Gamma^{NS \, \textrm{mag}} (x_{10}^2, x_{21}^2,  z')  \Bigg\} \notag \\ 
& \approx  - \frac{\as \, N_c}{2 \pi} \, {\un x}_{10} \cdot {\un S}_P \, \int\limits_{\max \left\{ \frac{\Lambda^2}{s}, \frac{1}{s x_{10}^2} \right\}}^z \frac{d z'}{z'} \,  \int\limits^{x_{10}^2}_\frac{1}{z' s} \frac{d x^2_{21}}{x^2_{21}} \, \Gamma^{NS \, \textrm{mag}} (x_{10}^2, x_{21}^2,  z') .
\end{align}

Assembling everything together we arrive at \eq{F_mag_eq_4} in the main text.


\section{Solution of the Evolution Equations and the Small-$x$ Asymptotics of the Flavor Non-Singlet Boer-Mulders function}

\label{sec:app_sol}

In this Appendix we solve Eqs.~\eqref{HS_evol} for two different forms of the impact-parameter integrated amplitudes. We first solve Eqs.~\eqref{HS_evol77} following the strategy presented in \cite{Kovchegov:2017jxc}. Differentiating the two equations yields
\begin{subequations}\label{HG1}
\begin{align}
& \pd_{\zeta} H (\zeta) = - \int\limits_0^\zeta d \xi' \, \Gamma (\zeta, \xi'), \label{HG1a} \\
& \pd_{\zeta} \Gamma (\zeta, \zeta') = - \int\limits_0^{\zeta'} d \xi' \, \Gamma (\zeta, \xi'). \label{HG1b}
\end{align}
\end{subequations}
Introducing the inverse Laplace transforms 
\begin{align}\label{Laplace}
H (\zeta) = \int \frac{d \omega}{2 \pi i} \, e^{\omega \zeta} \, H_\omega, \ \ \ \Gamma (\zeta, \zeta') = \int \frac{d \omega}{2 \pi i} \, e^{\omega \zeta'} \, \Gamma_\omega (\zeta),
\end{align}
where the contours for integration are to the right of all the poles in $H_{\omega}$ and $\Gamma_{\omega}$, we can take the Laplace transform of \eq{HG1b} to find
\begin{align}\label{G1}
\pd_{\zeta} \Gamma_\omega (\zeta) = - \frac{1}{\omega} \,  \Gamma_\omega (\zeta) . 
\end{align}
The solution of \eq{G1} is
\begin{align}\label{Gsol1}
\Gamma_\omega (\zeta) = e^{-\frac{\zeta}{\omega}} \Gamma_{\omega} (0) ,
\end{align}
with $\Gamma_{\omega} (0)$ an initial condition parameter which we will determine from the initial condition $H (0)$. We can combine the solution \eqref{Gsol1} with Eqs.~\eqref{Laplace} and the $\Gamma (\zeta, \zeta ) = H (\zeta)$ condition to arrive at
\begin{subequations}\label{HG2}
\begin{align}
& H (\zeta) = \int \frac{d \omega}{2 \pi i} \, e^{\left( \omega - \frac{1}{\omega} \right) \, \zeta} \, \Gamma_\omega (0) , \\
&  \Gamma (\zeta, \zeta') = \int \frac{d \omega}{2 \pi i} \, e^{ \omega \zeta' - \frac{\zeta}{\omega}} \, \Gamma_{\omega} (0). \label{HG2b}
\end{align}
\end{subequations}
Substituting Eqs.~\eqref{HG2} into both sides of Eq.~\eqref{HG1b} we arrive at the constraint 
\begin{align}\label{gamconst}
\int \frac{d \omega}{2 \pi i} \, e^{ - \frac{\zeta}{\omega} } \, \frac{1}{\omega} \,  \Gamma_{\omega} = 0  .
\end{align}
Equation \eqref{HG1a} similarly yields another constraint
\begin{align} \label{gamconst2}
    \int \frac{d \omega}{2 \pi i} \, e^{\left( \omega - \frac{1}{\omega} \right) \, \zeta} \, \omega \, \Gamma_\omega (0) = 0 .
\end{align}
To satisfy these, we make a power series ansatz
\begin{align}
    \omega \Gamma_{\omega} (0) = \sum_{n=-\infty}^\infty c_n \, \omega^n .
\end{align}
Following  \cite{Kovchegov:2017jxc}, we insert our power series ansatz into the constraints. The constraint in \eq{gamconst2} yields
\begin{align}
     \int \frac{d \omega}{2 \pi i} \, e^{\left( \omega - \frac{1}{\omega} \right) \, \zeta} \, \sum_{n=-\infty}^\infty c_n \omega^n &= \sum_{m = 0}^{\infty} \frac{(-\zeta)^m}{m!} \sum_{n = -\infty}^{\infty} \int \frac{d \omega}{2 \pi i} \, e^{ \omega \zeta} \frac{c_n}{\omega^{m - n}} \\
     &= \sum_{n = -\infty}^{\infty} c_n \sum_{m = n+1}^{\infty} \frac{(-1)^m (\zeta)^{2m -n -1}}{m! (m-n-1)!} \\
     &= \sum_{n = -\infty}^{\infty} c_n \, (-1)^{n+1} \, J_{n+1} \left(2 \zeta \right) = 0  \notag
\end{align}
The Bessel functions satisfy $J_{-n} (x) = (-1)^n J_n (x)$, so we have
\begin{align}
    c_{-1} J_0 \left( 2 \zeta \right) + \sum_{n=1}^\infty  \left( c_{n-1} (-1)^n - c_{-n-1} \right) J_{n} \left( 2 \zeta \right) = 0 ,
\end{align}
which, in turn, requires that $c_{-1} = 0$ and $c_{n} = (-1)^n c_{-n-2}$. We can now rewrite the ansatz as
\begin{align}
    \omega \Gamma_{\omega} (0) = \sum_{n = 0}^{\infty} c_n \left[ \omega^n + \frac{(-1)^n}{\omega^{n+2}} \right] ,
\end{align}
such that the constraint in \eq{gamconst} yields
\begin{align}
     \int \frac{d \omega}{2 \pi i} \, e^{ - \frac{\zeta}{\omega}} \, \frac{1}{\omega} \Gamma_{\omega} (0) = \int \frac{d \omega}{2 \pi i} \, e^{ - \frac{\zeta}{\omega}} \, \sum_{n = 0}^{\infty} c_n \left[ \omega^{n - 2} + \frac{(-1)^n}{\omega^{n+4}} \right] = 0 .
\end{align}
The integral only converges for the $n=0$ term, so only $c_0$ can be nonzero. We have
\begin{align}
     \int \frac{d \omega}{2 \pi i} \, e^{ - \frac{\zeta}{\omega}} \, c_0 \left[ \omega^{-2} + \omega^{-4} \right] =  0 .
\end{align}
Therefore,
\begin{align}
\Gamma_\omega (0) = \frac{c_0}{\omega^3} \, \left( 1 + \omega^2 \right) .
\end{align}
 We thus obtain
\begin{align}\label{Hsol}
    H ( \zeta ) &= \int \frac{\dd{\omega}}{2 \pi i} e^{\left( \omega - \frac{1}{\omega} \right) \zeta} \frac{c_0}{\omega^3} \, \left( 1 + \omega^2 \right) \\
    &= c_0 \sum_{m = 0}^{\infty} \frac{(-\zeta)^m}{m!} \int \frac{\dd{\omega}}{2 \pi i} e^{\omega \zeta} \, \left( \frac{1}{\omega^{m+3}} + \frac{1}{\omega^{m+1}} \right) \notag \\
    &= c_0 \sum_{m = 0}^{\infty} \frac{(-1)^m}{m!} \left[ \frac{\zeta^{2m + 2}}{(m+2)!} + \frac{\zeta^{2m}}{m!} \right] \notag \\
    &= c_0 \left[ J_0 \left( 2 \zeta \right) + J_2 \left( 2 \zeta \right) \right]  = c_0 \, \frac{J_1 (2 \zeta)}{\zeta}. \notag
\end{align}
Requiring that $H(0) = 1$ sets $c_0 = 1$, giving the result in \eq{Hsolved} of the main text.

To test the conclusion of no correction to the intercept of the flavor non-singlet Boer-Mulders function we derived from this result in the main text, let us assume that
\begin{subequations}\label{HS_defs6}
\begin{align}
& \int d^2 b_\perp \, H^{1}_{10} (z) = \frac{{\un x}_{10} \times {\un S}}{x_{10}^2} \, H^{1} (x_{10}^2, z ) ,  \label{HS_defs6a} \\
& \int d^2 b_\perp \, H^{2}_{10} (z) = \frac{{\un x}_{10} \cdot {\un S}}{x_{10}^2} \, H^{2} (x_{10}^2, z ) \label{HS_defs6b} 
\end{align}
\end{subequations}
instead of Eqs.~\eqref{HS_defs4}. Using \eq{HS_defs6a} in \eq{HS_evol_a}, and constructing the corresponding evolution equation for the neighbor dipole amplitude, we get (for either $H^{1}_{10} (z)$ or $H^{2}_{10} (z)$, along with the corresponding neighbor amplitude)
\begin{subequations}\label{HS_evol5}
\begin{align}
& H (x_{10}^2, z ) = H^{(0)} (x_{10}^2, z ) + \frac{\as \, N_c}{2 \pi} \,  \int\limits_\frac{1}{s x_{10}^2}^z \frac{dz'}{z'} \int\limits_{1/z' s}^{x_{10}^2} \frac{d x^2_{21}}{x_{21}^2} \, \left[ H (x_{21}^2, z' )  - \Gamma (x_{10}^2 , x_{21}^2 , z') \right] , \\
& \Gamma (x_{10}^2 , x_{21}^2 , z') = H^{(0)} (x_{10}^2, z' ) + \frac{\as \, N_c}{2 \pi} \,  \int\limits_\frac{1}{s x_{10}^2}^{z'} \frac{dz''}{z''} \int\limits_{1/z'' s}^{\min \left\{ x_{10}^2, x_{21}^2 \frac{z'}{z''} \right\} } \frac{d x^2_{32}}{x_{32}^2} \, \left[ H (x_{32}^2, z'' )  - \Gamma (x_{10}^2 , x_{32}^2 , z'') \right]  .
\end{align}
\end{subequations}

In the scaling form this yields
\begin{subequations}\label{HS_evol6}
\begin{align}
& H (\zeta) = 1 + \int\limits_0^\zeta d \xi \int_0^\xi d \xi' \left[ H (\xi') - \Gamma (\xi, \xi') \right], \label{HS_evol6a} \\
& \Gamma (\zeta, \zeta') = 1 + \int\limits_0^{\zeta'} d \xi \int_0^\xi d \xi' \left[ H (\xi') - \Gamma (\xi, \xi') \right] + \int\limits_{\zeta'}^\zeta d \xi \int_0^{\zeta'} d \xi' \left[ H (\xi') - \Gamma (\xi, \xi') \right] \\
& = H (\zeta') + \int\limits_{\zeta'}^\zeta d \xi \int_0^{\zeta'} d \xi' \left[ H (\xi') - \Gamma (\xi, \xi') \right]  . \notag
\end{align}
\end{subequations}

Differentiating we get
\begin{subequations}\label{HS_evol7}
\begin{align}
& \pd_\zeta H  (\zeta) = \int_0^\zeta d \xi' \left[ H (\xi') - \Gamma (\zeta, \xi') \right], \label{HS_evol7a} \\
&\pd_\zeta  \Gamma (\zeta, \zeta') = \int_0^{\zeta'} d \xi' \left[ H (\xi') - \Gamma (\zeta, \xi') \right]  . \label{HS_evol7b}
\end{align}
\end{subequations}
Performing the Laplace transform
\begin{align}
    H (\zeta) = \int \frac{d\omega}{2 \pi i} \, e^{\omega \zeta} \, H_\omega, \ \ \ \Gamma (\zeta, \zeta') = \int \frac{d\omega}{2 \pi i} \, e^{\omega \zeta'} \, \Gamma_\omega (\zeta), 
\end{align}
gives
\begin{align}
\pd_\zeta \Gamma_\omega (\zeta) = \frac{1}{\omega} \left[ H_\omega - \Gamma_\omega (\zeta) \right]
\end{align}
such that
\begin{align}
\Gamma_\omega (\zeta) = H_\omega + e^{-\zeta/\omega} \left[ \Gamma_\omega (0) - H_\omega \right]
\end{align}
and 
\begin{align}\label{H1_sol_1/2}
\Gamma (\zeta, \zeta') = H  (\zeta') + \int \frac{d\omega}{2 \pi i} \, e^{\omega \zeta' - \frac{\zeta}{\omega}} \left[ \Gamma_\omega (0) - H_\omega \right]. 
\end{align}
Since $\Gamma (\zeta', \zeta') = H (\zeta')$ we use the above technique (see also \cite{Kovchegov:2017jxc}) to conclude that
\begin{align}\label{series111}
\Gamma_\omega (0) - H_\omega = \sum_{n=0}^\infty c_n \left[ \omega^n + \frac{(-1)^n}{\omega^{n+2}}. \right]
\end{align}
Plugging \eq{H1_sol_1/2} into \eq{HS_evol7b} yields
\begin{align}\label{series222}
0 = \int \frac{d\omega}{2 \pi i} \, e^{- \frac{\zeta}{\omega}} \, \frac{1}{\omega} \left[ \Gamma_\omega (0) - H_\omega \right] = \sum_{n=0}^\infty c_n  \int \frac{d\omega}{2 \pi i} \, e^{- \frac{\zeta}{\omega}} \, \frac{1}{\omega} \left[ \omega^n + \frac{(-1)^n}{\omega^{n+2}}. \right],
\end{align}
where we have used \eq{series111} in the last step. Since none of the integrals on the right of \eq{series222} is convergent, we conclude that all $c_n =0$ for $n \ge 0$. We arrive at $\Gamma_\omega (0) - H_\omega = 0$ and, consequently,
\begin{align}
    H (\zeta') = \Gamma (\zeta, \zeta') = 1
\end{align}
as the solution of Eqs.~\eqref{HS_evol6}. In retrospect, clearly 
\begin{align}
    H (x_{21}^2, z ) = \Gamma (x_{10}^2 , x_{21}^2 , z) = H^{(0)} (x_{21}^2, z )
\end{align}
is a valid non-trivial solution of Eqs.~\eqref{HS_evol5}. These functions simply do not evolve.

The corresponding flavor non-singlet Boer-Mulders function \eqref{BM_NS_2} is going to be proportional to 
\begin{align}\label{h1_perp_aympt_2}
    h_1^{\perp \, \textrm{NS}} (x \ll 1, k_T^2) \sim x \, \int\limits_\frac{\Lambda^2}{s}^1 \frac{dz}{z} \, H^{(0)} (x_{10}^2, z ),
\end{align}
which, for $H^{(0)} (x_{10}^2, z ) = 1$, gives $h_1^{\perp \, \textrm{NS}} (x \ll 1, k_T^2) \sim x \, \ln (1/x)$, where the power of $x$ is in agreement with \eq{bm_evolved} in the main text (\eq{bm_evolved} does not include the slowly-varying functions of $x$ such as the powers of $\ln (1/x)$ which may depend on the initial condition, among other factors). Since the initial condition $H^{(0)} (x_{10}^2, z )$, in general, should not contain powers of $z s$, and should typically be a lower-order polynomial in $\ln (zs/\Lambda^2)$ \cite{Kovchegov:2015pbl,Kovchegov:2016zex}, the leading power of $x$ resulting from \eq{h1_perp_aympt_2} is $h_1^{\perp \, \textrm{NS}} (x \ll 1, k_T^2) \sim x$, again in agreement with \eq{bm_evolved}.


\section{Mass-Dependent Corrections to the Flavor Non-Singlet Boer-Mulders Function st Small $x$}

\label{sec:app_mass}

In this Appendix we explore the quark mass corrections to the flavor non-singlet Boer-Mulders function at small $x$. Once again, we use \eq{h1q2}
with the Dirac matrix element of the $\pm$-reversed BL spinors now including the light-quark mass terms \cite{Kovchegov:2018zeq} (cf. \eq{mel})
\begin{align}\label{mel2}
{\bar v}_{\chi_2} (k_2)  \thalf \gamma^5 \gamma^+ \gamma^1 v_{\chi_1} (k_1) = \frac{1}{2 \, \sqrt{k_1^- \, k_2^-}} \, \Big[ & \, - \chi_1 \, \delta_{\chi_1 \chi_2} \, (2 \, {\un S} \cdot {\un k}_1 \, {\un S} \cdot {\un k}_2 - {\un k}_1 \cdot {\un k}_2 - m^2) + i \, m \,  \delta_{\chi_1 \chi_2} \, {\un S} \times ({\un k}_1 - {\un k}_2)  \\ 
& - i \, \chi_1 \, \delta_{\chi_1, \, - \chi_2} \, ({\un S} \times {\un k}_1 \, {\un S} \cdot {\un k}_2 + {\un S} \cdot {\un k}_1 \, {\un S} \times {\un k}_2 ) - m \,  \delta_{\chi_1, \, - \chi_2} \, {\un S} \cdot ({\un k}_1 + {\un k}_2) \Big] ,  \notag
\end{align}
where, again, ${\un S} = {\hat x}$ is the unit vector in the direction of the proton spin and $m$ denotes the (light) quark mass. Employing the replacement \eqref{V_repl} we concentrate on the mass-dependent correction,
\begin{align}\label{BMm}
& \frac{k^y}{M_P} \, h_1^{\perp \, q} (x, k_T^2) \bigg|_{m-\textrm{dependent}} \subset  -\frac{2 p_1^+ \, m}{2 (2 \pi)^3} \int d^2 {\zeta_{\perp}}  d^2 {w_{\perp}} d^2 z_\perp \, \frac{ d^2{k_{1 \perp}} d{k_1^-}}{(2\pi)^3} e^{i \underline{k}_1 \cdot (\un{w} - \un{\zeta}) + i \underline{k} \cdot (\un{z} - \un{\zeta})} \theta (k_1^-) \notag \\
& \times \frac{1}{(x p_1^+ k_1^- + \underline{k}_1^2 ) (x p_1^+ k_1^- + \underline{k}^2)} 
 \sum_{\chi_1 , \chi_2} \Big[  i \,  \delta_{\chi_1 \chi_2} \, {\un S} \times ({\un k}_1 + {\un k})   -  \delta_{\chi_1, \, - \chi_2} \, {\un S} \cdot ({\un k}_1 - {\un k}) \Big]   \Big{\langle} \tord \tr \left[ V_{\underline{\zeta}} \, V^{\textrm{pol} \, \dagger}_{{\un z}, {\un w}; \chi_2 , \chi_1} \right] \Big{\rangle}  + \mbox{c.c.} .
\end{align}
We now have the $\delta_{\chi_1 \chi_2}$ and $\delta_{\chi_1, \, - \chi_2}$ structures, which can be eikonal or sub-eikonal \cite{Kovchegov:2021iyc}. The light quark mass dependence, which would likely result in the $m/k_T$ suppression factor, may be offset by the fact that the $x$-dependence of the accompanying operators is eikonal or sub-eikonal, instead of the sub-sub-eikonal one in \eq{bm_evolved}. Let us investigate this possibility. (In \eq{BMm} we neglect the squares of the quark masses in the propagator denominators as the order $m^2/k_T^2$ corrections, which are even smaller than the ones considered here.)

We start with the potential eikonal contribution to the Boer-Mulders function, which, as follows from \eq{BMm}, is
\begin{align}\label{BMm_eik}
& \frac{k^y}{M_P} \, h_1^{\perp \, q} (x, k_T^2) \bigg|_{m-\textrm{dependent}, \, \textrm{eikonal}} \subset  -\frac{4 p_1^+ \, m}{2 (2 \pi)^3} \int d^2 {\zeta_{\perp}}  d^2 {w_{\perp}}  \, \frac{ d^2{k_{1 \perp}} d{k_1^-}}{(2\pi)^3} e^{i (\underline{k}_1 + \underline{k}) \cdot (\un{w} - \un{\zeta}) } \theta (k_1^-) \notag \\
\times 
&  \, \frac{i \,   {\un S} \times ({\un k}_1 + {\un k}) }{(x p_1^+ k_1^- + \underline{k}_1^2 ) (x p_1^+ k_1^- + \underline{k}^2)}  \,  \Big{\langle} \tord \tr \left[ V_{\underline{\zeta}} \, V^{\dagger}_{{\un w}} \right] - \atord \tr \left[ V_{\underline{\zeta}} \, V^{\dagger}_{{\un w}} \right]  \Big{\rangle}  .
\end{align}
However, the eikonal Wilson lines obey the reflection property with respect to the final state cut, which implies that \cite{Mueller:2012bn,Kovchegov:2018znm,Kovchegov:2018zeq}
\begin{align}
\Big{\langle} \tord \tr \left[ V_{\underline{\zeta}} \, V^{\dagger}_{{\un w}} \right] - \atord \tr \left[ V_{\underline{\zeta}} \, V^{\dagger}_{{\un w}} \right]  \Big{\rangle} = 0 . 
\end{align}
Hence, the Boer-Mulders function receives no eikonal contribution, even for the mass-dependent terms.

We proceed by studying the sub-eikonal light quark mass-dependent contribution. Employing \eq{Vphase&mag} we rewrite \eq{BMm} as
 \begin{align}\label{BMm2}
& \frac{k^y}{M_P} \, h_1^{\perp \, q} (x, k_T^2) \bigg|_{m-\textrm{dependent}, \, \textrm{sub-eikonal}} \subset  -\frac{4 p_1^+ \, m}{2 (2 \pi)^3} \int d^2 {\zeta_{\perp}}  d^2 {w_{\perp}} \, \frac{ d^2{k_{1 \perp}} d{k_1^-}}{(2\pi)^3} e^{i (\underline{k}_1 + \underline{k}) \cdot (\un{w} - \un{\zeta})} \theta (k_1^-)  \notag \\
\times 
& \frac{1}{(x p_1^+ k_1^- + \underline{k}_1^2 ) (x p_1^+ k_1^- + \underline{k}^2)} \, \Big[  i \,  {\un S} \times ({\un k}_1 + {\un k}) \, \Big{\langle} \tord \tr \left[ V_{\underline{\zeta}} \, V^{\textrm{phase} \, \dagger}_{{\un w}; \un{k}, \un{k}_1} \right] - \atord \tr \left[ V_{\underline{w}}^\dagger \, V^{\textrm{phase}}_{{\un \zeta}; \un{k}, \un{k}_1} \right] \Big{\rangle}  \notag \\ 
& -  {\un S} \cdot ({\un k}_1 - {\un k}) \, \Big{\langle} \tord \tr \left[ V_{\underline{\zeta}} \, V^{\textrm{mag} \, \dagger}_{{\un w}} \right] + \atord \tr \left[ V_{\underline{w}}^\dagger \, V^{\textrm{mag}}_{{\un \zeta}} \right] \Big{\rangle} \Big]  .
\end{align}

Since we are calculating the Boer-Mulders function, the proton is unpolarized: hence, the averaging $\langle \ldots \rangle$ does not contain any preferred direction in the transverse plane. We thus conclude that the impact parameter integral gives
\begin{align}
\int d^2 b_\perp \, \Big{\langle} \tord \tr \left[ V_{\underline{\zeta}} \, V^{\textrm{mag} \, \dagger}_{{\un w}} \right] + \atord \tr \left[ V_{\underline{w}}^\dagger \, V^{\textrm{mag}}_{{\un \zeta}} \right] \Big{\rangle} = f (|{\underline{\zeta}} - {\underline{w}} |^2)
\end{align}
with some scalar function $f$, whose exact form is not important, and ${\un b} = ({\un w} + {\un \zeta})/2$. We see that the last line of \eq{BMm2} does not contain the two-dimensional Levi-Civita symbol $\epsilon^{ij}$ needed to obtain $k^y = \epsilon^{ij} S^i k^j$ on the left-hand side. Therefore the last line of \eq{BMm2} does not contribute to the Boer-Mulders function. We, therefore, discard it and write, with the help of \eq{Vkk1},
\begin{align}\label{BMm3}
& \frac{k^y}{M_P} \, h_1^{\perp \, q} (x, k_T^2) \bigg|_{m-\textrm{dependent}, \, \textrm{sub-eikonal}} \subset  -\frac{4 p_1^+ \, m}{2 (2 \pi)^3} \int d^2 {\zeta_{\perp}}  d^2 {w_{\perp}} \, \frac{ d^2{k_{1 \perp}} d{k_1^-}}{(2\pi)^3} e^{i (\underline{k}_1 + \underline{k}) \cdot (\un{w} - \un{\zeta})} \theta (k_1^-)   \\
\times 
& \frac{i \,  {\un S} \times ({\un k}_1 + {\un k})}{(x p_1^+ k_1^- + \underline{k}_1^2 ) (x p_1^+ k_1^- + \underline{k}^2)} \, \Big[ (k -k_1)^i \,  \Big{\langle} \tord \tr \left[ V_{\underline{\zeta}} \, V^{i \, \dagger}_{{\un w}} \right] - \atord \tr \left[ V_{\underline{w}}^\dagger \, V^{i}_{{\un \zeta}} \right] \Big{\rangle}  + \Big{\langle} \tord \tr \left[ V_{\underline{\zeta}} \, V^{[2] \, \dagger}_{{\un w}; \un{k}, \un{k}_1} \right] - \atord \tr \left[ V_{\underline{w}}^\dagger \, V^{[2]}_{{\un \zeta}; \un{k}, \un{k}_1} \right] \Big{\rangle} \Big] . \notag
\end{align}

Repeating the above steps for the anti-quark distribution we obtain
\begin{align}\label{BMm4}
& \frac{k^y}{M_P} \, h_1^{\perp \, {\bar q}} (x, k_T^2) \bigg|_{m-\textrm{dependent}, \, \textrm{sub-eikonal}} \subset  -\frac{4 p_1^+ \, m}{2 (2 \pi)^3} \int d^2 {\zeta_{\perp}}  d^2 {w_{\perp}} \, \frac{ d^2{k_{1 \perp}} d{k_1^-}}{(2\pi)^3} e^{i (\underline{k}_1 + \underline{k}) \cdot (\un{w} - \un{\zeta})} \theta (k_1^-)   \\
\times 
& \frac{i \,  {\un S} \times ({\un k}_1 + {\un k})}{(x p_1^+ k_1^- + \underline{k}_1^2 ) (x p_1^+ k_1^- + \underline{k}^2)} \Big[ (k -k_1)^i \,  \Big{\langle} \tord \tr \left[ V_{\underline{\zeta}}^\dagger \, V^{i}_{{\un w}} \right] - \atord \tr \left[ V_{\underline{w}} \, V^{i \, \dagger}_{{\un \zeta}} \right] \Big{\rangle}  + \Big{\langle} \tord \tr \left[ V_{\underline{\zeta}}^\dagger \, V^{[2]}_{{\un w}; \un{k}, \un{k}_1} \right] - \atord \tr \left[ V_{\underline{w}} \, V^{[2] \, \dagger}_{{\un \zeta}; \un{k}, \un{k}_1} \right] \Big{\rangle} \Big] . \notag
\end{align}

Subtracting \eq{BMm4} from \eq{BMm3} we see that the mass correction to the flavor non-singlet Boer-Mulders distribution is
\begin{align}\label{BMmNS}
& \frac{k^y}{M_P} \, h_1^{\perp \, \textrm{NS}} (x, k_T^2) \bigg|_{m-\textrm{dependent}, \, \textrm{sub-eikonal}} \subset  -\frac{4 p_1^+ \, m}{2 (2 \pi)^3} \int d^2 {\zeta_{\perp}}  d^2 {w_{\perp}} \, \frac{ d^2{k_{1 \perp}} d{k_1^-}}{(2\pi)^3} e^{i (\underline{k}_1 + \underline{k}) \cdot (\un{w} - \un{\zeta})} \theta (k_1^-)   \\
\times 
& \frac{i \,  {\un S} \times ({\un k}_1 + {\un k})}{(x p_1^+ k_1^- + \underline{k}_1^2 ) (x p_1^+ k_1^- + \underline{k}^2)} \, \Big[ (k -k_1)^i \,  \Big{\langle} \tord \tr \left[ V_{\underline{\zeta}} \, V^{i \, \dagger}_{{\un w}} \right] - \atord \tr \left[ V_{\underline{w}}^\dagger \, V^{i}_{{\un \zeta}} \right] - \tord \tr \left[ V_{\underline{\zeta}}^\dagger \, V^{i}_{{\un w}} \right] + \atord \tr \left[ V_{\underline{w}} \, V^{i \, \dagger}_{{\un \zeta}} \right] \Big{\rangle}  \notag \\ 
& + \Big{\langle} \tord \tr \left[ V_{\underline{\zeta}} \, V^{[2] \, \dagger}_{{\un w}; \un{k}, \un{k}_1} \right] - \atord \tr \left[ V_{\underline{w}}^\dagger \, V^{[2]}_{{\un \zeta}; \un{k}, \un{k}_1} \right] - \tord \tr \left[ V_{\underline{\zeta}}^\dagger \, V^{[2]}_{{\un w}; \un{k}, \un{k}_1} \right] + \atord \tr \left[ V_{\underline{w}} \, V^{[2] \, \dagger}_{{\un \zeta}; \un{k}, \un{k}_1} \right] \Big{\rangle} \Big] . \notag
\end{align}

Similar to the Sivers function, we define the dipole amplitudes 
\begin{subequations}\label{HNS_defs}
\begin{align}
& H^{NS \, i}_{10} (z) = \frac{1}{2 N_c} \, \mbox{Re} \, \llangle \tord  \tr \left[ V_{\underline 0} \, V^{i \, \dagger}_{{\un 1}} \right] - \tord  \tr \left[ V^{i}_{{\un 1}} \, V_{\underline{0}}^\dagger \right]  \rrangle, \\
& H^{NS \, [2]}_{10} (z) = \frac{1}{2 N_c} \, \mbox{Im} \, \llangle \tord  \tr \left[ V_{\underline{0}} \, V^{[2] \, \dagger}_{{\un 1}; \un{k}, \un{k}_1} \right] - \tord  \tr \left[ V^{[2]}_{{\un 1}; \un{k}, \un{k}_1} \, V_{\underline{0}}^\dagger \right]  \rrangle . 
\end{align}
\end{subequations}
Since the two-dimensional Levi-Civita symbol $\epsilon^{ij}$ is now present both on the left- and right-hand sides of \eq{BMmNS}, we conclude that these dipole amplitudes, integrated over all impact parameters, should not contain $\epsilon^{ij}$. Only the transverse vector ${\un x}_{10}$ can carry the $i$ index for the impact-parameter integrated $H^{NS \, i}_{10} (z)$. We, thus, write
\begin{subequations}\label{Hdecomp}
\begin{align}
& \int d^2 b_\perp H^{NS \, i}_{10} (z) = x_{10}^i \, H^{NS} (x_{10}^2, z) , \label{HdecompA} \\
& \int d^2 b_\perp H^{NS \, [2]}_{10} (z) = H^{NS \, [2]} (x_{10}^2, z).
\end{align}
\end{subequations}
Employing Eqs.~\eqref{Hdecomp} in \eq{HNS_defs} we arrive at the following expression for the mass correction to the flavor non-singlet sub-eikonal Boer-Mulders function, 
\begin{align}\label{BMmNS2}
& \frac{k^y}{M_P} \, h_1^{\perp \, \textrm{NS}} (x, k_T^2) \bigg|_{m-\textrm{dependent}, \, \textrm{sub-eikonal}} =  -\frac{16 N_c  \, m}{2 (2 \pi)^3} \int d^2 x_{10} \, \frac{ d^2{k_{1 \perp}}}{(2\pi)^3} e^{i (\underline{k}_1 + \underline{k}) \cdot \un{x}_{10} } \int\limits_\frac{\Lambda^2}{s}^1 \frac{dz}{z}  \\
\times 
& \frac{{\un S} \times ({\un k}_1 + {\un k})}{(x p_1^+ k_1^- + \underline{k}_1^2 ) (x p_1^+ k_1^- + \underline{k}^2)} \, \Big[ i \,  ({\un k} -{\un k}_1) \cdot {\un x}_{10} \,  H^{NS} (x_{10}^2, z) - H^{NS \, [2]} (x_{10}^2, z) \Big] . \notag
\end{align}

Next we need to construct the DLA evolution equations for the amplitudes $H^{NS} (x_{10}^2, z)$ and $H^{NS \, [2]} (x_{10}^2, z)$. Just like for the Sivers function, in anticipation of the large-$N_c$ limit, we will be working with the gluon sector only. The evolution for $H^{NS \, i}_{10} (z)$ obeys the same equation \eqref{F_NS_1} as $F^{NS \, i}_{10} (z)$. There is one difference though: since the mixing between $F^{NS \, i}_{10} (z)$ and $F^{NS \, \textrm{mag}}_{10} (z)$ in \eq{F_NS_1} involves $\epsilon^{ij}$, similar terms cannot contribute to the evolution of $H^{NS \, i}_{10} (z)$, which, after integration over impact parameters, should not contain $\epsilon^{ij}$ in order to contribute to the Boer-Mulders function (see \eqref{HdecompA}). Dropping those terms we see that the relevant evolution equation for $H^{NS \, i}_{10} (z)$ is
\begin{align}\label{H_NS_1}
& H^{NS \, i}_{10} (z) = H^{NS \, (0) \, i}_{10} (z) + \frac{\as N_c}{4 \pi^2}  \int\limits_{\frac{\Lambda^2}{s}}^z \frac{d z'}{z'}  \int d^2 x_2 \, \Bigg\{  \left[ \delta^{ij} \left( \frac{3}{x_{21}^2} -  2 \, \frac{{\un x}_{20} \cdot {\un x}_{21}}{x_{20}^2 \, x_{21}^2} - \frac{1}{x_{20}^2} \right)  - 2 \frac{x_{21}^i \, x_{20}^j}{x_{21}^2 \, x_{20}^2} \left( 2 \frac{{\un x}_{20} \cdot {\un x}_{21}}{x_{20}^2} + 1 \right)  \right. \notag \\ 
& \left. + 2 \frac{x_{21}^i \, x_{21}^j}{x_{21}^2 \, x_{20}^2} \left( 2 \frac{{\un x}_{20} \cdot {\un x}_{21}}{x_{21}^2} + 1 \right) + 2 \frac{x_{20}^i \, x_{20}^j}{x_{20}^4} - 2 \frac{x_{21}^i \, x_{21}^j}{x_{21}^4}   \right]  \,  \left[ - H^{NS \, j}_{21} (z') + H^{NS \, j}_{20} (z')  \right]  \Bigg\} \notag \\
&  + \frac{\as \, N_c}{2 \pi^2} \, \int\limits_{\frac{\Lambda^2}{s}}^z \frac{d z'}{z'} \, \int d^2 x_2 \, \frac{x_{10}^2}{x_{21}^2 \, x_{20}^2} \, \Bigg\{ H^{NS \, i}_{12} (z') - \Gamma^{NS \, i}_{10, 21} (z')  \Bigg\} 
\end{align}
with the corresponding neighbor dipole amplitude $\Gamma^{NS \, i}_{10, 21}$. Integrating \eq{H_NS_1} over impact parameters while employing \eq{HdecompA} yields
\begin{align}\label{H_NS_2}
& x_{10}^i \, H^{NS} (x_{10}^2, z) = x_{10}^i \, H^{NS \, (0)} (x_{10}^2, z)  + \frac{\as N_c}{4 \pi^2}  \int\limits_{\frac{\Lambda^2}{s}}^z \frac{d z'}{z'}  \int d^2 x_2 \, \Bigg\{  \left[ \delta^{ij} \left( \frac{3}{x_{21}^2} -  2 \, \frac{{\un x}_{20} \cdot {\un x}_{21}}{x_{20}^2 \, x_{21}^2} - \frac{1}{x_{20}^2} \right)    \right. \notag \\ 
& \left. - 2 \frac{x_{21}^i \, x_{20}^j}{x_{21}^2 \, x_{20}^2} \left( 2 \frac{{\un x}_{20} \cdot {\un x}_{21}}{x_{20}^2} + 1 \right) + 2 \frac{x_{21}^i \, x_{21}^j}{x_{21}^2 \, x_{20}^2} \left( 2 \frac{{\un x}_{20} \cdot {\un x}_{21}}{x_{21}^2} + 1 \right) + 2 \frac{x_{20}^i \, x_{20}^j}{x_{20}^4} - 2 \frac{x_{21}^i \, x_{21}^j}{x_{21}^4}   \right] \notag \\ 
& \times \,  \left[ - x_{21}^j \, H^{NS} (x_{21}^2, z')  + x_{20}^j \, H^{NS} (x_{20}^2, z')   \right]  \Bigg\} \notag \\
&  + \frac{\as \, N_c}{2 \pi^2} \, \int\limits_{\frac{\Lambda^2}{s}}^z \frac{d z'}{z'} \, \int d^2 x_2 \, \frac{x_{10}^2}{x_{21}^2 \, x_{20}^2} \, \Bigg\{  - x_{21}^i \, H^{NS} (x_{21}^2, z') - x_{10}^i \, \Gamma^{NS} (x^2_{10}, x^2_{21}, z')  \Bigg\}  ,  
\end{align}
where the decomposition \eqref{HdecompA} is valid for the neighbor dipole amplitude as well,
\begin{align}
\int d^2 b_\perp \Gamma^{NS \, i}_{10, 21} (z) = x_{10}^i \, \Gamma^{NS} (x^2_{10}, x^2_{21}, z').
\end{align}

We begin with the first term in the kernel of \eq{H_NS_2}: it contains no UV $2 \to 0$ logarithm. There is no IR logarithm at $x_{21} \sim x_{20} \gg x_{10}$ either.  The logarithmic UV divergence at $2 \to 1$ gives 
\begin{align}\label{term1}
\frac{\as N_c}{2 \pi} \int\limits_{\frac{1}{s \, x_{10}^2}}^z \frac{d z'}{z'}  \int\limits^{x_{10}^2}_\frac{1}{z' s} \frac{d x_{21}^2}{x_{21}^2} \, x_{10}^i \, \Gamma^{NS} (x^2_{10}, x^2_{21}, z').
\end{align}

The second (eikonal) term in the kernel of \eq{H_NS_2} also contains no UV $2 \to 0$ logarithm and no IR logarithm. When $2 \to 1$, we get
\begin{align}\label{term2}
- \frac{\as N_c}{2 \pi} \int\limits_{\frac{1}{s \, x_{10}^2}}^z \frac{d z'}{z'}  \int\limits^{x_{10}^2}_\frac{1}{z' s} \frac{d x_{21}^2}{x_{21}^2} \, x_{10}^i \, \Gamma^{NS} (x^2_{10}, x^2_{21}, z'). 
\end{align}

We see that the terms in Eqs.~\eqref{term1} and \eqref{term2} cancel, and the DLA evolution following from \eq{H_NS_2} becomes trivial,
\begin{align}\label{H_NS_2_triv}
H^{NS} (x_{10}^2, z) = H^{NS \, (0)} (x_{10}^2, z) .
\end{align}
Hence the amplitude $H^{NS} (x_{10}^2, z)$ does not evolve, at least not at the DLA level. It contributes to the Boer-Mulders function only if $H^{NS \, (0)} (x_{10}^2, z) \neq 0$. Our preliminary calculation indicates that $H^{NS \, (0)} (x_{10}^2, z) = 0$ at the level of two- and three-gluon exchanges with an unpolarized target: hence, it is possible that $H^{NS} (x_{10}^2, z) =0$ and does not contribute to the Boer-Mulders function. A more detailed determination of the initial conditions is left for future work.

Last, we have to consider the DLA evolution for the amplitude $H^{NS \, [2]} (x_{10}^2, z)$. Using the decomposition \eqref{V2_full}, where, similar to the case of the Sivers function, we only keep the second (sub-eikonal) term on the right, along with the propagator \eqref{aaa2} we construct the following evolution equation
\begin{align}\label{H_NS_3}
& H^{NS \, [2]} (x_{10}^2, z) = H^{NS \, [2] \, (0)} (x_{10}^2, z) - \frac{\as}{2 N_c \pi^2}  \int\limits_{\frac{\Lambda^2}{s}}^z \frac{d z'}{z'}   \int d^2 x_2 \, d^2 b_\perp \, \left[ 2 \, \frac{({\un x}_{21} \cdot {\un x}_{20})^2}{x^2_{21} \, x^4_{20}} - \frac{1}{x_{21}^2}  - \frac{1}{x_{20}^2}  \right]  \\ 
& \times \textrm{Im} \llangle \left\{ \tord \tr \left[ t^b V_{\un 1} t^a V_{\un 0}^\dagger \right] -  \tord \tr \left[ t^b V_{\un 0} t^a V_{\un 1}^\dagger \right]  \right\}  \, \left( U_{\un 2}^{[2]} \right)^{ba}\rrangle (z') \\
& - \frac{\as}{4 N_c \pi^2}  \int\limits_{\frac{\Lambda^2}{s}}^z \frac{d z'}{z'}   \int d^2 x_2 \, d^2 b_\perp \left[ 2 \, x_{21}^i \,  \frac{{\un x}_{21} \cdot {\un x}_{20}}{x^2_{21} \, x^2_{20}} - \frac{x_{21}^i}{x_{21}^2}  - \frac{x_{20}^i}{x_{20}^2}  \right] \notag \\
& \times \textrm{Im} \llangle \left\{ \tord \tr \left[ t^b (\pd_1^i V_{\un 1}) t^a V_{\un 0}^\dagger \right] -  \tord \tr \left[ t^b V_{\un 0} t^a (\pd_1^i V_{\un 1}^\dagger) \right]  \right\}  \, \left( U_{\un 2}^{[2]} \right)^{ba}\rrangle (z') \notag \\ 
& + \frac{\as \, N_c}{2 \pi^2}  \int\limits_\frac{\Lambda^2}{s}^z \frac{dz'}{z'} \int d^2 x_2 \,  \frac{x_{10}^2}{x_{21}^2 \, x_{20}^2} \, \left[ H^{NS \, [2]} (x^2_{12}, z') - \Gamma^{NS \, [2]} (x^2_{10} , x^2_{21} , z') \right], \notag
\end{align}
where, for simplicity, the last term is taken in the large $N_c$ limit and is linearized (by putting $S=1$). (We have also dropped the delta-function terms like $\delta^2 ({\un x}_{21})$ in the kernel, since they do not contribute to small-$x$ evolution \cite{Cougoulic:2022gbk}.) 

One can readily see that the first kernel in 
\eq{H_NS_3}, 
\begin{align}
    2 \, \frac{({\un x}_{21} \cdot {\un x}_{20})^2}{x^2_{21} \, x^4_{20}} - \frac{1}{x_{21}^2}  - \frac{1}{x_{20}^2},
\end{align}
contains neither IR ($x_{21} \sim x_{20} \gg x_{10}$) nor UV ($2 \to 0$, $2 \to 1$) logarithms. Hence it does not contribute to the DLA evolution. To assess the second kernel, we note that the derivative $\pd_1^i$ in the Wilson lines multiplying it would end up acting on the amplitude for the dipole $12$: in the linearized approximation this dipole amplitude has to be ``polarized", otherwise it will be put to 1 ($S=1$) and will be zero after differentiation. The impact-parameter integrated dipole amplitude for the dipole $12$ will be a function of ${\un x}_{21}$ only. We thus can replace $\pd_1^i \to - \pd_2^i$ and integrate by parts, making this derivative act on the kernel in the square brackets. We will get a kernel proportional to   
\begin{align}
   \pd_2^i  \left[ 2 \, x_{21}^i \,  \frac{{\un x}_{21} \cdot {\un x}_{20}}{x^2_{21} \, x^2_{20}} - \frac{x_{21}^i}{x_{21}^2}  - \frac{x_{20}^i}{x_{20}^2}  \right] = 2 \, \frac{({\un x}_{21} \cdot {\un x}_{20})^2}{x^2_{21} \, x^4_{20}} - \frac{{\un x}_{21} \cdot {\un x}_{20}}{x_{21}^2 \, x_{20}^2}  - \frac{1}{x_{20}^2}
\end{align}
which also has neither IR ($x_{21} \sim x_{20} \gg x_{10}$) nor UV ($2 \to 0$, $2 \to 1$) logarithms and, therefore, does not contribute to the DLA evolution. 

We conclude that only the last (eikonal) term contributes at DLA, giving us the following equations for the amplitude $H^{NS \, [2]} (x_{10}^2, z)$ and for its corresponding neighbor amplitude:
\begin{subequations}\label{HS_NS_4}
\begin{align}
& H^{NS \, [2]} (x_{10}^2, z ) = H^{NS \, [2] \, (0)} (x_{10}^2, z ) + \frac{\as \, N_c}{2 \pi} \,  \int\limits_\frac{1}{s x_{10}^2}^z \frac{dz'}{z'} \int\limits_{1/z' s}^{x_{10}^2} \frac{d x^2_{21}}{x_{21}^2} \, \left[ H^{NS \, [2]} (x_{21}^2, z' )  - \Gamma^{NS \, [2]} (x_{10}^2 , x_{21}^2 , z') \right] , \\
& \Gamma^{NS \, [2]} (x_{10}^2 , x_{21}^2 , z') = H^{NS \, [2] \, (0)} (x_{10}^2, z' ) + \frac{\as \, N_c}{2 \pi}  \int\limits_\frac{1}{s x_{10}^2}^{z'} \frac{dz''}{z''} \!\!\!\!\!\!\!\! \int\limits_{1/z'' s}^{\min \left\{ x_{10}^2, x_{21}^2 \frac{z'}{z''} \right\} } \frac{d x^2_{32}}{x_{32}^2} \, \left[ H^{NS \, [2]} (x_{32}^2, z'' )  - \Gamma^{NS \, [2]} (x_{10}^2 , x_{32}^2 , z'') \right]  .
\end{align}
\end{subequations}
These equations are equivalent to Eqs.~\eqref{HS_evol5}, which have been solved in Appendix~\ref{sec:app_sol} with the solution being given by the inhomogenous terms/initial conditions. 
\begin{align}\label{HS_NS_5}
    H^{NS \, [2]} (x_{10}^2, z ) = \Gamma^{NS \, [2]} (x_{10}^2 , x_{21}^2 , z') = H^{NS \, [2] \, (0)} (x_{10}^2, z ). 
\end{align}
We see that the last amplitude $H^{NS \, [2]} (x_{10}^2, z)$ in the mass correction to the Boer-Mulders function also does not evolve at DLA. 

Additionally, similar to the sub-eikonal contribution to the Sivers function \cite{Kovchegov:2021iyc}, one can show that the lowest-order contribution to $H^{NS \, [2] \, (0)}$ is zero in the gluon sector. However, it may be non-zero in the quark-exchange sector \cite{Meissner:2007rx}. Therefore, we can conclude that while $H^{NS \, [2]} (x_{10}^2, z )$ may or may not be zero, it certainly does not evolve under the DLA large-$N_c$ evolution. 


\providecommand{\href}[2]{#2}\begingroup\raggedright\endgroup


\end{document}